\documentclass[aip,reprint,amsmath,amssymb]{revtex4-1} %,linenumbers

\usepackage[utf8]{inputenc}
\usepackage[T1]{fontenc}
\usepackage{mathptmx}
\usepackage{etoolbox}

\usepackage{graphicx}% Include figure files
\usepackage{bm}% bold math

\usepackage{txfonts}
\usepackage{bbold}
\usepackage{relsize}
\usepackage{mathtools}
\usepackage{bigints}
\usepackage{sidecap}
\usepackage[caption=false]{subfig}
\usepackage{setspace}
\usepackage[usenames,dvipsnames]{color}
\usepackage{soul}
\usepackage{comment}
\usepackage[normalem]{ulem}
\usepackage{multirow}
\usepackage{booktabs}

\usepackage{macros_ub}

\usepackage{hyperref}

\makeatletter
\def\@email#1#2{%
 \endgroup
 \patchcmd{\titleblock@produce}
  {\frontmatter@RRAPformat}
  {\frontmatter@RRAPformat{\produce@RRAP{*#1\href{mailto:#2}{#2}}}\frontmatter@RRAPformat}
  {}{}
}%
\makeatother
\begin{document}

\preprint{AIP/123-QED}

\title[Non-thermal particle acceleration in kinetic plasmas]{Non-thermal particle acceleration in multi-species kinetic plasmas: universal power-law distribution functions and temperature inversion in the solar corona}
% Force line breaks with \\
\author{Uddipan Banik}
\email{uddipanbanik@ias.edu}
\affiliation{Institute for Advanced Study, Einstein Drive, Princeton, NJ 08540, USA}
 \altaffiliation[Also at ]{Department of Astrophysical Sciences, Princeton University, 112 Nassau Street, Princeton, NJ 08540, USA \\ Perimeter Institute for Theoretical Physics, 31 Caroline Street N., Waterloo, Ontario, N2L 2Y5, Canada}%Lines break automatically or can be forced with \\
\author{Amitava Bhattacharjee}%
 \email{amitava@princeton.edu}
\affiliation{ 
Department of Astrophysical Sciences, Princeton University, 112 Nassau Street, Princeton, NJ 08540, USA%\\This line break forced with \textbackslash\textbackslash
}%

\date{\today}% It is always \today, today,

\begin{abstract}

Non-thermal power-law distribution functions are ubiquitous in astrophysical, space, and laboratory kinetic plasmas, but their origin remains unclear. A related puzzle is the temperature inversion of the solar corona. We show that these phenomena are deeply connected by developing a self-consistent quasilinear theory for electromagnetically driven, unmagnetized kinetic plasmas. The theory yields a multi-species Fokker--Planck equation containing drive-induced diffusion, due to direct acceleration by broad-band turbulent or narrow-band wave-like fields and indirect acceleration by the waves excited, together with Balescu--Lenard diffusion and drag from self-generated Debye-scale fluctuations and Coulomb collisions. For a super-Debye turbulent electric-field spectrum, $|{\bf E}_{\bf k}|^2\propto k^{-\alpha}$, both electron and ion distributions relax toward a universal $f(v)\propto v^{-5}$, or $N(E)\propto E^{-2}$, attractor, equivalent to the high-energy fall-off of a $\kappa=1.5$ distribution, provided $\alpha\ge5$. This universality follows from Debye screening: large-scale fields accelerate unscreened fast particles but not screened slow ones. For shallower spectra, $\alpha<5$, the tail scales as $v^{-\alpha}$; incomplete relaxation and anisotropy can further break universality. Anisotropic wave drives produce branch- and spectrum-dependent exponents. Collisions do not efficiently decelerate suprathermal particles, so high-velocity tails resist Maxwellianization. In the solar atmosphere, such tails may be generated by chromospheric convection or nanoflares despite collisional and radiative losses. Direct wave heating preferentially energizes electrons through Landau-resonant interactions with whistler and electron-cyclotron waves, while ions may be accelerated by turbulent ambipolar fields. The resulting $\kappa\simeq1.5$--$3$ distributions naturally produce an abrupt upper-chromosphere/lower-corona temperature transition and velocity-filtration-driven inverted profiles, yielding coronal temperatures $\sim10^6\,{\rm K}$.

%The suprathermal particles escape sun's gravity (velocity filtration), inverting the temperature profile and raising it to $10^6$ K. A proper analysis of velocity filtration with a $\kappa \approx 1.5-2$ distribution inspired by QLT provides a reasonable fit to the spectroscopic data of heavy ions and explains the abrupt temperature rise, a consequence of the divergence of pressure in the $\kappa \to 1.5$ limit.

%Significance statement
%Non-thermal power-law tails are widely observed in kinetic plasmas, yet their physical origin remains unresolved. We show that Debye screening of large-scale electromagnetic fluctuations naturally produces robust power-law distribution functions. For sufficiently steep super-Debye turbulent spectra, both electrons and ions relax toward a universal $f(v)\propto v^{-5}$ tail, whereas shallower spectra or anisotropic wave drives produce non-universal exponents. In the solar atmosphere, these tails survive collisional relaxation and enable velocity filtration, offering a kinetic explanation for both the abrupt temperature rise from the upper chromosphere to the lower corona and the inverted coronal temperature profile.

\end{abstract}

\maketitle

%\begin{quotation}
%The ``lead paragraph'' is encapsulated with the \LaTeX\ 
%\verb+quotation+ environment and is formatted as a single paragraph before the first section heading. 
%(The \verb+quotation+ environment reverts to its usual meaning after the first sectioning command.) 
%Note that numbered references are allowed in the lead paragraph.
%
%The lead paragraph will only be found in an article being prepared for the journal \textit{Chaos}.
%\end{quotation}

\section{\label{sec:intro}Introduction}

How collisionless plasmas relax to specific distribution functions (DFs) has been a persistent mystery. Collisions can drive a plasma towards a universal thermal/Maxwellian DF, a consequence of the Boltzmann H-theorem or the second law of thermodynamics. Collisionless (kinetic) plasmas, on the other hand, are governed by the Vlasov equation that admits infinitely many steady-state solutions. However, even in kinetic plasmas, while the fine-grained DF follows the Vlasov equation, the coarse-grained one can be shown to follow a (Balescu-Lenard type) kinetic equation with an effective collision operator whose steady-state solution is a Maxwellian. In other words, an effective H-theorem arises from coarse-graining. Yet, kinetic plasmas in nature and numerical simulations commonly harbor non-thermal DFs with Maxwellian cores but extended power-law tails. A particular power-law DF, $f_0(v)\sim v^{-5}$, where $v$ is the (non-relativistic) particle velocity, with a corresponding $E^{-2}$ energy ($E$) distribution \citep[][]{Fisk.Gloeckler.14}, has been inferred from heliospheric measurements in the solar wind by ACE and Ulysses \citep[][]{Fisk.Gloeckler.12} and in the heliosheath by Voyager \citep[][]{Krimigis.etal.77,Stone.etal.77}. More generally, direct fits to in-situ electron DFs show that $\kappa$-like suprathermal tails are widespread in the solar wind: Ulysses measurements established that many electron DFs are well fit by kappa distributions \citep[][]{Maksimovic.etal.97}, while subsequent Helios, Wind, Cluster, and Ulysses studies tracked the radial evolution and non-thermal properties of the core, halo, and strahl populations from $\sim 0.3$ to several AU \citep[][]{Maksimovic.etal.05,Stverak.etal.08,Stverak.etal.09,Pierrard.etal.16}. Recent Parker Solar Probe measurements further show that non-Maxwellian electrons are already present deep in the inner heliosphere, and quasi-thermal-noise spectroscopy reveals that the fitted $\kappa$ index varies systematically over $12$--$76\,R_\odot$ with three distinct radial regimes \citep[][]{Zheng.etal.24}. At much larger heliocentric distances, the smallest $\kappa$ values appear to occur in the outer heliosphere and inner heliosheath: IBEX-based analyses find that most inferred $\kappa$ indices are in the range $\sim 1.5$--$2.5$ ($f_0(v) \sim v^{-5}-v^{-7}$ at suprathermal $v$) \citep[][]{Livadiotis.etal.11,Livadiotis.etal.22}, consistent with the transport analysis of \citet[][]{Livadiotis.McComas.23}. Related evidence now also comes from the Jovian system, where Juno/JADE measurements in Jupiter's magnetodisc and near the Galilean moons are well described by bi-$\kappa$ or anisotropic $\kappa$ populations \citep[][]{Pelcener.etal.24,Sarkango.etal.25,LeLiboux.etal.25,Ellis.etal.26}. In particular, \citet[][]{Sarkango.etal.25} find that the cold and hot magnetodisc electron populations around Ganymede are predominantly close to $\kappa=1.5$, while that around Callisto is fit well by a single $\kappa\approx 1.5$ distribution. \citet[][]{Pelcener.etal.24} obtain bi-$\kappa$ fits at Ganymede's orbit with a cold component near $\kappa \approx 1.5$. These observations suggest that hard suprathermal tails with $\kappa$ of order a few, and in some environments approaching $3/2$ ($f_0(v)\sim v^{-5}$ at large $v$), are a generic feature of weakly collisional space plasmas including the solar wind.

The theory proposed to explain the universality of such DFs \citep[][]{Fisk.Gloeckler.14} in the solar wind has been questioned on both theoretical and observational grounds \citep[][]{Jokipii.Lee.10}, especially because the observed DFs show some variability, either due to collisions or a variety of plasma waves and instabilities not considered in the theory. Despite the ubiquity of these $\kappa$-like and power-law distributions across the heliosphere and planetary magnetospheres, there is still no rigorous first-principles explanation for why weakly collisional plasmas so often develop such hard suprathermal tails. This paper is one of a sequence of recent papers dedicated towards uncovering the rich physics of the relaxation of kinetic plasmas and explaining the apparent near-universality of non-thermal power-law tails.

Our recent \textit{self-consistent} treatment of electrostatic kinetic plasmas demonstrates the presence an attractor DF, $f_0(v)\sim v^{-5}$, in a plasma relaxing under large-scale electric field fluctuations. Using quasilinear theory (QLT), we showed that this DF is an outcome of the self-consistent plasma response or Debye shielding that naturally yields a $v$ dependent particle acceleration mechanism. Earlier test-particle treatments \citep[][]{Fisk.Gloeckler.14,Jokipii.Lee.10} did not include the self-consistent Maxwell's equations. Whether this asymptotic steady state is measured by \textit{in situ} spacecrafts might depend on the collisional age of the plasma and electromagnetic (EM) effects, but the general agreement between our theoretical result and the spacecraft measurements suggests that our treatment captures essential aspects of kinetic plasma relaxation despite certain simplifying assumptions. This power-law tail also emerges in a kinetic plasma subject to electrostatic turbulence arising from the two-stream instability, as recently demonstrated by a particle-in-cell (PIC) simulation \citep[][]{Ewart.etal.25}. Simulations of magnetic reconnection and collisionless shocks show similar power-law tails \citep[][]{Sironi.Spitkovsky.14,Hesse.etal.18,Hoshino.22,Gupta.etal.24,Wong.etal.25}.  

The implications of non-thermal power-law DFs are potentially far reaching. Such plasma kinetic effects may play an important role in the abrupt transition from the solar chromosphere to the corona and in coronal temperature inversion. Regarding the coronal heating problem, the following questions have puzzled heliophysicists for a long time. Why is the corona significantly hotter than the chromosphere? How is an inverted temperature profile sustained despite the flow of energy from low to high temperatures in apparent violation of the second law of thermodynamics? Why is the temperature gradient so steep in the transition zone between the chromosphere and the corona? In other words, why is the transition region as thin as a few hundred kilometers? This is much thinner than the solar tachocline that separates the radiative and convective zones. In standard fluid models, the coronal transition region is often understood as arising from the interplay of strong radiative cooling and field-aligned heat conduction\citep[][]{Klimchuk.06}. However, a simple local conduction--radiation balance suffers from a breakdown of the local conductivity closure\citep[][]{Spitzer.Harm.53} especially near the transition region with steep temperature gradients\citep[][]{Luciani.etal.83,Arber.etal.23}. This motivates exploring a kinetic alternative. To explain the origin of temperature inversion in the corona, a physical mechanism called velocity filtration was proposed by \citet[][]{Scudder.92a,Scudder.92b}, which suggests that the solar gravitational field can filter out the high velocity particles from a non-thermal $\kappa$ type distribution at the coronal base and preferentially allow these suprathermal particles to escape outwards. While a Maxwellian distribution entails an isothermal plasma, a $\kappa$ distribution with finite $\kappa$ results in an inverted (effective) temperature profile through velocity filtration \citep[][]{Scudder.92a,Scudder.92b,Scudder.94}. Although the underlying assumption of a $\kappa$ distribution is consistent with spacecraft measurements across the heliosphere \citep[][]{Meyer-Vernet.07}, sustaining such a distribution in the moderately collisional environment of the corona is apparently difficult because the collisional mean free path is around two orders of magnitude smaller than the pressure scale height, implying that $\kappa$ distributions might Maxwellianize in this locale. Theory and idealized kinetic simulations \citep[][]{Anderson.94,Landi.Pantellini.01} demonstrate that, unless shallow power-law tails ($\kappa < 4$) are assumed at the coronal base \citep[][]{Scudder.94}, which is difficult to justify from the perspective of conventional wisdom, the temperature inversion cannot be explained.

It is evident from the discussion above that the issue of steady-state distributions of kinetic plasmas is closely connected to the solar transition region and coronal temperature inversion. Motivated by this connection, we develop a general quasilinear theory (QLT)\citep[][]{Diamond_Itoh_Itoh_2010,Banik.Bhattacharjee.24a,Banik.Bhattacharjee.25}\footnote{The self-consistent QLT we have developed so far applies to (electron only) electrostatic kinetic plasmas as well as collisionless self-gravitating systems like cold dark matter halos.} for the relaxation of electromagnetically driven kinetic plasmas. We investigate the simultaneous, self-consistent evolution of multiple species (electrons and ions) using a Balescu-Lenard (BL)-type framework for kinetic plasma turbulence and weak binary collisions. We study the relaxation of unmagnetized kinetic plasmas (evolving from a drive-free initial state), primarily under an isotropic turbulent drive acting on super-Debye scales. We also briefly discuss an anisotropic coherent-wave extension, in which EM waves excited in the presence of a guide field on super-Larmor scales cascade down and heat the plasma on sub-Larmor scales. It should be borne in mind though that only the quiescent region of the corona strictly falls within the regime of validity of QLT; violent instabilities and highly turbulent scenarios are not properly captured by this approach. Using this theory, we show that dielectric polarization or Debye shielding is the key phenomenon behind producing a $v^{-5}$ tail in the DF of both electrons and ions and the $E^{-2}$ tail in their energy distribution when the electric-field power spectrum of the drive is sufficiently steep, ${|\bE_\bk|}^2\propto k^{-\alpha}$ with $\alpha\geq 5$. The Debye screening of large-scale (super-Debye) fields suppresses particle heating at low velocities, which yields a $v^4$ diffusion coefficient and a steady-state DF, $f_0(v)\sim v^{-5}$ (beyond the thermal speed). For shallower spectra ($\alpha<5$), $f_0(v)\sim v^{-\alpha}$ (at large $v$) in the steady state. Steep $\alpha\geq 5$ spectra are not generic to all turbulence, but are possible in special regimes, such as in strong Langmuir turbulence\citep[][]{Sun.etal.22} and around the Debye scale \citep[][]{Nastac.etal.23,Nastac.etal.25,Ewart.etal.25}. For coherent anisotropic wave drives, the tail depends on the wave branch and anisotropic spectral indices. The power-law tail thus produced survives even in the presence of weak Coulomb collisions, since the collision frequency falls off at large $v$.

This helps address a major challenge for velocity filtration in the solar atmosphere, namely how hard suprathermal tails can be maintained despite collisional relaxation. Our focus here is not on velocity filtration itself, which \citet[][]{Scudder.92a,Scudder.92b,Scudder.94} has worked out in great detail. Rather, this paper provides a robust mechanism for generating and sustaining a non-thermal power-law tail at the coronal base, which is crucial for filtration to operate, but the origin of which has not been explained by \citet[][]{Scudder.92a,Scudder.92b,Scudder.94}. One central question, though, is: what triggers these high energy suprathermal tails? In other words, what heats the corona? We find that, for typical solar parameters, the electrons may be heated by resonant interactions with electron-scale whistler and electron-cyclotron waves (Landau resonance), which could be associated with turbulent reconnection in the form of nano-flares \citep[][]{Parker.88}. But the protons and heavier ions generally cannot be directly heated by these waves. They may instead be accelerated by the turbulent ambipolar electric field generated as the electrons are displaced relative to the ions, which acts in addition to, and can be much stronger than, the weak DC ambipolar field that keeps the gravitationally bound background plasma quasi-neutral. Ion cyclotron heating is another possible mechanism but is beyond the scope of the unmagnetized treatment of the plasma response in this paper. For fiducial low-coronal parameters, we show that the quasilinear tail-formation time is typically much shorter than the characteristic transport time across a pressure scale height as well as the radiative cooling time, which implies that such tails can indeed develop in the lower corona.

This paper is organized as follows. In section~\ref{sec:resp_theory}, we discuss the linear and quasilinear theories for the relaxation of kinetic plasmas. Using this theory, we derive in section~\ref{sec:trans_eq} a general transport equation for non-thermal particle acceleration (NTPA) under isotropic turbulent and anisotropic coherent-wave drives, and present the criteria for the emergence of the non-thermal tail. We then explore the implications of this theory for (1) steep transition from the chromosphere to the corona, and (2) coronal temperature inversion through velocity filtration in a stratified solar atmosphere, in section~\ref{sec:corona}. We summarize our findings in section~\ref{sec:conclusion}. For convenience, Table~\ref{tab:notation} summarizes the notation used most frequently in the paper.

\begin{table*}[t]
\small
\centering
\renewcommand{\arraystretch}{1.15}
\caption{Frequently used notation.}
\label{tab:notation}
\begin{ruledtabular}
\begin{tabular}{ll}
{\bf Symbol} & {\bf Meaning} \\
\hline
$f_\rms(\bx,\bv,t)$ & Distribution function of species $\rms$. \\
$f_{\rms 0},\,f_{\rms 1},\,f_{\rms 2}$ & Mean, linear, and second-order parts of $f_\rms$. \\
$q_\rms,\;m_\rms,\;n_\rms$ & Charge, mass, and number density of species $\rms$. \\
$\epsilon,\;\Lambda$ & Small expansion/quasilinear ordering parameter, and plasma parameter. \\
$\bE^{(\rmP)},\;\bB^{(\rmP)}$ & Drive electric and magnetic fields. \\
$\sigma_\rms,\;\sigma_{\rms 0}$ & Thermal speed of species $\rms$, and its reference value at $r_0$. \\
$\omega_{\rmP\rms}$ & Plasma frequency of species $\rms$. \\
$\lambda_{\rmD e}$ & Electron Debye length. \\
$\lambda_{\rmc \rms}$ & Larmor radius of species $\rms$. \\
$\nu_{\rmc \rms}$ & Collision frequency of species $\rms$. \\
$\beta_\rms$ & Plasma beta for species $\rms$. \\
$\varepsilon_{\bk\parallel},\;\varepsilon_{\bk\perp}$ & Longitudinal and transverse dielectric components. \\
$D_{ij}^{(\rms)},\;D_{p\,ij}^{(\rms)},\;D_{w\,ij}^{(\rms)}$ & Full, direct, and wave-mediated drive diffusion tensors. \\
$\calD_{ij}^{(\rms)},\;\calD_i^{(\rms)}$ & BL diffusion and drag tensors. \\
$D^{(\rms)}(v),\;D_p^{(\rms)}(v),\;D_w^{(\rms)}(v)$ & Isotropic drive diffusion coefficients. \\
$\calD_1^{(\rms)}(v),\;\calD_2^{(\rms)}(v)$ & Isotropic BL drag and diffusion coefficients. \\
$\varmathbb{P}_{\bk\,ij},\;\varmathbb{Q}_{\bk\,ij}$ & Kernels in the direct and wave-mediated diffusion tensors. \\
$\omega_\bk,\;\eta_\bk,\;\gamma_\bk$ & Least-damped mode frequency and its real and imaginary parts. \\
$\omega_k,\;\eta_k,\;\gamma_k$ & Isotropic least-damped mode frequency and its real and imaginary parts. \\
$\calE_{ij}(\bk),\;\calE(k)$ & Tensor and isotropic drive power spectra. \\
$\calC_t(t-t'),\;\calC_\omega(\omega)$ & Temporal correlation function and its Fourier transform. \\
$t_\rmc$ & Correlation time of turbulent drive. \\
$\alpha$ & Spectral index of the turbulent drive, $\calE(k)\sim k^{-\alpha}$. \\
$k_{\rm min},\;k_\rmc$ & Minimum and cutoff drive wavenumbers. \\
$\kappa$ & $\kappa$-distribution index. \\
$r_0$ & Reference radius, usually the coronal base. \\
$\Phi_\rmG,\;\Phi_\rmE,\;\Phi_{\rm eff}^{(\rms)}$ & Gravitational, electrostatic, and effective potentials. \\
$v_{\rm esc},\;v_{\rm max},\;v_{{\rm max},\rms}$ & Escape speed, generic truncation speed, and species-dependent truncation speed. \\
$E_{\rm drive},\;E_{\rm int}$ & Drive and internal electric-field amplitudes. \\
$t_{\rm diff}^{(\rms)},\;t_{\rm tr}^{(\rms)}$ & Quasilinear tail-formation time and transport time. \\
$t_{\rm ff}^{(\rms)},\;t_{\rm line}^{(\rms)},\;t_{e\rms}^{(\rm E)}$ & Free-free cooling, line-cooling, and electron--ion energy-exchange times. \\
$v_\parallel,\;v_\perp,\;\xi_\parallel,\;\xi_\perp$ & Guide-field velocity components and anisotropy parameters. \\
$\alpha_\parallel,\;\alpha_\perp,\;\delta_\parallel,\;\delta_\perp$ & Spectral indices of the anisotropic wave drive. \\
$D^{(\rms)}_{\parallel\parallel},\;D^{(\rms)}_{\perp\perp},\;D^{(\rms)}_{\parallel\perp}$ & Components of the anisotropic wave-drive diffusion tensor. \\
\end{tabular}
\end{ruledtabular}
\end{table*}

\section{\label{sec:resp_theory}Relaxation theory for weakly collisional plasmas} 

\subsection{Governing equations}

A plasma is characterized by the DF or phase space ($\bx,\bv$) density of particles of the ${\rms}^{\rm th}$ charged species, $f_\rms(\bx,\bv,t)$. The governing equations for a weakly collisional plasma are the Boltzmann-Maxwell equations. The Boltzmann equation,

\begin{align}
&\frac{\partial f_\rms}{\partial t} + \bv\cdot{\bf \nabla} f_\rms + \frac{q_\rms}{m_\rms}\,{\bf \nabla}_\bv f_\rms \cdot \left[\left(\bE^{(\rmP)}+\bE\right) + \frac{\bv}{c}\times \left(\bB^{(\rmP)}+\bB\right)\right] \nonumber\\
&= C\left[f_\rms\right],
\label{Vlasov_eq}
\end{align}
evolves the DF of each charged species (electrons and ions), with $q_\rms$ and $m_\rms$ the electric charge and mass of each species. Here, $\bE$ and $\bB$ are the self-generated electric and magnetic fields, sourced by the DF via Maxwell's equations,

\begin{align}
&\nabla\cdot\bE = {4 \pi} \sum_{\rms} q_\rms \int \rmd^3 v\, f_\rms,\nonumber\\
&\nabla \times \bB = \frac{4\pi}{c} \sum_{\rms} q_\rms \int \rmd^3 v\, \bv f_\rms + \frac{1}{c}\frac{\partial \bE}{\partial t},\nonumber\\
&\nabla\cdot\bB = 0,\quad \nabla\times\bE = -\frac{1}{c}\frac{\partial \bB}{\partial t}.
\label{Maxwell_eqs}
\end{align}
We assume charge neutrality in equilibrium, i.e., the total equilibrium charge density is zero. $\bE^{(\rmP)}$ and $\bB^{(\rmP)}$ are perturbing electric and magnetic fields that we hereafter refer to as the drive. These fields may be sourced externally to the plasma under consideration, or may arise self-consistently from super-Debye-scale structures and fluctuations acting on its smooth distribution. Since plasma is an open system, it is nearly impossible to switch off such fields entirely. Background turbulence spontaneously arising from nonlinear wave activity or coherent structures (electrostatic, e.g., Bernstein-Greene-Kruskal (BGK) modes \citep[][]{Bernstein.etal.57} or electromagnetic, e.g., plasmoids \citep[][]{Loureiro.etal.07,Bhattacharjee.etal.09,Comisso.etal.16}), copiously present within the solar plasma, can act as part of this drive. We assume the collision operator to be of the Balescu-Lenard form,
\begin{align}
C[f] = \frac{\partial}{\partial v_i}\left(\calD^{(\rms)}_{ij}\left(\bv\right)\frac{\partial f_\rms}{\partial v_j} + \calD^{(\rms)}_i\left(\bv\right) f_\rms\right),
\end{align}
with the diffusion ($\calD^{(\rms)}_{ij}$) and drag ($\calD^{(\rms)}_i$) coefficients given by equation~(\ref{BL_coeff_app}). In the present context, the BL term captures the cumulative effect of self-generated small-scale (Debye-scale) fluctuations together with Coulomb collisions, whereas the drive refers to super-Debye-scale fields. We do not include an equilibrium guide magnetic field. More precisely, our analysis is a local one that is restricted to scales smaller than the electron Larmor radius $\lambda_{\rmc e} = m_e \sigma_e c /e B_0$, where $\sigma_e$ is the electron thermal speed and $B_0$ is the strength of the guide field. In other words, the coarse-graining scale we later on adopt to compute the smoothed DF is $\lesssim \lambda_{\rmc e}$. The EM drive can, however, be generated on large super-Larmor scales, and cascade down to small sub-Larmor scales.

The nonlinear Boltzmann-Maxwell equations are difficult to solve in their full generality, and demand the use of perturbation theory to make analytical progress. If the strength of the electrostatic potential, $\Phi^{(\rmP)} = -\int \bE^{(\rmP)}\cdot \rmd \bx$, is smaller than $m_\rms\sigma^2_\rms/|q_\rms|$, where $\sigma_\rms$ is the velocity dispersion or thermal velocity of the $\rms^{\rm th}$ species in the unperturbed near-equilibrium system, then the perturbation in $f_\rms$ can be expanded as a power series in the small perturbation parameter, $\epsilon \sim \max_{\rms} \{|q_\rms\Phi^{(\rmP)}|/m_\rms\sigma^2_\rms\}$, i.e., $f_\rms = f_{\rms 0} + \epsilon f_{\rms 1} + \epsilon^2 f_{\rms 2} + ...\,$; $\bE$ and $\bB$ can also be expanded accordingly, assuming $\bE^{(\rmP)}$ and $\bB^{(\rmP)}$ to be $\calO(\epsilon)$. We perform a Fourier transform with respect to $\bx$ and Laplace transform with respect to $t$ of $f_{\rms i}$, $\bE_i$, $\bE^{(\rmP)}$, $\bB_i$ and $\bB^{(\rmP)}$ in the ($i^{\rm th}$ order) perturbed Vlasov-Maxwell equations, to derive the response of the system order by order. 

\subsection{Linear theory}\label{sec:LT}

The Fourier-Laplace coefficients of the linear response for a weakly collisional plasma, in the large mean free path limit of $\nu_{\rmc\rms} \ll k\sigma_\rms$, where $\nu_{\rmc\rms} \sim \omega_{\rmP\rms} \ln\Lambda/\Lambda$ is the collision frequency ($\omega_{\rmP\rms}$ is the plasma frequency and $\Lambda$ is the plasma parameter or number of particles within the Debye sphere), can be expressed as

\begin{align}
&\Tilde{f}_{\rms 1\bk}(\bv,\omega) = -\frac{iq_\rms}{m_\rms} \, \frac{\left(\Tilde{\bE}^{(\rmP)}_{\bk}(\omega) + \Tilde{\bE}_{1\bk}(\omega)\right)\cdot {\partial f_{\rms 0}}/{\partial \bv}}{\omega - \bk\cdot\bv} + \frac{i f_{\rms 1 \bk}\left(\bv,0\right)}{\omega - \bk\cdot\bv},\nonumber\\
&\Tilde{E}_{\bk i}(\omega) = \Tilde{E}^{(\rmP)}_{\bk i}(\omega) + \Tilde{E}_{1\bk i}(\omega) = \varepsilon^{-1}_{\bk\, ij}(\omega)\,\left[{\Tilde{\bE}^{(\rmP)}_{\bk j}(\omega)} + g_{\bk j}(\omega)\right],
\label{lin_resp_eq_app}
\end{align}
with the dielectric tensor ${\varepsilon}_{\bk}(\omega) = {\rm diag}\left(\varepsilon_{\bk \perp},\varepsilon_{\bk \perp},\varepsilon_{\bk \parallel}\right)$ given by

\begin{align}
&\varepsilon_{\bk \parallel}(\omega) = 1 + \sum_{\rms}\frac{\omega^2_{\rmP\rms}}{k^2} \int \rmd^3 v\, \frac{\bk\cdot{\partial f_{\rms 0}}/{\partial \bv}}{\omega - \bk\cdot\bv},\nonumber\\
&\varepsilon_{\bk \perp}(\omega) = 1 - 
\frac{\omega}{c^2k^2-\omega^2} \sum_\rms {\omega^2_{\rmP\rms}} \int \rmd^3 v\, \frac{v_{\perp}{\partial f_{\rms 0}}/{\partial v_\perp}}{\omega - \bk\cdot\bv},
\label{eps_k_app}
\end{align}
and the vector $g_{\bk}(\omega) = \left(g_{\bk\perp},g_{\bk\perp},g_{\bk\parallel}\right)$, corresponding to the initial perturbation, given by

\begin{align}
&g_{\bk\parallel}(\omega) = \frac{4 \pi}{k} \sum_{\rms} {n_\rms q_\rms} \int \rmd^3 v\, \frac{f_{\rms 1\bk}\left(\bv,0\right)}{\omega - \bk\cdot\bv},\nonumber\\
&g_{\bk\perp}(\omega) = -\frac{4 \pi\omega}{c^2 k^2 - \omega^2} \sum_{\rms} {n_\rms q_\rms} \int \rmd^3 v\, \frac{v_\perp f_{\rms 1\bk}\left(\bv,0\right)}{\omega - \bk\cdot\bv}.
\label{g}
\end{align}
Here $\omega_{\rmP\rms} = \sqrt{4\pi n_\rms q^2_\rms/m_\rms}$ is the plasma (electron Langmuir) frequency or that of the ion Langmuir waves, $n_\rms$ being the number density of the charged species. The subscript $\bk$ stands for the Fourier transform in $\bx$, and the tilde represents the Laplace transform in $t$. Without loss of generality, we have assumed $\bk = k\,\hat{\bz}$ with the perpendicular space spanned by $\hat{\bx}$ and $\hat{\by}$, and $v_{\perp}$ denoting the component of $\bv$ perpendicular to $\bk$, i.e., either $v_x$ or $v_y$. We have assumed that the DF $f_{\rms 0}$ is isotropic in $\bv$, i.e., $f_{\rms 0}(\bv) = f_{\rms 0}(v)$, such that $\partial f_{\rms 0}/\partial \bv = \partial f_{\rms 0}/\partial v \, \hat{\bv}$ and therefore the magnetic Lorentz force is zero at linear order. Thus we ignore all plasma dynamics related to DF anisotropies, e.g., the Weibel instability, which we leave for future investigation.

The dielectric tensor represents (i) the longitudinal polarization of the medium that manifests as Debye shielding/screening of the electric field as well as longitudinal plasma waves (electron Langmuir and ion acoustic and Langmuir waves) parallel to $\bk$ and (ii) the transverse EM or light waves, perpendicular to $\bk$, that are modulated by plasma oscillations. The real part of the longitudinal component $\varepsilon_{\bk \parallel}$ universally scales as $1 - \omega^2_{\rmP e}/\omega^2$ for $\omega \gg k\sigma_e$, independent of the detailed functional form of $f_{\rms 0}$, a property that plays a crucial role in NTPA, as we will demonstrate shortly. 

The zeros of the dielectric tensor denote the Landau modes or waves that oscillate and (Landau) damp due to collective wave-particle interactions. The electron Langmuir waves, with the dispersion relation $\omega^2 \approx \omega^2_{\rmP e} + 3 k^2\sigma^2_e$, and the ion Langmuir and ion acoustic waves, with $\omega^2 \approx k^2 c^2_\rms/(1+k^2\lambda^2_{\rmD e})$, emerge from the zeros of $\varepsilon_{\bk \parallel}\left(\omega\right)$. Here, $\omega_{\rmP e} = \sqrt{4\pi n_e e^2/m_e}$ is the electron plasma frequency ($n_e=$ electron number density), $c_\rms = \sqrt{k_\rmB Z T_e/m_i}$ is the ion sound speed ($T_e$ = electron temperature, $Z$ the atomic number of the dominant ionic species), and $\lambda_{\rmD e}=\sigma_e/\omega_{\rmP e}$ is the electron Debye length. The electron Langmuir and ion acoustic and Langmuir waves are longitudinal waves, with electric field oscillations parallel to $\bk$, that get Landau damped via wave-particle interactions. The transverse light waves, with EM oscillations perpendicular to $\bk$, arise from the zeros of $\varepsilon_{\bk \perp}\left(\omega\right)$, follow the dispersion relation $\omega^2 \approx c^2 k^2 + \sum_\rms \omega^2_{\rmP\rms}\left(1+\sigma^2_\rms/c^2\right)$, and do not undergo Landau damping due to the absence of superluminal particles and therefore wave-particle resonances ($\omega = \bk\cdot\bv$) that might damp them.

In the small mean free path limit $(\nu_{\rmc e} \gg k\sigma_\rms)$, the response can still be described by equations~(\ref{lin_resp_eq_app}) for $v \gg \nu_{\rmc e}/k$. The functional form of the dielectric tensor is, however, modified. Assuming a simplified Lenard-Bernstein form for the collision operator \citep[][]{Lenard.Bernstein.58}, where the drag and diffusion coefficients are respectively $\calD_1^{(\rms)}(v) = \nu_{\rmc\rms} v$ and $\calD_2^{(\rms)}(v) = \nu_{\rmc\rms} \sigma^2_\rms$ (the Balescu-Lenard coefficients scale the same way at $v \lesssim \sigma_\rms$), the longitudinal component, in the limit of $\omega \gg k\sigma_e$, can be written as follows \citep[][]{Lenard.Bernstein.58,Banik.Bhattacharjee.24b}:

\begin{align}
\varepsilon_{\bk\parallel}(\omega) \approx 1 - \frac{\omega^2_{\rmP e}}{\omega^2}\dfrac{1}{1 + \dfrac{i\nu_{\rmc e}}{\omega}}.
\end{align}
Here we have neglected the sub-dominant ion term. For $\omega \gg \nu_{\rmc e} \gg k\sigma_e$, the real part scales as $1 - \omega^2_{\rmP e}/\omega^2$, just as in the large mean free path limit. Hence, the dielectric factor $\varepsilon_{\bk\parallel}(\bk\cdot\bv)$ that multiplies the acceleration of a charged particle with velocity $\bv$ in a plasma, scales as $1 - \omega^2_{\rmP e}/{\left(\bk\cdot\bv\right)}^2$ for the high $v$ particles, independent of whether the mean free path is small or large compared to the scale of the perturbation, $k^{-1}$. This is because, for these very fast particles, the free-streaming scale $v/\nu_{\rmc e}$ always exceeds $k^{-1}$, even if the mean free path or the average free-streaming scale is small. In other words, the fastest particles are unaffected by collisions.

\subsection{Quasilinear theory}\label{sec:QLT}

Linear theory tells us how small perturbations propagate as undulations on a smooth background DF $f_{\rms 0}$. However, the non-linear coupling between these perturbations itself modifies the mean DF. The evolution of the mean DF of each species, $f_{\rms 0} = {\left(2\pi\right)}^3 f_{\rms 2\,\bk=0}/V$, averaged over a volume $V$ of the bulk plasma, can be studied by computing the second order response, $f_{\rms 2\,\bk}$, taking the $\bk \to 0$ limit and ensemble averaging the response over the random phases of the linear fluctuations (see Appendix~A.2 of \citet[][]{Banik.Bhattacharjee.24a}). This yields the following quasilinear equation for each species:

\begin{align}
\frac{\partial f_{\rms 0}}{\partial t} &= -\frac{{\left(2\pi\right)}^3 q_\rms}{m_\rms V} \int \rmd^3 k \, \left<\left(\bE_{\bk}^{\ast} + \frac{\bv}{c}\times \bB_{\bk}^{\ast} \right) \cdot \nabla_{\bv} f_{\rms 1\bk}\right>,
\label{quasilin_resp_eq}
\end{align}
where $\bE_{\bk} = \bE_{1\bk} + \bE_{\bk}^{(\rmP)}$ and $\bB_{\bk} = \bB_{1\bk} + \bB_{\bk}^{(\rmP)}$, and the subscript $1$ denotes linear perturbations. We have used the reality condition, $\bE_{1,-\bk} = \bE_{1\bk}^{\ast}$, and similarly for the other quantities.

Now, we need to make assumptions about the temporal correlation of the perturbing EM fields, $E_{\bk i}^{(\rmP)}(t)$ and $B_{\bk i}^{(\rmP)}(t)$, where the subscript $i$ denotes the $i^{\rm th}$ component. Faraday's law (fourth of Maxwell's equations~[\ref{Maxwell_eqs}]) dictates that the electric and magnetic field perturbations are related by 
$\Tilde{\bB}_{\bk}^{(\rmP)}\left(\omega\right) = \frac{c}{\omega} \left(\bk\times\Tilde{\bE}_{\bk}^{(\rmP)}\left(\omega\right)\right)$. We assume that the perturbing electric field $E_{\bk i}^{(\rmP)}(t)$ is a generic turbulent field with a spatial power-spectrum and a temporal correlation (red noise):

\begin{align}
\left<E_{\bk i}^{(\rmP)\ast}(t) E_{\bk j}^{(\rmP)}(t')\right> &= \calE_{ij}\left(\bk\right)\,\calC_t\left(t-t'\right),
\label{white_noise_t}
\end{align}
where $\calC_t$ is the temporal correlation function. For white noise, this is simply $\delta\left(t-t'\right)$, whose Fourier transform is $\calC_{\omega}\left(\omega\right) = 1$. We shall now study the quasilinear evolution of the plasma due to this turbulent drive. In section~\ref{sec:wave_drive_diff}, we briefly discuss the case of a coherent wave drive.

Substituting the expressions for the linear quantities, $\bE_{\bk}(t)$ and $f_{\rms 1\bk}(\bv,t)$, obtained by performing the inverse Laplace transform of equations~(\ref{lin_resp_eq_app}) in the quasilinear equation~(\ref{quasilin_resp_eq}) above, and using the noise spectrum for the perturbing electric field given in equation~(\ref{white_noise_t}), we obtain a simplified form for the quasilinear transport equation for each charged species,

\begin{align}
\frac{\partial f_{\rms 0}}{\partial t} &= \frac{\partial}{\partial v_i}\left[\left(D_{ij}^{(\rms)}(\bv) +  \calD^{(\rms)}_{ij}(\bv)\right)\frac{\partial f_{\rms 0}}{\partial v_j} + \calD^{(\rms)}_i\left(\bv\right)\, f_{\rms 0} \right],
\label{quasilin_resp_FP_eq}
\end{align}
which is a Fokker-Planck equation with the diffusion tensor $D_{ij}^{(\rms)} + \calD_{ij}^{(\rms)}$ and the drag/friction tensor $\calD_i^{(\rms)}$. At this stage, $D_{ij}^{(\rms)}$ denotes the full diffusion tensor sourced by the turbulent EM drive and therefore, in general, contains both electric and magnetic contributions. However, in the isotropic setup adopted for computation, the explicit magnetic contribution drops out of the reduced transport coefficients, so that the diffusion is governed by the longitudinal electric fields (akin to electrostatic turbulence). The coefficients $\calD_{ij}^{(\rms)}$ and $\calD_i^{(\rms)}$ are the BL coefficients sourced by self-generated small-scale fluctuations and collisions. We discuss the detailed velocity dependence of these coefficients as follows.

%\begin{figure}[t!]
%\centering
%\includegraphics[width=0.85\textwidth]{eps_vs_k_e_ion.jpeg}
%\caption{Dielectric constant vs $v$}
%\label{fig:eps_vs_k}
%\end{figure}

\subsubsection{Turbulent drive diffusion}\label{sec:drive}

The drive diffusion tensor $D_{ij}^{(\rms)}$ is given by 
%(see Appendix~\ref{App:LT} for a discussion of when and why the Landau term can be neglected)

\begin{align}
&D_{ij}^{(\rms)}(\bv) \approx D_{p\,ij}^{(\rms)}(\bv) + D_{w\,ij}^{(\rms)}(\bv,t),
\label{diffusion_tensor}
\end{align}
where $D_{p\,ij}^{(\rms)}$ denotes the direct diffusion of the dressed (Debye shielded) particles by the EM drive and is given by
\begin{align}
&D_{p\,ij}^{(\rms)}(\bv) \approx \frac{8\pi^4 q^2_\rms}{m^2_\rms V} \int \rmd^3 k\, {\left[ \varepsilon^{-1}_{\bk}\left(\bk\cdot\bv\right)\, \varmathbb{P}_{\bk}\left(\bk\cdot\bv\right)\, \varepsilon^{-1\dagger}_{\bk}\left(\bk\cdot\bv\right) \right]}_{ij},\nonumber\\
&\varmathbb{P}_{\bk\, ij}\left(\bk\cdot\bv\right) = \frac{k_i v_l\, \calE_{lj}\left(\bk\right) \calC_{\omega}\left(\bk\cdot\bv\right)}{\bk\cdot\bv},
\label{diffusion_tensor_part}
\end{align}
and $D_{w\,ij}^{(\rms)}$ denotes the diffusion mediated by the Landau modes or waves excited by the drive and is given by
\begin{align}
&D_{w\,ij}^{(\rms)}(\bv,t) \approx \frac{8\pi^4 q^2_\rms}{m^2_\rms V} \int \rmd^3 k\, \frac{\left|\gamma_\bk\right|}{{\left(\eta_\bk - \bk\cdot\bv\right)}^2 + \gamma^2_\bk} \exp{\left[2\gamma_\bk t\right]} \nonumber\\
&\times {\varepsilon'^{-1}_{\bk\, im}\left(\omega_\bk\right)\, \varmathbb{Q}_{\bk\, mn}\, \varepsilon'^{-1\dagger}_{\bk\, nj}\left(\omega_\bk\right)},\nonumber\\
&\varmathbb{Q}_{\bk\, ij} = \left|\calC'_{\omega}(\omega_\bk)\right| \left[\calE_{ij}(\bk)\left(1 - \frac{\bk\cdot\bv}{\omega_\bk}\right) + \frac{k_i v_l}{\omega_\bk}\calE_{lj}(\bk)\right].
\label{diffusion_tensor_wave}
\end{align}
Note that both direct and wave-mediated diffusion are governed by the spatial power-spectrum $\calE_{ij}(\bk)$ and the temporal power-spectrum $\calC_\omega$. Here, $\omega_\bk = \eta_\bk + i\gamma_\bk$ is the frequency of the least damped Landau mode, prime denotes the derivative of a quantity with respect to its argument, and the dielectric constant is given by $\varepsilon_{\bk}\left(\omega\right) = {\rm diag}\left(\varepsilon_{\bk \perp},\varepsilon_{\bk \perp},\varepsilon_{\bk \parallel}\right)$ with its components given by equations~(\ref{eps_k_app}). 

The longitudinal component of $\varepsilon_{\bk}$ scales universally as $1 - \omega^2_{\rmP e}/{\left(\bk\cdot\bv\right)}^2$ ($\omega_{\rmP e} = \sqrt{4\pi n_e e^2/m_e}$ is the plasma frequency) over a large velocity range, $\sigma_e < v < \omega_{\rmP e}/k$, for super-Debye fields $\left(k\lambda_{\rmD e} \ll 1\right)$, independent of the detailed functional form of $f_{\rms 0}$. This implies that the particles do not experience the bare field. Rather, they are Debye-shielded or dressed, the slower ones even more so than the faster ones. Particles near-resonant with the Langmuir waves $(\omega_{\rmP e} \approx \bk\cdot\bv)$ experience the self-consistent polarizing field synchronously with the bare field, and are excited the most. The slower particles $(\omega_{\rmP e} > \bk\cdot\bv)$ are less heated, since the polarizing field Debye screens the large-scale field. The faster ones $(\omega_{\rmP e} < \bk\cdot\bv)$, on the other hand, are unscreened and directly heated by the bare field. This velocity-dependent diffusion of dressed/Debye screened particles by the turbulent EM drive is described by the direct, drive diffusion tensor $D_{p\,ij}^{(\rms)}$ given in equation~(\ref{diffusion_tensor_part}).

Contrary to equation~(\ref{diffusion_tensor_part}) that describes the direct diffusion of particles by the drive, equation~(\ref{diffusion_tensor_wave}) denotes the wave diffusion tensor that describes the indirect heating of particles by the waves excited by the drive. If the plasma is stable to perturbations, then the waves Landau-damp, the wave diffusion tensor $D_{w\,ij}^{(\rms)}$ dies away and only the dressed particle diffusion tensor $D_{p\,ij}^{(\rms)}$ survives at long time. This is strictly only true if the Landau damping timescale is shorter than that of quasilinear relaxation, which typically holds, as we show in Appendix~\ref{App:Landau_vs_QL}. If, on the other hand, the plasma is unstable or marginally stable to perturbations, i.e., there exit plateaus and/or bumps in the DF, then $\gamma_\bk \approx 0$ for the least damped mode (once an unstable bump has saturated to a plateau), for $k\lambda_{\rmD e} \lesssim 1$, and the waves keep contributing to diffusion along with direct heating. In this case, we have

\begin{align}
&D_{w\,ij}^{(\rms)}(\bv,t) \to D_{w\,ij}^{(\rms)}(\bv) \nonumber\\
&\approx \frac{8\pi^5 q^2_\rms}{m^2_\rms V} \int \rmd^3 k\, \delta\left(\eta_\bk - \bk\cdot\bv\right) \varepsilon'^{-1}_{\bk\, im}\left(\eta_\bk\right)\, \varmathbb{Q}_{\bk\, mn}\, \varepsilon'^{-1\dagger}_{\bk\, nj}\left(\eta_\bk\right),\nonumber\\
&\varmathbb{Q}_{\bk\, ij} = \frac{k_i v_l\, \calE_{lj}\left(\bk\right) \left|\calC'_{\omega}\left(\eta_\bk\right)\right|}{\eta_\bk}.
\label{diffusion_tensor_wave_long_time}
\end{align}
This is the classic expression for the diffusion tensor that describes the quasilinear saturation of instabilities or the `heating' of particles due to the Landau damping of waves. Notably, only the resonant particles with $\omega_\bk \approx \eta_\bk = \bk\cdot\bv$ significantly exchange energy with the waves, as indicated by the $\delta\left(\eta_\bk - \bk\cdot\bv\right)$ factor above. The total energy of the waves is, however, distributed equally between the resonant and non-resonant particles. While the electromagnetic potential energy of the waves is exchanged with the resonant particles, the kinetic energy is exchanged with the non-resonant particles of the bulk. Although each resonant particle exchanges much more energy with the waves than a non-resonant one, the much greater abundance of non-resonant than resonant particles implies that an equipartition of energy exchanged is possible between the resonant and non-resonant species. The quasilinear evolution of the mean DF occurs hand in hand with the linear Landau damping (growth) of the waves in the marginally stable (unstable) regime. $D^{(\rms)}_{w\,ij}$ therefore denotes the heating of particles due to wave-particle interactions.

\subsubsection{Wave drive diffusion}\label{sec:wave_drive_diff}

We now consider the direct diffusion produced by a narrow set of coherent waves (in the long-time limit) as opposed to the turbulent drive explored so far. We consider EM waves that are generated on super-Larmor scales and are typically anisotropic with respect to a guide field, $\bB_0=B_0\hat{\bz}$. The responding plasma itself is still treated as unmagnetized, i.e., we study the sub-Larmor response of the plasma to super-Larmor waves (that cascade down via wave-wave interactions). In the non-relativistic regime of interest, the transverse light-wave branch does not contribute to secular resonant diffusion, because the resonance condition $\omega_\bk=\bk\cdot\bv$ cannot be satisfied on that branch. The relevant contribution therefore comes only from the longitudinal projection of the drive electric field,
\[
\calE_L^{(\rmP)}(\bk)=\calE_{ij}^{(\rmP)}(\bk)\,\hat{k}_i\hat{k}_j.
\]

Denoting the parallel and perpendicular directions relative to $\bB_0$ by $\parallel$ and $\perp$ respectively, and writing $\bv=(v_\perp,0,v_\parallel)$ and $\bk=(k_\perp\cos\phi,k_\perp\sin\phi,k_\parallel)$, so that $\bk\cdot\bv=k_\parallel v_\parallel+k_\perp v_\perp\cos\phi$, the long-time direct diffusion tensor becomes
\begin{align}
D_{ij}^{(\rms)}(\bv)
\approx
\frac{8\pi^5 q_\rms^2}{m_\rms^2 V}
\int \rmd^3k\,
\hat{k}_i\hat{k}_j\,
\frac{\calE_L^{(\rmP)}(\bk)}
{|\varepsilon_{\bk\parallel}(\omega_\bk)|^2}\,
\delta\!\left(\omega_\bk-\bk\cdot\bv\right).
\label{eq:Dij_wave_drive_main}
\end{align}
This expression is closely related to equation~(\ref{diffusion_tensor_wave_long_time}), since both describe diffusion due to resonant wave-particle interactions in the long-time limit. While equation~(\ref{eq:Dij_wave_drive_main}) describes diffusion due to a wave drive, equation~(\ref{diffusion_tensor_wave_long_time}) describes that due to the self-consistent plasma waves generated by a broad-band turbulent drive. After averaging over the gyrophase, we obtain explicit expressions for
$D_{\parallel\parallel}^{(\rms)}$, $D_{\perp\perp}^{(\rms)}$, and
$D_{\parallel\perp}^{(\rms)}=D_{\perp\parallel}^{(\rms)}$.
We defer these expressions to Appendix~\ref{App:wave_drive}. 

\subsubsection{Balescu-Lenard diffusion and drag}\label{sec:BL}

The quasilinear equation~(\ref{quasilin_resp_FP_eq}) not only describes particle diffusion due to the drive and the waves excited by it, but also the diffusion and drag due to self-generated small-scale fluctuations and Coulomb collisions. The latter is described by the BL diffusion and drag coefficients, given by

\begin{align}
&\calD_{ij}^{(\rms)}\left(\bv\right) = \frac{\pi}{m^2_\rms}{\left(4\pi q_\rms q_{\rms'}\right)}^2 \sum_{\rms'} \int \frac{\rmd^3 k}{{\left(2\pi\right)}^3} \frac{k_i k_j}{k^4} \nonumber\\
&\times \frac{1}{{\left|\varepsilon_{\bk\parallel}\left(\bk\cdot\bv\right)\right|}^2} \int \rmd^3 v'\, \delta\left(\bk\cdot\left(\bv-\bv'\right)\right) f_{\rms'0}\left(\bv'\right),\nonumber\\
&\calD_i^{(\rms)}\left(\bv\right) = -\frac{\pi}{m_\rms}{\left(4\pi q_\rms q_{\rms'}\right)}^2  \sum_{\rms'} \frac{1}{m_{\rms'}} \int \frac{\rmd^3 k}{{\left(2\pi\right)}^3} \frac{k_i k_j}{k^4} \nonumber\\
&\times \frac{1}{{\left|\varepsilon_{\bk\parallel}\left(\bk\cdot\bv\right)\right|}^2} \int \rmd^3 v'\, \delta\left(\bk\cdot\left(\bv-\bv'\right)\right) \frac{\partial f_{\rms' 0}}{\partial v'_j}.
\label{BL_coeff_app}
\end{align}
Note that only the mutually resonant particles with $\bk\cdot\bv = \bk\cdot\bv'$ exchange energy and momentum efficiently via collisions. Only the longitudinal modes contribute to BL diffusion and drag under isotropic conditions in a non-relativistic plasma. In the relativistic regime, the transverse modes also contribute \citep[][]{Silin.61}. While the diffusion tensor describes relaxation due to the exchange of energy between the particles via collisions and self-generated small-scale fluctuations, the drag tensor describes the exchange of momentum. As we shall see, the BL coefficients generally dominate at small $v$, whereas the drive diffusion coefficient dominates at large $v$ for a strong enough super-Debye scale drive.

%The $v$ dependence of the BL coefficients are compared to that of the drive diffusion coefficient in Fig.~\ref{fig:D_vs_v}. Note that, at high enough $v$, a large-scale drive always wins.

\section{\label{sec:trans_eq}A general transport equation for particle acceleration in unmagnetized plasmas}

We shall now discuss the quasilinear relaxation of an isotropic plasma $\left(f_{\rms 0}(\bv) = f_{\rms 0}(v)\right)$. If we also assume a spatially isotropic turbulent drive\footnote{Note that the assumption of isotropic fluctuations, $\calE_{ij}(\bk) = \calE(k)\,\hat{k}_i\hat{k}_j$, is different from that in \citet[][]{Banik.Bhattacharjee.24a}. This, however, does not make any qualitative difference in the main results of the paper.}, i.e.,

\begin{itemize}
    \item Isotropic turbulent drive: $\calE_{ij}(\bk) = \calE(k)\,\hat{k}_i\hat{k}_j$
    \item Isotropic DF: $f_{\rms 0}(\bv) = f_{\rms 0}(v)$,
\end{itemize}
with $\hat{k}_i = k_i/k$, then the dimensionality of the problem is greatly reduced. An isotropic DF is justified on scales smaller than the Larmor radius; an isotropic drive can be used to model electrostatic turbulence (see section~\ref{sec:aniso_wave} and Appendix~\ref{App:wave_drive} for a discussion of anisotropic EM fluctuations). The multi-dimensional transport equation~(\ref{quasilin_resp_FP_eq}) then becomes the following one-dimensional equation in $v$:

\begin{align}
&\frac{\partial f_{\rms 0}}{\partial t} = \frac{1}{v^2}\frac{\partial}{\partial v}\left[v^2 \left( \left(D^{(\rms)}(v) + \calD^{(\rms)}_2(v)\right)\frac{\partial f_{\rms 0}}{\partial v} + \calD^{(\rms)}_1(v) f_{\rms 0} \right)\right].
\label{QL_eq}
\end{align}
The drive diffusion coefficient $D^{(\rms)}$ is equal to $D^{(\rms)}_p + D^{(\rms)}_w$ with $D^{(\rms)}_p$ the dressed particle diffusion coefficient and $D^{(\rms)}_w$ the wave diffusion coefficient. $\calD^{(\rms)}_2$ and $\calD^{(\rms)}_1$ denote the BL diffusion and drag coefficients respectively. In what follows, we discuss and compare the velocity dependence of these different coefficients. 

\subsection{Velocity dependence of the coefficients}\label{sec:coeff_compare}

\begin{figure*}
\centering
\includegraphics[width=0.9\textwidth]{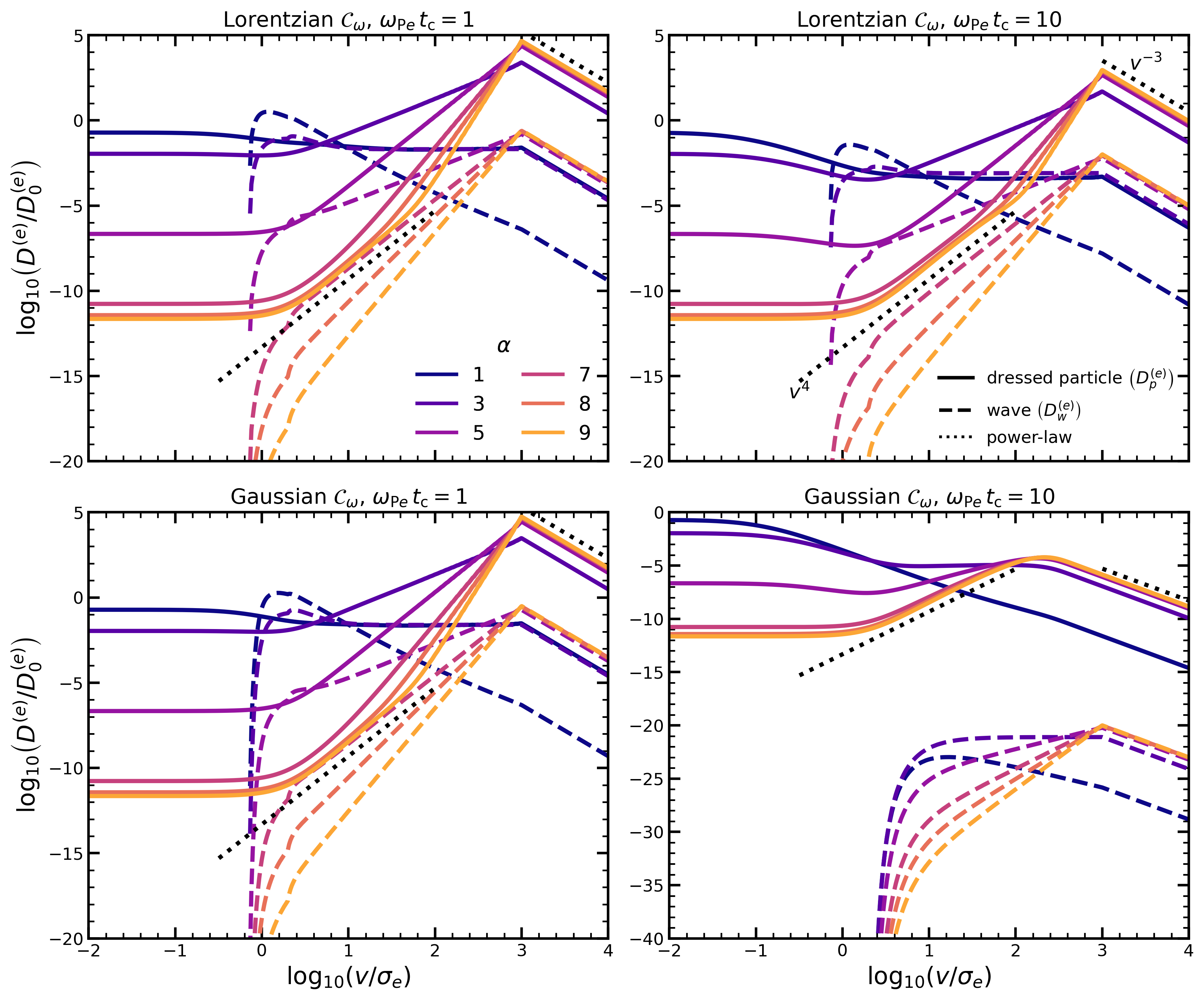}
\caption{Electron diffusion coefficient due to direct dressed-particle heating by the drive, $D_p^{(e)}(v)$ (solid lines), and indirect heating by the waves, $D_w^{(e)}(v)$ (dashed lines), in units of $D^{(e)}_0 = \left(32\pi^5 e^2 k^3_\rmc \calE_0/m_e^2 V\right)\left[(3-\alpha)/\left(1-{\left(k_{\rm min}/k_\rmc\right)}^{3-\alpha}\right)\right]$, as functions of $v$ for different values of $\alpha$, with the electric-field power spectrum of the drive given by $\calE(k)\sim k^{-\alpha}$ (equation~[\ref{Schechter_func}]). The top and bottom rows correspond to Lorentzian and Gaussian temporal correlation spectra, respectively, while the left and right columns correspond to $\omega_{\rmP e} t_\rmc = 1$ and $10$. The plasma is assumed to be marginally stable, and we adopt $T_e=T_i$ and $k_{\rm min}\lambda_{\rmD e}=10^{-3}$. The ion diffusion coefficients scale similarly with $v$. The dressed-particle coefficient typically exceeds the wave coefficient. $D_p^{(e)}(v)$ scales as $v^{\alpha-1}$ at $\sigma_e<v<\omega_{\rmP e}/k_{\rm min}$ for $\alpha < 5$, while, for $\alpha \ge 5$, it scales as $v^4$ at intermediate velocities and as $v^{\alpha-1}$ up to $v\sim \omega_{\rmP e}/k_{\rm min}$. $D_w^{(e)}(v)$ is zero below $\sigma_e$ due to the absence of wave-particle resonance, and scales as $v^{\alpha-3}$ beyond, up to $\omega_{\rmP e}/k_{\rm min}$. Both $D_p^{(e)}(v)$ and $D_w^{(e)}(v)$ fall off as $v^{-3}$ beyond this scale. The Gaussian correlation smooths out the turnovers and pushes the $v^4$ scaling of $D_p^{(e)}$ to $\alpha\geq 7$. The black dotted lines indicate the reference scalings $v^4$ and $v^{-3}$; for clarity, the lower-right panel is shown with a larger vertical range.}
\label{fig:D_vs_v_alpha}
\end{figure*}

The dressed particle diffusion coefficient $D^{(\rms)}_p$ can be obtained from equation~(\ref{diffusion_tensor_part}) with the assumption of isotropy, and is given by

\begin{align}
D^{(\rms)}_p(v) &= \frac{32\pi^5 q^2_\rms}{m^2_\rms V} \nonumber\\
&\times \int_0^{\infty} \rmd k\, k^2 \calE(k) \int_0^1 \rmd\cos{\theta}\,\cos^2\theta\, \frac{ \calC_{\omega}\left(k v \cos{\theta}\right)}{{\left|\varepsilon_{k\parallel}\left(kv\cos\theta\right)\right|}^2},
\label{Dp_iso}
\end{align}
with $\theta$ the angle between $\bk$ and $\bv$, and $\varepsilon_{k\parallel}$ the parallel component of the dielectric tensor that represents the Debye screening of large-scale EM perturbations. We assume a red noise with correlation time $t_\rmc$, i.e.,

\begin{align}
&\calC_t\left(t-t'\right) = \frac{1}{2 t_\rmc} \exp{\left[-\left|t-t'\right|/t_\rmc\right]}\nonumber\\
&\implies \calC_\omega(\omega) = \frac{1}{1 + \omega^2 t^2_\rmc},
\label{red_noise_model}
\end{align}
Later on we explore the case of a Gaussian correlation, $\calC_t\sim \exp{\left[-0.5 (t/t_\rmc)^2\right]}$, for which $\calC_\omega\sim \exp{\left[-0.5 \omega^2 t^2_\rmc\right]}$. Substituting $\varepsilon_{k\parallel}\left(kv\cos\theta\right) = 1 - \omega^2_{\rmP e}/k^2v^2\cos^2\theta$, we can perform the $\cos\theta$ integral and obtain a closed analytic form for $D^{(\rms)}_p(v)$ (derived in Appendix~\ref{App:Dp_calc}):

\begin{align}
&D^{(\rms)}_p(v) = \frac{32\pi^5 q^2_\rms}{m^2_\rms V} \int_0^{\infty} \rmd k\, k^2 \calE(k)\,\calI\left(kvt_\rmc,\omega_{\rmP e}t_\rmc\right),\nonumber\\
&\calI\left(kvt_\rmc,\omega_{\rmP e}t_\rmc\right) = \frac{1}{{\left(k v t_\rmc\right)}^2} \left[1 - \frac{1}{{\left(1+{\omega_{\rmP e}^2 t_\rmc^2}\right)}^2} \frac{\tan^{-1}\left(k v t_\rmc\right)}{k v t_\rmc} \right.\nonumber\\
&\left.+ \frac{\omega^2_{\rmP e}t^2_\rmc}{1 + \omega^2_{\rmP e}t^2_\rmc} \left(\frac{1}{2} \frac{\omega^2_{\rmP e}}{\omega^2_{\rmP e}-k^2 v^2} - \frac{5 + 3\omega^2_{\rmP e}t^2_\rmc}{4\left(1 + \omega^2_{\rmP e}t^2_\rmc\right)}\frac{\omega_{\rmP e}}{k v}\ln{\left(\left|\frac{\omega_{\rmP e} + kv}{\omega_{\rmP e} - kv}\right|\right)} \right)\, \right].
\end{align}
It is not hard to see that $D^{(\rms)}_p(v)$ scales as $\sim {\left|\varepsilon_{k\parallel}\right|}^{-2}$ and therefore develops a $v^4$ scaling in the range, $\sigma_e < v < \omega_{\rmP e}/k$ ($\calI \sim {\left(kv/\omega_{\rmP e}\right)}^4/7$), for a white noise drive with $\omega_{\rmP e} t_\rmc < 1$. Even for a generic red noise with $\omega_{\rmP e} t_\rmc > 1$, the $v^4$ scaling persists in the range, $\sigma_e < v < 1/k t_\rmc$. In this case, $D^{(\rms)}_p(v)$ harbors a shallower $v^2$ dependence at $1/k t_\rmc < v < \omega_{\rmP e}/k$. At larger $v$, $D^{(\rms)}_p(v)$ falls off as $v^{-3}$. The detailed asymptotic scalings with the prefactors are provided in equation~(\ref{Dp_asymptotic}).

Equation~(\ref{diffusion_tensor_wave}) describes the wave diffusion coefficient, which can be expressed as

\begin{align}
D^{(\rms)}_w(v) &= \frac{16\pi^6 q^2_\rms}{m^2_\rms V} \nonumber\\
&\times \int_0^{\infty} \rmd k\, k^2 \calE(k) \frac{\eta^2_k \left|\calC'_\omega\left(\omega_k\right)\right|}{{\left|\varepsilon'_{k\parallel}\left(\omega_k\right)\right|}^2}\, \calF_k\left(\omega_k,v\right)\, \exp{\left[2\gamma_k t\right]},
\label{Dw_gen_iso}
\end{align}
with $\calF_k\left(\omega_k,v\right)$ given by

\begin{align}
&\calF_k\left(\omega_k,v\right) = \left|\gamma_k\right| \int_{-1}^1 \rmd\cos\theta \frac{\cos^2\theta}{{\left(\eta_k - k v \cos\theta\right)}^2 + \gamma^2_k} \nonumber\\
&= \frac{1}{{\left(kv\right)}^3} \left[\left(\eta^2_k - \gamma^2_k\right)\left(\tan^{-1}\left(\frac{\eta_k + kv}{\left|\gamma_k\right|}\right) - \tan^{-1}\left(\frac{\eta_k - kv}{\left|\gamma_k\right|}\right)\right)\right.\nonumber\\
&\left.- \left|\gamma_k\right|\eta_k\ln{\left(\frac{{\left(\eta_k + kv\right)}^2 + \gamma^2_k}{{\left(\eta_k - kv\right)}^2 + \gamma^2_k}\right)}\right] + \frac{2\left|\gamma_k\right|}{{\left(kv\right)}^2}.
\end{align}
In the marginally stable limit $\gamma_k \to 0$, or $\eta_k + kv \gg \left|\gamma_k\right|$, which is typically true over a large $v$ range for $k\lambda_{\rmD e} \lesssim 1$,

\begin{align}
\calF_k\left(\omega_k,v\right) \to \frac{\pi\eta^2_k}{{\left(kv\right)}^3} \Theta\left(k - \frac{\eta_k}{v}\right),
\end{align}
and we have

\begin{align}
D^{(\rms)}_w(v) &\approx \frac{16\pi^6 q^2_\rms}{m^2_\rms V} \nonumber\\
&\times v^{-3} \int_0^{\infty} \rmd k\, \frac{\calE(k)}{k} \frac{\eta^2_k \left|\calC'_\omega\left(\eta_k\right)\right|}{{\left|\varepsilon'_{k\parallel}\left(\eta_k\right)\right|}^2} \Theta\left(k - \frac{\eta_k}{v}\right),
\label{Dw_iso}
\end{align}
where $\Theta$ denotes the heaviside function. This can also be obtained directly from the long-time limit, given by equation~(\ref{diffusion_tensor_wave_long_time}), when the resonant particles are the ones that dominantly exchange energy with the waves.

\begin{figure}
\centering
\includegraphics[width=0.48\textwidth]{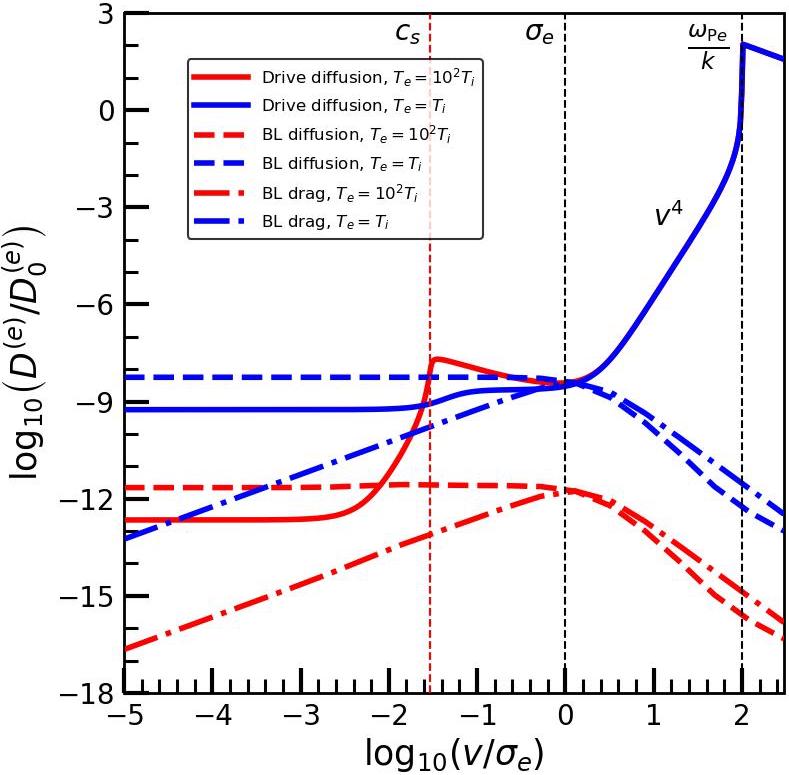}
\caption{Diffusion and drag coefficients vs $v$ for electrons, all normalized by the reference scale $D^{(e)}_0 = 32\pi^5 e^2 k^2 \calE_0/m_e^2 V$, assuming $\calE(k') = \calE_0\,\delta(k'-k)$, $k\lambda_{\rmD e} = 10^{-2}$ and $\omega_{\rmP e}t_\rmc = 1$. Here $\calE_0$ is chosen such that $D^{(e)}(v\ll \sigma_e)=10\,\calD^{(e)}_2(v\ll \sigma_e)$. The ion coefficients scale similarly with $v$. Solid, dashed and dot-dashed lines indicate the drive diffusion, BL diffusion and BL drag coefficients respectively. Blue (red) lines represent $T_e =  T_i$ $(T_e = 10^2 T_i)$. Note that the drive diffusion coefficient $D^{(e)}(v)$ scales as $v^4$ in the range $\sigma_e = \sqrt{k_\rmB T_e/m_e}<v<\omega_{\rmP e}/k$ for $T_e = T_i$ (also at $T_i > T_e$), and similarly at $\sigma_i=\sqrt{k_\rmB T_i/m_i}<v<c_\rms = \sqrt{k_\rmB T_e/m_i}$ for $T_i < T_e$. The BL diffusion (drag) coefficient is constant (scales as $v$) at small $v$ and scales as $v^{-3}$ $(v^{-2})$ at large $v$. Note that the drive coefficient always wins over the BL ones at large $v$ for a super-Debye scale drive.}
\label{fig:D_vs_v}
\end{figure}

Although $D_{ij}^{(\rms)}$ in equation~(\ref{quasilin_resp_FP_eq}) is defined generally, the transverse magnetic contribution does not show up in the reduced diffusion coefficients derived here, since the fluctuations in the unmagnetized plasma have been assumed to be isotropic. In this case the particle acceleration is governed by longitudinal electric fields (parallel to $\bk$) and the EM problem of quasilinear relaxation reduces to an electrostatic one \citep[][]{Banik.Bhattacharjee.24a,Chavanis.23}. And, since magnetic perturbations are always transverse $\left(\bk\cdot \bB_{\rm \bk} = 0\right)$, the acceleration in the isotropic setup is caused by electric field perturbations that are perpendicular to magnetic ones, akin to Fermi acceleration \citep[][]{Fermi.49}. If anisotropy is included in the problem, e.g., through spatially anisotropic perturbations or temperature anisotropies or a strong guide field, then the transverse component of the dielectric tensor would also show up in the diffusion coefficients. We briefly discuss anisotropic diffusion in section~\ref{sec:aniso_wave} but leave a detailed analysis for future work.

Let us briefly discuss the BL coefficients that describe the relaxation due to internal turbulence and collisions. In the isotropic scenario, equation~(\ref{BL_coeff_app}) dictates the following expression for the diffusion ($\calD^{(\rms)}_2$) and drag ($\calD^{(\rms)}_1$) BL coefficients:

\begin{align}
\calD^{(\rms)}_i(v) &= \frac{\omega_{\rmP\rms}^4}{n_\rms} \int \frac{\rmd k}{k} \int_0^1 \rmd\cos{\theta}\,\frac{\cos^2\theta\; \calF_{\rms i}\left(v\cos{\theta}\right)}{{\left|\varepsilon_{k\parallel}\left(kv\cos\theta\right)\right|}^2},
\end{align}
where $i=1,2$, $\calF_{\rms 1}\left(v\cos{\theta}\right) = v f_{\rms 0}\left(v\cos{\theta}\right)$, and $\calF_{\rms 2}\left(v\cos{\theta}\right) = F_{\rms 0}\left(v\cos{\theta}\right)$, $F_{\rms 0}(v) = 2\pi \int \rmd v'\,v' f_{\rms 0}\left(\sqrt{v^2 + v'^2}\right)$ being the one-dimensional DF. In these derivations, we have neglected the cross-terms denoting inter-species interactions. This is justified in an isotropic setup, if one of the species is significantly heavier, as is the case for a hydrogen or helium plasma, since electron-ion interactions scatter electrons off the inert ions, which only significantly alters the angular but not the $v$ dependence of their DFs.

Let us now compare the $v$ dependence of the various coefficients. For simplicity, we assume that the plasma is composed of electrons and ions of a single species. We assume the drive to be a temporal red noise ($\omega_{\rmP e} t_\rmc > 1$ with $t_\rmc$ the noise correlation time) that dominates on super-Debye scales ($k \lambda_{\rmD e} \ll 1$). To be fairly general, we assume that the spatial and temporal power spectra of the fluctuating drive are given by

\begin{align}
&\calE(k) = \calE_0\, {\left(\frac{k}{k_\rmc}\right)}^{-\alpha} \exp{\left[-k/k_\rmc\right]}\,\Theta\left(k-k_{\rm min}\right),\nonumber\\
&\calC_\omega(\omega) = \frac{1}{1 + \omega^2 t^2_\rmc},\quad \exp{\left[-0.5\omega^2 t^2_\rmc\right]},
\label{Schechter_func}
\end{align}
with $k_{\rm min}$ the minimum wavenumber of the drive and $k_\rmc$ the maximum cutoff in $k$. The three-dimensional electric-field spectrum is defined by $\calE(k)=|\bE_\bk|^2$. Standard sub-ion kinetic-Alfv\'en turbulence is comparatively shallow, with the electric spectrum typically expected to scale as $\calE(\bk)\sim k_\perp^{-1/3}$, where $k_\perp$ is the component of $\bk$ perpendicular to $\bB_0$ \citep[][]{Howes.etal.08}. Imbalanced turbulence with helicity barrier yields steeper spectra in the transition range: recent theory and simulations find $\calE(\bk)\sim k_\perp^{-4}$ near $k_\perp\lambda_{\rmc i}\sim 1$, followed by a reflattening at smaller sub-ion scales \citep[][]{Zhang.etal.25,Adkins.etal.25,McIntyre.etal.25}. At electron scales, the situation is less clear: there is evidence that the magnetic spectrum steepens near $\lambda_{\rmc e}$, plausibly due to electron Landau damping, but a universal electric-field scaling below the electron Larmor radius has not yet been established \citep[][]{Sahraoui.etal.13,Chen.etal.19,Zhou.etal.23}. By contrast, in electrostatic Langmuir turbulence the electric spectrum can be as steep as $\calE(k)\sim k^{-5}$ \citep[][]{Sun.etal.22}. Around the Debye scale, the spatial power-spectrum can be even steeper, scaling as $\calE(k)\sim k^{-(2d+2)}$ in $d$ dimensions, i.e., $k^{-8}$ in 3D \citep[][]{Ewart.etal.25,Nastac.etal.23,Nastac.etal.25,Ginat.etal.25}\footnote{The same power-spectrum has been shown to arise in sub-Debye Vlasov turbulence in kinetic plasmas and small-scale Vlasov turbulence for cold dark matter.}. 

With the power-law $\calE(k)\sim k^{-\alpha}$ and the Lorentzian $\calC_\omega$, let us look at the asymptotic scalings of $D_p^{(\rms)}$. For $\sigma_e<v<v_*=\omega_{\rmP e}/k_{\min}$, it becomes
\begin{align}
D_p^{(\rms)}(v)&\approx
\frac{\pi}{5}\,\frac{3-\alpha}{1-\left(k_{\min}\lambda_{\rmD e}\right)^{3-\alpha}}\,
Z_\rms^2\left(\frac{m_e}{m_\rms}\right)^2
\sigma_e^2\omega_{\rmP e}
\left(\frac{\langle E^2\rangle}{E_{\rm int}^2}\right)\nonumber\\
&\times
\begin{cases}
\dfrac{1}{3-\alpha}\left(\dfrac{v}{\sigma_e}\right)^{\alpha-1},
& 0<\alpha<3,\\[8pt]
\ln\!\left(\omega_{\rmP e}t_\rmc\right)\left(\dfrac{v}{\sigma_e}\right)^2,
& \alpha=3,\\[8pt]
\dfrac{\left(\omega_{\rmP e}t_\rmc\right)^{\alpha-4}}{\alpha-3}
\left(\dfrac{v}{\sigma_e}\right)^{\alpha-1},
& 3<\alpha<5,\\[8pt]
\dfrac{\left(\omega_{\rmP e}t_\rmc\right)^{\alpha-4}}{\alpha-3}
\left(\dfrac{v_{\rm tr}}{\sigma_e}\right)^{\alpha-1}
\left(\dfrac{v}{v_{\rm tr}}\right)^4,
& \alpha>5,\ \sigma_e<v<v_{\rm tr},\\[8pt]
\dfrac{\left(\omega_{\rmP e}t_\rmc\right)^{\alpha-4}}{\alpha-3}
\left(\dfrac{v}{\sigma_e}\right)^{\alpha-1},
& \alpha>5,\ v_{\rm tr}<v<v_*,
\end{cases}
\label{eq:Dp_subturnover_final}
\end{align}
where the transition velocity for $\alpha>5$ is obtained by matching the $v^4$ and $v^{\alpha-1}$ branches,
\begin{align}
v_{\rm tr}\approx \frac{1}{k_{\min}t_\rmc}
\left[\frac{5(\alpha-3)}{7(\alpha-5)}\right]^{1/(\alpha-5)}.
\label{eq:vtr_final}
\end{align}
The case $\alpha=5$ is marginal and acquires logarithmic corrections. At $v>v_\ast = \omega_{\rmP e}/k_{\min}$, $D_p^{(\rms)}$ scales as $v^{-3}$ just as the BL diffusion coefficient. The various scalings are listed in Table~\ref{tab:Dp_v_scalings}.

Similarly, from equation~(\ref{Dw_iso}), one has
$D_w^{(\rms)}(v)\propto v^{-3}\int_{\eta_k/v}^{\infty}\rmd k\,k^{-1-\alpha}$. Let us consider kinetic instabilities such as the bump on tail instability, for which $\eta_k\approx \omega_{\rmP e}$. For $v<v_*=\omega_{\rmP e}/k_{\min}$, the lower limit $\omega_{\rmP e}/v$ lies above $k_{\min}$, so the integral is controlled by its lower limit and scales as $(\omega_{\rmP e}/v)^{-\alpha}\propto v^\alpha$, which gives $D_w^{(\rms)}\propto v^{\alpha-3}$. By contrast, for $v>v_*$ the lower limit falls below $k_{\min}$, so the integral is cut off at $k_{\min}$ and becomes $v$-independent; the prefactor $v^{-3}$ then implies $D_w^{(\rms)}\propto v^{-3}$.

In Fig.~\ref{fig:D_vs_v_alpha} we plot $D^{(\rms)}_p$ and $D^{(\rms)}_w$ as functions of $v$ for different values of $\alpha$ as indicated, for both Lorentzian and Gaussian temporal correlations and for $\omega_{\rmP e}t_\rmc=1$ and $10$. We adopt $T_e = T_i$, $k_{\rm min}\lambda_{\rmD e} = 10^{-3}$, $k_\rmc\lambda_{\rmD e} = 1$, and $1/\omega_{\rmP e} \le t_\rmc < 1/k_{\rm min}\sigma_e$. We also assume marginal stability, i.e., $\gamma_k\to 0$, and that the (saturated) instability is kinetic in origin, i.e., $\eta_k \sim \omega_{\rmP e}$. In particular, we assume that a classic bump-on-tail instability has occurred, saturated and carved out a plateau at the bump (at $v = v_\rmb$), thereby driving the plasma marginally stable with $\gamma_k \propto \left.\partial f_{\rms 0}/\partial v\right|_{v_\rmb \approx \omega_{\rmP e}/k} \approx 0$. As evident from Fig.~\ref{fig:D_vs_v_alpha}, the dressed particle coefficient $D^{(\rms)}_p$ shows the same qualitative behavior in the Lorentzian and the Gaussian $\omega_{\rmP e} t_\rmc = 1$ cases. For the steepest spectra shown, $\alpha=7,8,9$, $D^{(\rms)}_p$ scales as $v^4$ at intermediate velocities, $\sigma_e < v < 1/k_{\rm min}t_\rmc$, before turning over and eventually falling off as $v^{-3}$ at larger $v$. Consistent with Table~\ref{tab:Dp_v_scalings}, for $5 \leq \alpha < 7$ the $v^4$ range extends to $\omega_{\rmP e}/k_{\rm min}$, while for $0 < \alpha < 5$ one instead has $D^{(\rms)}_p \sim v^{\alpha - 1}$ over the corresponding intermediate-$v$ interval. At larger $v$, $D^{(\rms)}_p$ always drops off as $v^{-3}$. At large $\omega_{\rmP e}t_\rmc$, the Gaussian correlation predicts a smoother turnover, yields a $v^4$ scaling for $\alpha\geq 7$, gets rid of the steeper $v^{\alpha-1}$ scaling at $1/k_{\rm min}t_\rmc < v < \omega_{\rmP e}/k_{\rm min}$, and strongly suppresses the wave contribution. Ultimately, it is the power-law scaling of $D^{(\rms)}_p$ at intermediate $v$ that gives rise to power-law tails in the steady state $f_{\rms 0}$ (see Table~\ref{tab:Dp_v_scalings}). As we shall see, this power-law scaling is pronounced at $\sigma_i < v < \sigma_e$ for the ions, and at velocities beyond $\sigma_e$ for both electrons and ions. 

The wave diffusion coefficient $D^{(\rms)}_w$ drops off to zero at $v\lesssim \sigma_e$, since the $\eta_k = {\rm Re}\,\omega_k = kv$ wave-particle resonance is not possible at low $v$. At $\sigma_e < v < \omega_{\rmP e}/k_{\rm min}$, it scales as $\sim v^{\alpha - 3}$, and falls off as $v^{-3}$ beyond. All in all, for steep enough spectra with $\alpha \geq 5$, i.e., large-scale (super-Debye) electric field fluctuations, $D^{(\rms)}_p$ universally scales as $v^4$ in a large $v$ range. This is because the dielectric constant $\varepsilon_{k\parallel}$ scales as $\sim {\left(\omega_{\rmP e}/kv\right)}^2$ in this velocity range \citep[][]{Banik.Bhattacharjee.24a}. Physically, this implies that slower particles are more Debye shielded and less accelerated, while the very fast ones are unscreened and readily heated. As discussed above, such steep spectra are plausible in some regimes, e.g., in Langmuir turbulence, around the ion Larmor scale in imbalanced turbulence with helicity barrier, and in near-Debye-scale plasma turbulence. In almost all cases, the dressed particle coefficient dominates over the wave coefficient, as long as the plasma is stable or marginally stable. In unstable scenarios, the wave coefficient initially dominates, but once the instability has saturated and the bump has plateaued out, the dressed particle coefficient emerges as the key player in particle heating.

\renewcommand{\arraystretch}{1.25}

\begin{table*}[t!]
\centering
\begin{ruledtabular}
\begin{tabular}{cccc}
\textbf{\shortstack{Power-\\spectrum}}
  & \raisebox{0.7ex}{\textbf{Velocity Range}}
  & \raisebox{0.7ex}{$D_p^{(\rms)}(v)$}
  & \textbf{\shortstack{Steady-state\\$f_{\rms 0}(v)$} } \\
\hline

\multirow{3}{*}{$\alpha \geq 7$}
  & $\sigma_e < v < {1}/{k_{\min} t_{\rmc}}$
  & $v^{4}$
  & $v^{-5}$ \\

  & ${1}/{k_{\min} t_{\rmc}} < v < {\omega_{\rmP e}}/{k_{\min}}$
  & $v^{\alpha-1}$
  & $v^{-\alpha}$ \\

  & $v > {\omega_{\rmP e}}/{k_{\min}}$
  & $v^{-3}$
  & $\exp\!\left[-v^2/2\sigma'^2_\rms\right]$ \\

\hline

\multirow{2}{*}{$5 \leq \alpha < 7$}
  & $\sigma_e < v < {\omega_{\rmP e}}/{k_{\min}}$
  & $v^4$
  & $v^{-5}$ \\

  & $v > {\omega_{\rmP e}}/{k_{\min}}$
  & $v^{-3}$
  & $\exp\!\left[-v^2/2\sigma'^2_\rms\right]$ \\

\hline

\multirow{2}{*}{$0 < \alpha < 5$}
  & $\sigma_e < v < {\omega_{\rmP e}}/{k_{\min}}$
  & $v^{\alpha-1}$
  & $v^{-\alpha}$ \\

  & $v > {\omega_{\rmP e}}/{k_{\min}}$
  & $v^{-3}$
  & $\exp\!\left[-v^2/2\sigma'^2_\rms\right]$
\end{tabular}
\end{ruledtabular}

\caption{Velocity scalings of the dressed particle diffusion coefficient $D^{(\rms)}_p$ and the corresponding steady-state distribution $f_{\rms 0}(v)$, obtained from equation~(\ref{f0_ss}), for a drive with electric-field power spectrum $\mathcal{E}(k)\sim k^{-\alpha}$ and a Lorentzian $\calC_\omega$. We assume $T_e=T_i$ and $1/\omega_{\rmP e} < t_\rmc < 1/k_{\rm min}\sigma_e$. The thermal speed of the Maxwellian cut-off, $\sigma'_\rms$, is approximately equal to $\sigma_{\rms}/{\left(k_{\rm min}\lambda_{\rmD e}\right)}^n$, where $n$ depends on $\alpha$.}
\label{tab:Dp_v_scalings}
\end{table*}

\renewcommand{\arraystretch}{1.0}

In Fig.~\ref{fig:D_vs_v} we compare the drive diffusion coefficient, $D^{(\rms)} = D^{(\rms)}_p + D^{(\rms)}_w$, to the BL diffusion and drag coefficients as functions of $v$, for $T_e = T_i$ and $10^2 T_i$. For simplicity, we only compute these quantities at $k = k_0 = 10^{-2}\lambda^{-1}_{\rmD e}$, i.e., we assume $\calE(k)\sim \delta\left(k - k_0\right)$. As shown above, for steep enough spectra, $\calE(k)\sim k^{-\alpha}$ with $\alpha > 5$, the drive scales the same way with $v$ (till $\omega_{\rmP e}/k_{\rm min}$ instead of $\omega_{\rmP e}/k_0$). The BL diffusion (drag) coefficient $\calD_2^{(\rms)}$ ($\calD_1^{(\rms)}$) is constant (scales as $\sim v$) at $v\lesssim \sigma_\rms$ and scales as $\sim v^{-3}$ ($\sim v^{-2}$) at $v\gtrsim \sigma_\rms$. For $T_e \sim T_i$, the drive diffusion coefficient  $D^{(\rms)}$ is constant for $v \lesssim \sigma_e = \sqrt{k_\rmB T_e/m_e}$ and scales as $\sim v^4$ for $\sigma_e \lesssim v \lesssim \omega_{\rmP e}/k = \sigma_e/k\lambda_{\rmD e}$, as long as $k\lambda_{\rmD e} < 1$. For $T_e \gg T_i$, the $v^4$ scaling also appears between the ion thermal speed $\sigma_i = \sqrt{k_\rmB T_i/m_i}$ and the ion acoustic speed $c_\rms = \sqrt{k_\rmB T_e/m_i}$. The wave particle resonance between the particles and the Langmuir waves causes $D^{(\rms)}(v)$ to spike at $v = \omega_{\rmP e}/k$ for $T_e = T_i$. A similar resonance with the ion acoustic waves gives rise to a spike at $v = c_\rms$ when $T_e \gg T_i$, a necessary condition for the ion acoustic waves to be weakly damped. At $T_e = T_i$, these waves are strongly damped, which is why the large imaginary part of $\varepsilon_{k\parallel}$ smooths out $D^{(\rms)}(v)$ at $v = c_\rms$; this effect is known as resonance broadening \citep[][]{Diamond_Itoh_Itoh_2010}. On the other hand, Langmuir waves are always weakly damped for $k\lambda_{\rmD e} < 1$, resulting in a pronounced spike at $v = \omega_{\rmP e}/k$ irrespective of $T_i/T_e$. The drive diffusion coefficient always exceeds the BL coefficients at large enough $v$, independent of the strength of the drive, as long as it predominantly acts on super-Debye scales. This is because the drive coefficient is an increasing function of $v$ in a large $v$ range and falls off only beyond $v = \omega_{\rmP e}/k$, whereas the BL coefficients drop off beyond the thermal speed, which is much smaller than $\omega_{\rmP e}/k$. Hence, a large-scale drive always dominates over collisions or self-generated small-scale fluctuations at high $v$. In other words, the fastest particles are unscreened and readily accelerated without slowing down due to collisions.

\subsection{Quasilinear relaxation of the distribution function}\label{sec:f_evol_QL}

\begin{figure*}
\centering
\includegraphics[width=0.8\textwidth]{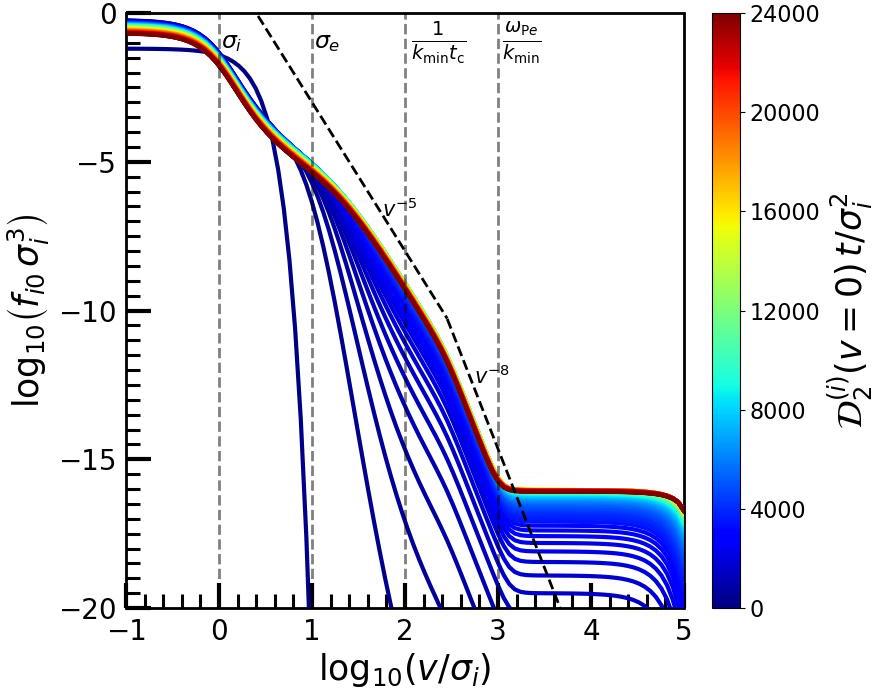}
\caption{Development of the non-thermal power-law tail in an initially Maxwellian ion distribution due to large-scale EM turbulence, obtained by solving equation~(\ref{QL_eq}). We adopt $D^{(i)}(v = 0) = 10^{-2} \calD^{(i)}_2(v = 0)$, $\sigma_i = 0.1\sigma_e$, $\alpha = 8$, $k_{\rm min} \lambda_{\rmD i} = k_{\rm min} \lambda_{\rmD e} = 10^{-3}$, and $\omega_{\rmP i} t_\rmc = 0.1 \omega_{\rmP e} t_\rmc = 1$. Note the Maxwellianization of the bulk around $\sigma_i$ and the emergence of the $v^{-5}$ tail beyond, followed by a $v^{-\alpha} \sim v^{-8}$ tail and a Maxwellian fall-off.}
\label{fig:fi_vs_t}
\end{figure*}

\begin{figure*}
\centering
\includegraphics[width=0.8\textwidth]{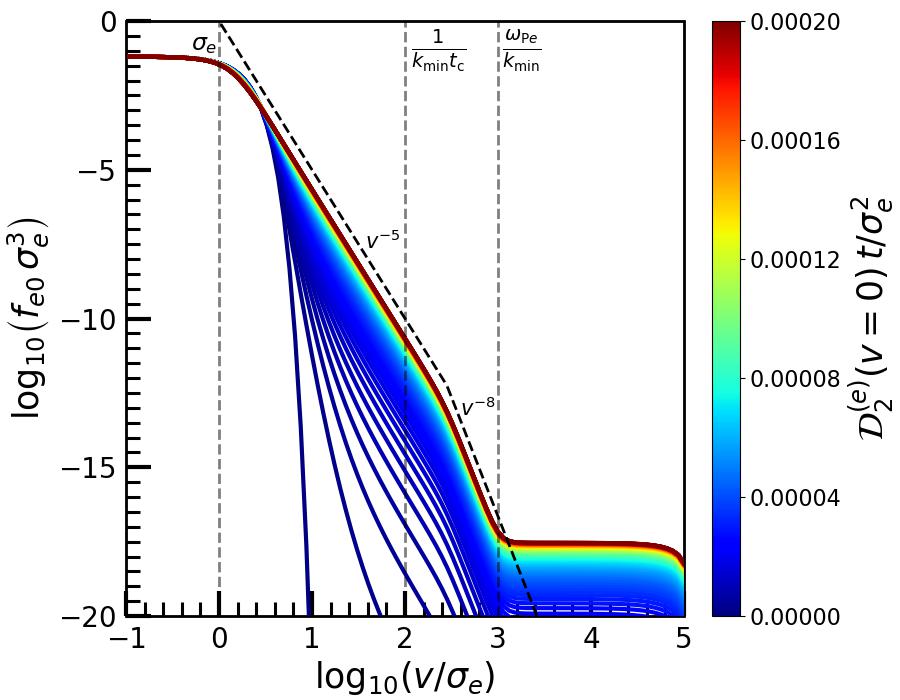}
\caption{Same as Fig.~\ref{fig:fi_vs_t} but for electrons. We adopt $D^{(e)}(v = 0) = \calD^{(e)}_2(v = 0)$, $\alpha = 8$, $k_{\rm min} \lambda_{\rmD e} = 10^{-3}$, and $\omega_{\rmP e} t_\rmc = 10$. Just as in the ions, a $v^{-5}$ tail develops, followed by a $v^{-\alpha} \sim v^{-8}$ tail and a Maxwellian fall-off.}
\label{fig:fe_vs_t}
\end{figure*}

\begin{figure*}
\centering
\includegraphics[width=0.8\textwidth]{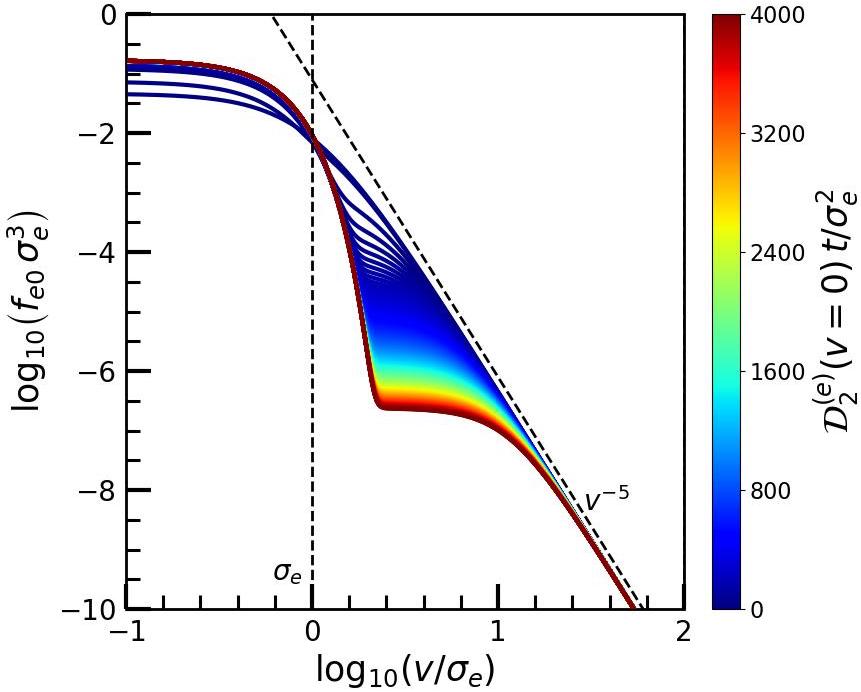}
\caption{Maxwellianization of the electron DF, initialized as a $\kappa = 1.5$ distribution ($v^{-5}$), after the drive has been switched off and the plasma relaxes due to turbulent and collisional relaxation through the BL equation (equation~[\ref{QL_eq}] without the drive). Note the continued presence of the $v^{-5}$ tail of runaway electrons.}
\label{fig:fe_vs_t_after_shock}
\end{figure*}

It is not surprising that the power-law scaling of $D^{(\rms)}(v)$ would translate to a power-law $f_{\rms 0}(v)$. To see this explicitly, we evolve the electron and ion DFs by numerically integrating equation~(\ref{QL_eq}) for each species, using a flux-conserving scheme detailed in Appendix~C.1 of \citet[][]{Banik.Bhattacharjee.24a}. We assume an isotropic turbulent drive with a power spectrum of the form given in equation~(\ref{Schechter_func}) with $\alpha = 8$, which is expected for Vlasov turbulence around the Debye length \citep[][]{Nastac.etal.25,Ewart.etal.25}. We adopt $\sigma_i = 0.1 \sigma_e$, $k_{\rm min}\lambda_{\rmD e} = 10^{-3}$, and $t_\rmc = 10/\omega_{\rmP e} = 1/\omega_{\rmP i}$. One may think of $k_{\rm min}$ as being set by the largest fluctuation scale below the relevant Larmor radius $\lambda_{\rmc}$, i.e., $k_{\rm min}\sim \lambda_{\rmc}^{-1}$. Likewise, the adopted correlation time should be of order the gyration period, $t_\rmc \sim \omega_{\rmc}^{-1}$, so that $\omega_{\rmP e} t_\rmc \sim 10$ serves as a representative fiducial choice, especially for electrons. We checked that $\omega_{\rmP e} t_\rmc \sim 100-1000$ yields similar results for steep enough power-spectra, so the present exercise is applicable to ions too. In this case, as evident from Fig.~\ref{fig:D_vs_v_alpha} and Table~\ref{tab:Dp_v_scalings}, $D^{(\rms)}_p(v)$ dominates over $D^{(\rms)}_w(v)$, and scales as $v^4$ for $\sigma_e < v < 1/k_{\rm min}t_\rmc$, as $v^{\alpha - 1} \sim v^7$ for $1/k_{\rm min}t_\rmc < v < \omega_{\rmP e}/k_{\rm min}$, and as $v^{-3}$ beyond. We normalize the coefficients such that, for the electrons, the drive diffusion coefficient $D^{(e)}(v)$ is equal to the BL diffusion coefficient $\calD_2^{(e)}(v)$ at $v \ll \sigma_e$, while for the ions, $D^{(i)}\left(v \ll \sigma_i\right) = 10^{-2} \calD_2^{(i)}\left(v \ll \sigma_i\right)$ (this is justified for $T_i/T_e \sim 10$ at low $Z$, since $D^{(i)}/\calD_2^{(i)} = Z^{3/2}\sqrt{m_e/m_i}\,(T_e/T_i)\, D^{(e)}/\calD_2^{(e)}$ at $v\to 0$). Due to its rising trend, $D^{(\rms)}(v)$ always exceeds $\calD_2^{(\rms)}(v)$ at $v > \sigma_e$.

We plot the resulting ion DF $f_{i 0}$ in Fig.~\ref{fig:fi_vs_t}. The electron DF has similar, albeit faster, evolution since $D^{(e)} \sim D^{(i)} {\left(m_i/Z m_e\right)}^2 \gg D^{(i)}$, and is plotted in Fig.~\ref{fig:fe_vs_t}. %The parameters have been chosen such that $D^{(e)}(v\to 0)/\calD_2^{(e)}(v\to 0) = 1$ and $D^{(i)}(v\to 0)/\calD_2^{(i)}(v\to 0) = 0.1$. 
The maximum value of the diffusion coefficient is much larger for electrons than ions in the current setup, which significantly reduces the required timestep (via the Courant condition) and the total integration time for the electrons. Evidently, an initial Maxwellian-like DF develops an extended $v^{-5}$ tail followed by a $v^{-8}$ ($v^{-\alpha}$) fall-off till $\omega_{\rmP e}/k_{\rm min}$, beyond which it forms a plateau and a Maxwellian fall-off. The BL diffusion (drag) increases (decreases) the particle velocities, which is evident from the evolution around $\sigma_i$ at earlier times. The drive diffusion coefficient gradually develops the $v^{-5}$ and $v^{-8}$ tails at higher $v$; the $v^4$ scaling of $D^{(\rms)}(v)$ yields the $v^{-5}$ tail, while the $v^{7}$ $(v^{\alpha-1})$ scaling gives rise to the $v^{-8}$ $(v^{-\alpha})$ one (see discussion below). The $v^{-3}$ fall-off of both drive and BL diffusion coefficients ensure a Maxwellian plateau at $v>\omega_{\rmP e}/k_{\min}$. Interestingly, the $v^{-5}$ tail appears in the ions even at $\sigma_i < v < \sigma_e$. This arises because (1) the drive diffusion coefficient shows a brief increasing trend with $v$ around the ion acoustic speed $c_\rms$ $(\approx \sigma_i)$ (see Fig.~\ref{fig:D_vs_v}, and) (2) we have adopted a representative drive strength such that $D^{(i)}\left(v \ll \sigma_i\right) = 10^{-2} \calD_2^{(i)}\left(v \ll \sigma_i\right)$, so the bulk of the ions at $v \lesssim \sigma_e$ tries to Maxwellianize; a stronger drive would tend to smooth out this feature. The origin of the power-law tails and the Maxwellian bulk and cut-off can be understood by working out the steady state solution to the quasilinear equation~(\ref{QL_eq}):

\begin{align}
&f_{\rms 0}(v) = N\,\exp{\left[-\int\rmd v'{\calD_1^{(\rms)}\left(v'\right)}\Big/\left({\calD_2^{(\rms)}\left(v'\right) + D^{(\rms)}\left(v'\right)}\right)\right]}\nonumber\\
&\times \left[1 + C \int \rmd v' \dfrac{\exp{\left[\int\rmd v''{\calD_1^{(\rms)}\left(v''\right)}\Big/\left({\calD_2^{(\rms)}\left(v''\right) + D^{(\rms)}\left(v''\right)}\right)\right]}}{v'^2\left(\calD_2^{(\rms)}\left(v'\right) + D^{(\rms)}\left(v'\right)\right)}\,\right],
\label{f0_ss}
\end{align}
with $N$ a normalization constant and $C$ an integration constant related to the drive diffusive flux, $-v^2 D^{(\rms)}(v)\, \partial f_{\rms 0}/\partial v$. The first term, a zero flux solution, can be shown to be a Maxwellian for $D^{(\rms)} = 0$. This represents the Maxwellian bulk of the particles around $\sigma_\rms$. The second term, a constant flux solution, is what gives rise to the power-law tail. At $v > \sigma_e$, $D^{(\rms)}$ exceeds both BL coefficients, which implies that $f_{\rms 0}$ scales as $\int_v^\infty \rmd v'/v'^2 D^{(\rms)}(v')$ at large $v$. Therefore, $f_{\rms 0}(v)\sim \int_v^\infty \rmd v'/v'^6 \sim v^{-5}$ at $\sigma_e < v < 1/k_{\rm min} t_\rmc$, and $f_{\rms 0}(v)\sim \int_v^\infty \rmd v' v'^{-1-\alpha} \sim v^{-\alpha} \sim v^{-8}$ at $1/k_{\rm min} t_\rmc < v < \omega_{\rmP e}/k_{\rm min}$. Beyond $\omega_{\rmP e}/k_{\rm min}$, both drive and BL diffusion coefficients scale as $v^{-3}$ and the BL drag as $v^{-2}$. Hence, $f_{\rms 0}(v)$ scales as $\exp{\left[-\int_0^v\rmd v'{\calD_1^{(\rms)}\left(v'\right)}\big/D^{(\rms)}\left(v'\right)\right]}$ (recall that the drive diffusion exceeds BL diffusion at large $v$), which is nothing but a Maxwellian, $\exp{\left[-v^2/2\sigma'^2_\rms\right]}$ with $\sigma'_\rms \approx \sigma_{\rms}/{\left(k_{\rm min}\lambda_{\rmD e}\right)}^2$. Overall, the DF is close to a $\kappa$ distribution \citep[][]{Livadiotis.McComas.13,Zhdankin.22a,Zhdankin.22b}, $f_{\rms 0}(v)\sim {(1+v^2/2\kappa\sigma^2_\rms)}^{-\left(1+\kappa\right)}$ with $\kappa = 1.5$ till $v \approx 1/k_{\rm min}t_\rmc$ and $\kappa = \alpha/2 - 1$ ($=3$ for $\alpha = 8$ in this case) at $1/k_{\rm min}t_\rmc < v < \omega_{\rmP e}/k_{\rm min}$. The broad Maxwellian fall-off at large $v$ ensures that all moments are finite. 

Figs.~\ref{fig:fi_vs_t} and \ref{fig:fe_vs_t} show that the DF relaxes to a $v^{-5}$ tail at long times (for $\alpha \geq 5$), but features steeper fall-offs at intermediate times. Therefore, beginning from a Maxwellian, $f_{\rms 0}$ asymptotically approaches a $\kappa$ distribution with $\kappa\approx 1.5$, although, incomplete relaxation, which can occur if the drive is switched off midway (mimicking an intermittent turbulent drive), would, however, yield larger values of $\kappa$. In the examples shown here, the $v^{-5}$ tail develops because we have adopted a sufficiently steep drive spectrum, $\alpha=8$; more generally, the same asymptotic $v^{-5}$ attractor arises whenever $\alpha\geq5$, whereas shallower spectra yield a $v^{-\alpha}$ tail instead. The parameter $k_{\rm min}\lambda_{\rmD e}$ and the relative amplitude of the drive and BL coefficients control the velocity range over which the power-law is realized. Decreasing $k_{\rm min}\lambda_{\rmD e}$ extends the suprathermal interval up to larger $v\sim \omega_{\rmP e}/k_{\rm min}$. Such steep spectra are plausible in various regimes of sub-Larmor-scale turbulence \citep[][]{Zhang.etal.25,Adkins.etal.24,McIntyre.etal.25,Sahraoui.etal.13,Zhou.etal.23,Chen.etal.19,Sun.etal.22}. For example, the power-spectrum $\calE(k)$ is expected to be quite steep on scales smaller than the electron Larmor radius and even as steep as $k^{-8}$ near the Debye scale \citep[][]{Nastac.etal.23,Nastac.etal.25}, which yields a $v^{-\alpha} \sim v^{-8}$ fall-off following the $v^{-5}$ tail, as shown in Figs.~\ref{fig:fi_vs_t} and \ref{fig:fe_vs_t}. Our theory therefore predicts a range of $\kappa = 1.5-3$ for a relaxed DF (provided $\alpha \geq 5$), which is in general agreement with heliospheric DF measurements. A focused analysis of these observations however requires the formulation of a self-consistent QLT for magnetized plasmas, which we leave for future work.

Despite possible variabilities such as those due to shallow power-spectra, incomplete relaxation or collisional Maxwellianization, the $v^{-5}$ tail appears to be a near-universal outcome of electromagnetically driven kinetic plasmas. As $f_{\rms 0}$ develops the $v^{-5}$ tail under the action of the drive, the energy ($E = m_\rms v^2/2$) distribution $N_\rms(E) = g(E)f_{\rms 0}(E)$, with $g(E)\sim v$ the density of states, approaches $v^{-4}\sim E^{-2}$ (see Appendix~\ref{App:entropy} for a discussion of entropy-based arguments \citep[][]{Livadiotis.McComas.13,Zhdankin.22a,Zhdankin.22b,Ewart.etal.22,Ewart.etal.23,Ewart.etal.25} to obtain this). A similar scaling is found in PIC simulations of collisionless shocks and reconnection \citep[][]{Sironi.Spitkovsky.14,Hesse.etal.18,Hoshino.22,Gupta.etal.24,Wong.etal.25}, where the plasma is often relativistic and magnetized. While the basic physics of Debye screening that gives rise to the $E^{-2}$ tail in our formalism is of course operant in these simulations, there could be other ways to obtain the same scaling. In fact, \citet[][]{Wong.etal.25} point out that the power-law tail in their simulation of relativistic plasma turbulence emerges from an interplay of the advection and diffusion coefficients (measured from the simulation) of an effective quasilinear transport equation, which is qualitatively different from our framework, where the tail is spawned by the drive diffusion coefficient alone. It would be interesting to see how the mechanism for generating power-law tails is modified in a self-consistent QLT for relativistic magnetized plasmas.

Once the turbulent drive is switched off, the electron DF, initialized as a $\kappa$ distribution with $\kappa = 1.5$ Maxwellianizes via turbulent and collisional relaxation through the BL coefficients, and does so much faster than the ion DF, as shown in Fig.~\ref{fig:fe_vs_t_after_shock}. The thermal bulk $(v \lesssim \sigma_e)$ gains energy and momentum from particles with intermediate velocities $(v \gtrsim \sigma_e)$ via the BL drag. BL diffusion, on the other hand, transfers energy and momentum from the core of the bulk $(v \ll \sigma_e)$ to farther out $(v \lesssim \sigma_e)$. The combined action of these two effects steepens the slope in the bulk and tends to Maxwellianize the DF. However, the $v^{-5}$ tail persists at high $v$, albeit suppressed with respect to the thermal bulk. This shows that, once the plasma has been heated by a super-Debye scale drive, the fastest particles cannot lose energy via collisions even after the drive has been switched off. In other words, a non-thermal power-law tail, once formed, is immune to collisions. This is especially true for heavy ions since their collisional relaxation time is significantly longer than that of the electrons. It is no surprise that the non-thermal tail survives collisions, since both BL diffusion and drag coefficients fall off at large $v$. Hence, a non-thermal population of runaway particles is a robust outcome of the relaxation of kinetic plasmas despite the presence of weak Coulomb collisions.

\subsection{Relaxation under anisotropic wave drive}\label{sec:aniso_wave}

Having studied the relaxation under an isotropic turbulent drive, let us now briefly discuss an anisotropic setup of a coherent drive that consists of EM waves. A fully self-consistent treatment of strongly magnetized, anisotropic plasmas is left for future work. For a narrow set of coherent waves, the quasilinear relaxation is naturally described in the axisymmetric velocity variables $(v_\parallel,v_\perp)$ defined with respect to a guide field:
\begin{align}
\frac{\partial f_{\rms0}}{\partial t}
&=
\frac{\partial}{\partial v_\parallel}
\left(
D^{(\rms)}_{\parallel\parallel}\frac{\partial f_{\rms0}}{\partial v_\parallel}
+
D^{(\rms)}_{\parallel\perp}\frac{\partial f_{\rms0}}{\partial v_\perp}
\right)
\nonumber\\
&\quad+
\frac{1}{v_\perp}\frac{\partial}{\partial v_\perp}
\left[
v_\perp
\left(
D^{(\rms)}_{\perp\parallel}\frac{\partial f_{\rms0}}{\partial v_\parallel}
+
D^{(\rms)}_{\perp\perp}\frac{\partial f_{\rms0}}{\partial v_\perp}
\right)
\right].
\label{eq:FP2D_wave_main}
\end{align}
We consider two idealized longitudinal drive spectra: a nearly parallel one,
$\calE^{(\rmP)}_{L,\parallel}\propto k_\parallel^{-\alpha_\parallel}k_\perp^{-\delta_\parallel}\Theta(\xi_\parallel k_\parallel-k_\perp)$ with $\xi_\parallel\ll1$, and a nearly perpendicular one,
$\calE^{(\rmP)}_{L,\perp}\propto k_\perp^{-\alpha_\perp}k_\parallel^{-\delta_\perp}\Theta(\xi_\perp k_\perp-k_\parallel)$ with $\xi_\perp\ll1$.

The nearly parallel case applies to the one-sided $(k_\parallel>0)$ $R$ branch and the ion-cyclotron branch; the small- and large-$k_\parallel d_e$ ($d_e=$ electron inertial scale) limits of the $R$ branch are the whistler (wh) and electron-cyclotron (EC) regimes. The nearly perpendicular case applies to the dispersive shear-Alfv\'en branch, whose sub-ion-Larmor limits are the kinetic-Alfv\'en (KAW) and inertial-Alfv\'en (InA) waves. For suprathermal particles resonant with low frequency EM waves, $kv\sim \omega_\bk \ll \omega_{\rmP e}$, the diffusion coefficients scale as $|\varepsilon_{\bk\parallel}(\omega_\bk)|^{-2}$, which is $\sim (\omega_\bk/\omega_{\rmP e})^4$ in the super-Debye regime. The detailed derivation is given in Appendix~\ref{App:wave_drive}.

For nearly parallel spectra, $k_\parallel \gg k_\perp$, the dominant coefficient is $D^{(\rms)}_{\parallel\parallel}$, so the evolution is effectively one-dimensional in $v_\parallel$ at leading order. For the ion-cyclotron branch, which has a single resonant root,
\begin{align}
D^{(\rms)}_{\parallel\parallel,{\rm IC}}
\propto
v_\parallel^{\,\alpha_\parallel+\delta_\parallel-3}.
\label{eq:Daniso_par_main}
\end{align}
For the $R$ branch, below $v_\parallel=\omega_{\rm c e}d_e/2=\sigma_e/\sqrt{2\beta_e}$ $(\beta_e = 8\pi n_e k_\rmB T_e/B^2_0)$, the resonance condition has two roots, one on the whistler side and one on the electron-cyclotron side, so the exact coefficient is the sum of the two contributions,
\begin{align}
D^{(\rms)}_{\parallel\parallel,{\rm R}}
\propto
v_\parallel^{\,9-\alpha_\parallel-\delta_\parallel}
+
v_\parallel^{\,\alpha_\parallel+\delta_\parallel-3}.
\label{eq:Daniso_par_R_main}
\end{align}
The $R$ branch is special because the resonance becomes tangent to the dispersion curve at $v_\parallel=\omega_{\rm c e}d_e/2=\sigma_e/\sqrt{2\beta_e}$. In other words, the group velocity $v_{\rm gr}(k_r)$ of the wave matches up with the particle velocity, i.e., $v_\parallel = v_{\rm gr}(k_r)$, $k_r$ being the resonant wavenumber at which the resonance condition $\omega_\bk(k_r) = k_r v_\parallel$ is satisfied. Thus, both phase and group velocities of the wave are equal to the particle velocity, $v_\parallel=\sigma_e/\sqrt{2\beta_e}$, something we refer to as resonant tangency. This produces a narrow resonant spike in $D^{(\rms)}_{\parallel\parallel}$ at $\sigma_e/\sqrt{2\beta_e}$. For $v_\parallel>\omega_{\rm c e}d_e/2=\sigma_e/\sqrt{2\beta_e}$, there is no resonant contribution from the one-sided ($k_\parallel>0$) $R$ branch waves, and $D^{(\rms)}_{\parallel\parallel}$ is zero.

For nearly perpendicular spectra, $k_\perp \gg k_\parallel$, the dominant coefficient is $D^{(\rms)}_{\perp\perp}$, and the evolution is one-dimensional in $v_\perp$ at leading order, with
\begin{align}
&D^{(\rms)}_{\perp\perp,{\rm KAW}}
\propto
v_\perp^{\,9-\alpha_\perp-\delta_\perp},
\quad
D^{(\rms)}_{\perp\perp,{\rm InA}}
\propto
v_\perp^{\,\alpha_\perp+\delta_\perp-3}.
\label{eq:Daniso_perp_main}
\end{align}
Unlike the $R$ branch, the KAW and InA branches do not develop a resonant spike in their respective sub-Larmor regimes.

The steady-state DFs follow from constant flux in the dominant direction. For nearly parallel waves, the diffusive flux,
$-D^{(\rms)}_{\parallel\parallel}\,\partial f_{\rms0}/\partial v_\parallel$ is constant. For the $R$ branch below the resonant tangency at $v_\parallel = \sigma_e/\sqrt{2\beta_e}$, the diffusion coefficient is a sum of the whistler-like and electron-cyclotron-like terms that scale as different power-laws, and one of them dominates over the other for some combination of $\alpha_\parallel$ and $\delta_\parallel$. The steady-state DF is given by
\begin{align}
&f_{e0,{\rm wh}}(v_\parallel)\propto
v_\parallel^{-(8-\alpha_\parallel-\delta_\parallel)},
& \alpha_\parallel+\delta_\parallel>6,
\nonumber\\
&f_{e0,{\rm EC}}(v_\parallel)\propto
v_\parallel^{\,4-\alpha_\parallel-\delta_\parallel},
& 4<\alpha_\parallel+\delta_\parallel<6,
\nonumber\\
&f_{i0,{\rm IC}}(v_\parallel)\propto
v_\parallel^{\,4-\alpha_\parallel-\delta_\parallel}.
\label{eq:fpar_main}
\end{align}
Evidently, the power-law exponents are greater than $-2$ for the R-branch, indicating very hard non-thermal tails in the electron distribution. Near the resonant tangency $v_\parallel = \sigma_e/\sqrt{2\beta_e}$, the spike in the $R$-branch diffusion coefficient produces a plateau in the corresponding $f_{\rms0}$. Beyond $v_\parallel=\sigma_e/\sqrt{2\beta_e}$, $D^{(\rms)}_{\parallel\parallel,{\rm R}}$ vanishes, so the DF reverts to a Maxwellian fall-off due to BL relaxation.

For nearly perpendicular waves, $v_\perp D^{(\rms)}_{\perp\perp}\,\partial f_{\rms0}/\partial v_\perp={\rm const}$, which gives
\begin{align}
&f_{i0,{\rm KAW}}(v_\perp)\propto
v_\perp^{-(9-\alpha_\perp-\delta_\perp)},
\nonumber\\
&f_{e0,{\rm InA}}(v_\perp)\propto
v_\perp^{\,3-\alpha_\perp-\delta_\perp}.
\label{eq:fperp_main}
\end{align}
Hence the steady-state tail depends on both the wave family and the anisotropic spectral indices. A detailed solution of the anisotropic diffusion equation~(\ref{eq:FP2D_wave_main}), along with a proper treatment of anisotropy-driven (or current driven) instabilities such as the Weibel instability, is left for future work.

\section{\label{sec:corona}Application to the solar corona}

So far, in our treatment, we have assumed that the coarse-graining scale is smaller than macroscopic scales, e.g., the gravitational scale height $h$ in a plasma under a gravitational field. To study the behavior of the plasma on scales $\sim h$, we must take into account the effect of the attractive gravitational potential. To describe the behavior of the macroscopic thermodynamic quantities such as density, pressure and temperature in the weakly collisional plasma of the solar corona, we can construct an effective fluid model from the kinetic description. Our goal here is to focus on the simplest, qualitative deviation from the original velocity filtration model \citep[][]{Scudder.92a,Scudder.92b,Scudder.94} that addresses the key issue of steep temperature rise from the chromosphere to the corona. As we show below, the non-thermal power-law tails predicted by our kinetic theory formalism play a crucial role in this.

The non-thermal tail in our theory is generated by electric field fluctuations from some large-scale broad-band turbulence or coherent EM waves as discussed above, which we refer to as the drive. The waves originate on super-Larmor scales but cascade down (via wave-wave resonant interactions) to sub-Larmor scales, where they heat the plasma through wave-particle interactions (Landau resonances). Below we discuss whether any of these agents have enough power to overcome collisions and generate a suprathermal tail.

\subsection{Criteria for emergence of the non-thermal tail}\label{sec:criteria}

A non-thermal tail forms once the drive-induced diffusion exceeds the internal Balescu--Lenard (BL) diffusion. A useful onset estimate is therefore obtained by comparing the drive coefficient to $D_{\rm BL}^{(\rms)}(v)$ at the thermal speed of each species (see Appendix~\ref{App:threshold_wave} for details). For a steep super-Debye turbulent spectrum ($\alpha\simeq 5$), the
thermal-velocity thresholds are
\begin{align}
\left(\frac{E_{\rm drive}}{E_{\rm int}}\right)_{{\rm thr},e}^{\rm turb}(\sigma_e)
&\approx
0.33\,
\left(\frac{\ln\Lambda}{6.95}\right)^{1/2}
\left(\frac{B_0}{100\,{\rm G}}\right)^{-1/2}
\left(\frac{T_e}{10^4\,{\rm K}}\right)^{-3/4}\nonumber\\
&\times\left(\frac{n_e}{10^{11}\,{\rm cm}^{-3}}\right)^{1/2},
\\[4pt]
\left(\frac{E_{\rm drive}}{E_{\rm int}}\right)_{{\rm thr},i}^{\rm turb}(\sigma_i)
&\approx
0.33\,
Z_i^{-3/4}
\left(\frac{T_i}{T_e}\right)^{1/2}
\left(\frac{m_i}{m_e}\right)^{1/4}
\left(\frac{\ln\Lambda}{6.95}\right)^{1/2}\nonumber\\
&\times\left(\frac{B_0}{100\,{\rm G}}\right)^{-1/2}
\left(\frac{T_e}{10^4\,{\rm K}}\right)^{-3/4}
\left(\frac{n_e}{10^{11}\,{\rm cm}^{-3}}\right)^{1/2}.
\end{align}
Thus the ion threshold at the ion thermal speed exceeds the electron threshold by
the factor $Z_i^{-3/4}(T_i/T_e)^{1/2}(m_i/m_e)^{1/4}$; for protons with
$T_i\simeq T_e$, this is $\simeq 6.5$.

For coherent waves, the lowest threshold is obtained on the electron-scale $R$ branch. Below $v_\parallel=\omega_{\rm c e}d_e/2=\sigma_e/\sqrt{2\beta_e}$, the exact coefficient contains both the whistler and electron-cyclotron roots. Evaluated at $v_\parallel=\sigma_e$, the resulting thermal threshold for a representative parallel spectrum with $p_\parallel=5$ is
\begin{align}
\left(\frac{E_{\rm drive}}{E_{\rm int}}\right)_{{\rm thr},e}^{R}(\sigma_e)
&\approx
2.1\,
\frac{\left[1-\left((k_{\min,\parallel}\lambda_{\rm c e})
(\omega_{\rm c e}/\omega_{\rm P e})\right)^2\right]^{1/2}}
{k_{\min,\parallel}\lambda_{\rm c e}} .
\end{align}
For $k_{\min,\parallel}\sim \lambda_{\rm c e}^{-1}$ this threshold is of order unity, but it is sharply reduced as the resonant spike at
$v_\parallel=\sigma_e/\sqrt{2\beta_e}$ is approached from below. 

By contrast, the other wave families require much larger amplitudes around thermal speeds. For electron-scale inertial-Alfv\'en waves one has, schematically,
\[
\left(\frac{E_{\rm drive}}{E_{\rm int}}\right)_{{\rm thr},e}^{\rm InA}(\sigma_e)
\sim
4.8\times10^2\,
\left(\frac{\xi_\perp v_{\rm A}}{\sigma_e}\right)^{p_\perp/2-3},
\]
while for ion-cyclotron and kinetic-Alfv\'en waves,
\[
\left(\frac{E_{\rm drive}}{E_{\rm int}}\right)_{{\rm thr},i}^{\rm IC}(\sigma_i)
\sim
2.9\,
Z_i^{-9/4}\left(\frac{T_i}{T_e}\right)^{1/2}
\left(\frac{m_i}{m_e}\right)^{7/4}
\left(\frac{v_{\rm A}}{\sigma_i}\right)^{(p_\parallel-3)/2},
\]
\[
\left(\frac{E_{\rm drive}}{E_{\rm int}}\right)_{{\rm thr},i}^{\rm KAW}(\sigma_i)
\sim
4.1\,
Z_i^{-9/4}\left(\frac{T_i}{T_e}\right)^{1/2}
\left(\frac{m_i}{m_e}\right)^{7/4}
Q_i^{(p_\perp-6)/4}
\]
\[
\times \left(\frac{\xi_\perp v_{\rm A}}{\sigma_i}\right)^{(6-p_\perp)/2}.
\]
The factor $v_{\rm A}/\sigma_i=\sqrt{2Z_iT_e/(\beta_e T_i)}$ is much larger than $1$ in the upper chromosphere due to the smallness of $\beta_e$, and increases the threshold. The Alfv\'enic anisotropy factor $k_\parallel/k_\perp\sim \xi_\perp\ll1$ tries to bring it down. But the large factor $(m_i/m_e)^{7/4}$ makes the InA, KAW, and ion-cyclotron thresholds much larger than the $R$-branch electron threshold in upper-chromospheric conditions. In practice, this means that, while electrons can be readily heated by broad-band turbulence or by electron-scale whistler and electron-cyclotron waves (through wave-particle Landau resonance), ions can only be heated by a turbulent drive in the current treatment of unmagnetized plasma response. Once self-consistent magnetic response is taken into account, cyclotron resonances can boost both ion and electron heating, something we leave for follow-up work.

Our quasilinear calculation hinges on the smallness of the expansion parameter $\epsilon$. For the solar application, it is therefore useful to estimate this parameter under typical conditions of the upper chromosphere (see Appendix~\ref{App:corona} for detailed calculations). For electron-heating, the estimate of the electric field carried by the right circularly polarized (R) branch, consisting of the whistler and electron-cyclotron waves gives $\epsilon \sim [2/(\beta_e+2)]\,g_\epsilon\,(\delta B/B_0)$, where $g_\epsilon=\max_\rms\{Z_\rms T_e/T_\rms\}$ is an $\calO(1)$ factor that ensures the same ordering parameter applies to all species. For typical conditions in the upper chromosphere or lower corona, where $\beta_e\ll 1$, we have $k_\parallel d_e = \sqrt{2/\beta_e} \gg 1$, so the R branch consists of electron-cyclotron rather than whistler waves. Thus we have $\epsilon \approx g_\epsilon\,(\delta B/B_0)$, so the smallness of $\epsilon$ is only consistent with sub-equipartition magnetic fluctuations, $\delta B/B_0<1$. By contrast, for electron-scale inertial Alfve\'n (InA) and ion-scale kinetic Alfv\'en waves, $\epsilon$ is suppressed by the anisotropy factor $k_\parallel/k_\perp\ll 1$, so that $\epsilon<1$ can in principle be satisfied even for $\delta B/B_0\sim 1$ if the turbulence is sufficiently anisotropic. For InAs, $\epsilon$ is also suppressed by the small $\sqrt{m_e/m_i}$ factor. On the other hand, for ion-cyclotron waves, for which $\epsilon \approx g_\epsilon(T_\rms/Z_\rms T_e)(\delta B/B_0)$, we require $\delta B/B_0<1$ for $\epsilon<1$, unless the charge-state or $T_\rms/T_e$ is high, in which case $\delta B/B_0\sim 1$ allows quasilinear ordering.

For the turbulent drive, we have $\epsilon\sim g_\epsilon (\lambda_{\rmc e}/\lambda_{\rmD e})(E_{\rm drive}/E_{\rm int})$, where we have assumed $k_{\rm min}\sim \lambda^{-1}_{\rmc e}$ and $t_\rmc \sim \omega^{-1}_{\rmc e}$. The choice of $k_{\rm min}$ is motivated by the fact that the turbulent power-spectrum is expected to significantly steepen below the electron Larmor radius (electron Landau damping and imbalanced turbulence with helicity barrier are known mechanisms responsible for this), and it is this steepening scale that guides the $v$ dependence of the quasilinear diffusion coefficient. The choice of $t_\rmc \sim \omega^{-1}_{\rmc e}$ is associated with $k_{\rm min}\sim \lambda^{-1}_{\rmc e}$. To beat BL diffusion at the thermal speed, the electrons require a threshold $E_{\rm drive}/E_{\rm int}\sim 0.18\, {(B_0/100 \, \rmG)}^{-1/2}$ for $\alpha = 5$ in upper chromospheric conditions ($n_e = 10^{10}\,{\rm cm}^{-3}$ and $T_e=10^4$ K), which implies $\epsilon \sim 0.26$. The protons, on the other hand, require $(m_p/m_e)^{1/4} \approx 6.54$ times larger threshold to beat collisions at their thermal speed, implying $\epsilon \sim 1.68$, which is formally outside the quasilinear regime. The threshold $E_{\rm drive}/E_{\rm int}$ for protons, however, falls off as $(v/\sigma_p)^{-3/2}$ beyond $\sigma_p$ (since the BL coefficient scales as $(v/\sigma_p)^{-3}$ and the drive coefficient that scales as ${(E_{\rm drive}/E_{\rm int})^2}$ weakly varies with $v$ at $\sigma_p < v < \sigma_e$). This means that an $E_{\rm drive}$ that is weaker by a factor of $1/1.68 \approx 0.6$ or more than the thermal threshold for protons would allow for $\epsilon\lesssim 1$ and still be able to develop a power-law tail at $v \gtrsim {(1.68)}^{2/3}\sigma_p\approx 1.4\sigma_p$. Therefore, the allowed strength of $E_{\rm drive}$ that can generate non-thermal tails substantially overlaps with the regime of validity of QLT under chromospheric conditions; the lower corona widens this regime even further.

\subsection{From micro- to macro-scales}

Now we discuss how to estimate the global form of the DF (as a function of both $\bx$ and $\bv$) from the local form (as a function of $\bv$ alone) calculated so far using QLT. The Vlasov equation dictates that, although the DF is locally a function of $v$, under a global potential $\Phi(\bx)$, it is a function of the energy $E = v^2/2 + \Phi(\bx)$. Therefore, under the combined action of the solar gravitational field and the ambipolar electric field (that keeps the background plasma quasi-neutral), the ion DFs, which locally scale as $v^{-2\left(1 + \kappa_\rms\right)}$, become \citep[][]{Scudder.92a}
\begin{align}
f_{\rms 0}(v,r)\sim {\left[1 + \frac{v^2 + 2\left(\Phi_{\rm eff}^{(\rms)}(r) - \Phi_{\rm eff}^{(\rms)}(r_0)\right)}{2\kappa_\rms\sigma^2_{\rms 0}}\right]}^{-\left(1+\kappa_\rms\right)},
\end{align}
where $r$ is the radial distance from the center, $r_0$ is some reference radius, $\sigma_{\rms 0}$ is the thermal speed at $r_0$, $\Phi_\rmG(r) \approx -G M_{\odot}/r$ is the gravitational potential ($M_\odot$ is the solar mass), and $\kappa_\rms$ is the value of $\kappa$ for the $\rms^{\rm th}$ ionic species. Here we have used the fact that the total specific potential for an ion of charge state $Z_\rms$ and mass number $A_\rms$ is
\begin{align}
\Phi_{\rm eff}^{(\rms)}(r) \equiv \Phi_\rmG(r) + \frac{Z_\rms e}{m_\rms}\Phi_\rmE(r)
\approx \left(1-\frac{Z_\rms}{2A_\rms}\right)\Phi_\rmG(r),
\end{align}
where $\Phi_\rmE(r) \approx -(m_p/2e)\,\Phi_\rmG(r)$ is the ambipolar electrostatic potential set primarily by the electron--proton plasma. Thus protons feel an effective specific potential $\Phi_\rmG/2$, while minor ions with $A_\rms/Z_\rms \approx 2$ feel approximately $3\Phi_\rmG/4$. As we showed earlier, kinetic turbulence implies a particular preference for $1.5 \leq \kappa_\rms \leq 3$ $(v^{-5}-v^{-8})$, which lies within a plausible range for protons and heavy ions in the fast solar wind (see \citet[][]{Pierrard.Lazar.10} and references therein).

There is a potential complication introduced to this picture of non-thermal tails by the presence of collisions in the plasma. The mean free path $\lambda_{\rm mfp}$ of both ions and electrons in the coronal plasma is equal to $\left(\Lambda/\ln\Lambda\right) \lambda_{\rmD e} \approx 1.3 \, {\left(T_e/10^4 K\right)^2 {\left(n_e/10^{10} {\rm cm}^{-3}\right)}^{-1}} {\rm cm}$ $= 13\, {\left(T_e/10^6 K\right)^2 {\left(n_e/10^8 {\rm cm}^{-3}\right)}^{-1}} {\rm km}$, which is significantly smaller than the typical pressure scale height $L \sim 10^3$ km. In other words, the Knudsen number $\lambda_{\rm mfp}/L$ is small. This may lead us to believe that, despite the weakly collisional nature of the plasma $(\Lambda \gg 1)$, the DF loses its kinetic power-law tails and Maxwellianizes on scales above $\lambda_{\rm mfp}$ as the particles traverse a few mean free paths. This, however, would be an incorrect conclusion, since the drive diffusion coefficient, responsible for spawning the power-law tail, exceeds the Maxwellianizing BL coefficients over a large velocity range, $\sigma_\rms < v < \omega_{\rmP e}/k_{\rm min}$, as long as large-scale EM turbulence drives the plasma. Even after the drive is switched off, a power-law tail can survive collisional relaxation since the BL coefficients are small at large $v$ (see Fig.~\ref{fig:fe_vs_t_after_shock}). The free-streaming scale of the non-thermal particles is $v^2/\calD_1 \sim \lambda_{\rm mfp}\,{\left(v/\sigma_\rms\right)}^4$ ($\calD_1\sim v^{-2}$ at $v > \sigma_\rms$), which can be quite large in the high $v$ tail, and exceeds the mean free path by orders of magnitude. Recall that inter-species collisions tend to isotropize the DF rather than alter its functional dependence on $v$ when the masses are vastly different (e.g., electrons and ions), which is why the intra-species BL coefficients are relevant here. Therefore, the power-law tail would survive over large distances in a kinetic plasma driven by large-scale EM fluctuations, as the fast unscreened particles get accelerated without dissipating their momentum and energy through collisions.

An additional complication is introduced by the potential presence of a strong guide magnetic field. Since the plasma response is unmagnetized, our theory for non-thermal tail formation is strictly only applicable to plasma on sub-Larmor radius scales, although the EM fluctuations responsible can in principle arise on much larger scales. Under typical conditions of the solar corona, the Debye length is $\lambda_{\rmD e} \approx 0.68\, {\left(T_e/10^6\,{\rm K}\right)^{1/2} {\left(n_e/10^8\,{\rm cm}^{-3}\right)}^{-1/2}}$ cm, the electron Larmor radius is $\lambda_{\rmc e} \approx 2.2\,{\left(T_e/10^6\,{\rm K}\right)^{1/2} {\left(B_0/10\,{\rm G}\right)}^{-1}}$ cm, and the ion Larmor radius is $\lambda_{\rmc i} \approx 94.7\,\mu^{1/2} Z^{-1} {\left(T_i/10^6\,{\rm K}\right)^{1/2} {\left(B_0/10\,{\rm G}\right)}^{-1}}$ cm. Thus the Debye and Larmor scales are well separated, and both are much smaller than the mean free path or the pressure scale height. This scale separation implies that there exists a broad range of sub-Larmor but super-Debye scales in the corona, where our theory is applicable.

\subsection{Constructing an effective fluid model from the distribution function}\label{sec:EOS}

So far, we have seen that the ion DFs relax to a $\kappa$ distribution,
\begin{align}
&f_{\rms 0}(v,r) = \calN_\rms\, {\left[1 + \frac{v^2 + 2\left(\Phi_{\rm eff}^{(\rms)}(r) - \Phi_{\rm eff}^{(\rms)}(r_0)\right)}{2\kappa_\rms\sigma^2_{\rms 0}}\right]}^{-\left(1+\kappa_\rms\right)},\nonumber\\
&\calN_\rms = \frac{1}{{\left(2\pi\kappa_\rms\right)}^{3/2}} \frac{\Gamma\left(\kappa_\rms+1\right)}{\Gamma\left(\kappa_\rms-1/2\right)},
\end{align}
with a preference for $1.5 \leq \kappa_\rms \leq 3$ and a Maxwellian truncation beyond $v_{\rm max} = \omega_{\rmP e}/k_{\rm min}$. The ion DF is often a broken power-law that can be expressed as a linear combination of two or more such $\kappa$ distributions. The local thermodynamic variables such as density, pressure and temperature are nothing but the velocity moments of the DF. The total mass density, $\rho_\rms(r) = 4\pi m_\rms \int \rmd v\, v^2 f_{\rms 0}(v,r)$, and the total pressure, $p_\rms(r) = (4\pi m_\rms/3) \int \rmd v\, v^4 f_{\rms 0}(v,r)$, can be expressed as
\begin{align}
\rho_\rms(r) &= 4\pi \calN_\rms m_\rms n_\rms \;B\left(\chi^2_\rms(r),\frac{3}{2},\kappa_\rms-\frac{1}{2}\right) \nonumber\\
&\times {\left[1 + \frac{\Phi_{\rm eff}^{(\rms)}(r) - \Phi_{\rm eff}^{(\rms)}(r_0)}{\kappa_\rms\sigma^2_{\rms 0}}\right]}^{1/2-\kappa_\rms},\nonumber\\
p_\rms(r) &= \frac{8\pi \calN_\rms m_\rms n_\rms \kappa_\rms\sigma^2_{\rms 0}}{3}\; B\left(\chi^2_\rms(r),\frac{5}{2},\kappa_\rms-\frac{3}{2}\right) \nonumber\\
&\times {\left[1 + \frac{\Phi_{\rm eff}^{(\rms)}(r) - \Phi_{\rm eff}^{(\rms)}(r_0)}{\kappa_\rms\sigma^2_{\rms 0}}\right]}^{3/2-\kappa_\rms},
\label{p_rho_app}
\end{align}
with
\begin{align}
\chi_\rms(r) = \left(\frac{v_{\rm max}}{\sqrt{v^2_{\rm max} + 2\left[\Phi_{\rm eff}^{(\rms)}(r) - \Phi_{\rm eff}^{(\rms)}(r_0) + \kappa_\rms\sigma^2_{\rms 0}\right]}}\right),
\end{align}
where $v_{\rm max} \approx {\rm min}\left(\omega_{\rmP e}/k_{\rm min},v_{\rm esc}\right)$, $v_{\rm esc}$ being the escape speed, and $B\left(z,a,b\right) = \int_0^z \rmd u\, u^{a-1} {\left(1-u\right)}^{b-1}$ is the incomplete beta function. From the above equations, we see that, for $\kappa_\rms>3/2$, the plasma has an effective equation of state (EOS) that is polytropic, with $p_\rms\propto \rho^{\gamma_\rms}_\rms$, with $\gamma_\rms = \left(\kappa_\rms-3/2\right)/\left(\kappa_\rms-1/2\right)$ \citep[c.f.][]{Scudder.92a}. The polytropic coefficient tends to $1$ at $\kappa_\rms \to \infty$, i.e., the plasma becomes isothermal for a Maxwellian DF, as expected. For any non-thermal power law tail, however, we have $\gamma_\rms < 1$, i.e., the EOS is always softer than an isothermal one for a kinetic plasma. This is because the collisionless nature of the plasma implies that the pressure does not sufficiently increase with increasing density.

\begin{figure*}
\centering
\includegraphics[width=1\textwidth]{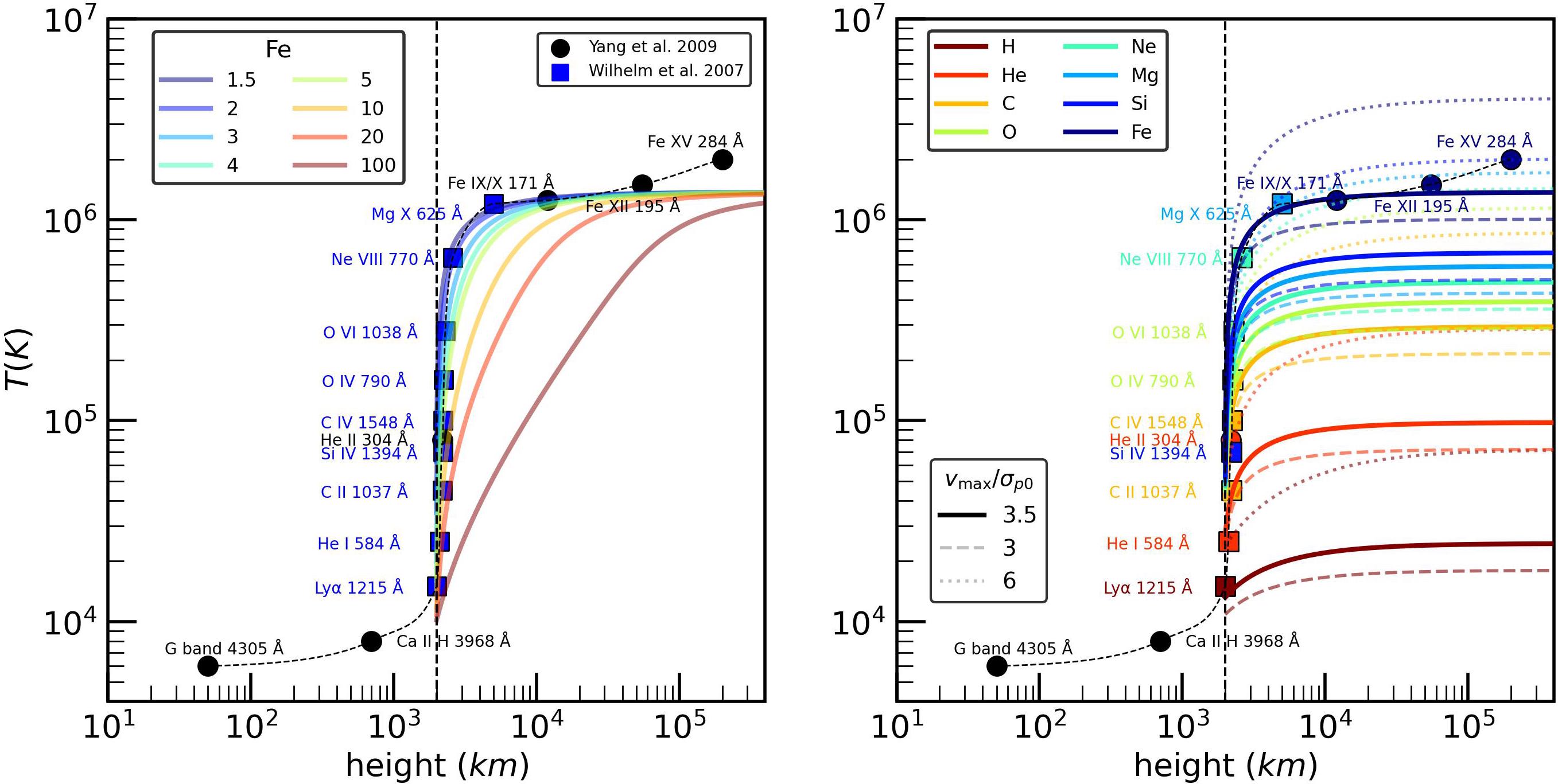}
\caption{Effective temperature profile (including Maxwellian bulk as well as non-thermal velocities) in the corona as a function of height above the photosphere, obtained from equations~(\ref{p_rho_app}) and (\ref{T_profile}) using a single-$\kappa$ distribution. Left panel: predictions of the velocity-filtration model for Fe for different values of $\kappa$, assuming $v_{\rm max}=3.5\,\sigma_{p0}$. Right panel: predictions for different species at fixed $\kappa=1.5$; the solid, dashed, and dotted curves correspond to $v_{\rm max}/\sigma_{p0}=3.5$, $3$, and $6$, respectively. The vertical dashed line marks the reference height $r_0-R_\odot=2\times10^3$ km, corresponding to the upper end of the chromosphere. The symbols denote characteristic effective temperatures inferred from UV/EUV spectroscopic line-width diagnostics of ionic species \citep[][]{Peter.01}, adapted from \citet[][]{Wilhelm.etal.07} and \citet[][]{Yang.etal.09}. The black dashed curve shows a smooth interpolation through the data points. The single-$\kappa$ filtration model predicts a strong mass dependence, with heavier ions reaching much higher post-filtration temperatures than lighter species. The narrow physical transition from the upper-chromospheric temperature $T_{\rms 0}$ to the pre-filtration value $T_\rms(r\to r_0^+)$ is not resolved in these profiles. Credit for data adapted from \citet[][]{Yang.etal.09}: Yang et al., ``Response of the solar atmosphere to magnetic field evolution in a coronal hole region'', A\&A, 501, 745--753, 2009, reproduced with permission \textcopyright~ESO. Credit for data adapted from \citet[][]{Wilhelm.etal.07}: Wilhelm et al., ``Observations of the Sun at Vacuum-Ultraviolet Wavelengths from Space. Part II: Results and Interpretations'', Space Science Reviews, 133, 103--179, 2007, Springer Nature.}
\label{fig:T_corona}
\end{figure*}

The effective temperature $T_\rms = \left(m_\rms/k_\rmB\right)\left(p_\rms/\rho_\rms\right)$ can be obtained from equations~(\ref{p_rho_app}) and is given by
\begin{align}
T_\rms(r) &= \frac{2\kappa_\rms}{3} \frac{B\left(\chi^2_\rms(r),\frac{5}{2},\kappa_\rms-\frac{3}{2}\right)}{B\left(\chi^2_\rms(r),\frac{3}{2},\kappa_\rms-\frac{1}{2}\right)}\nonumber\\
&\times\left[T_{\rms 0} \, + \, \frac{m_\rms}{\kappa_\rms k_\rmB}\left[\Phi_{\rm eff}^{(\rms)}(r) - \Phi_{\rm eff}^{(\rms)}(r_0)\right] \right],
\label{T_profile}
\end{align}
where we have defined $\sigma^2_{\rms 0} = k_\rmB T_{\rms 0}/m_\rms$, $T_{\rms 0}$ being the temperature at the reference radius $r_0$. This, in the limit of $v_{\rm max} \gg \sqrt{2\left[\Phi_{\rm eff}^{(\rms)}(r) - \Phi_{\rm eff}^{(\rms)}(r_0)\right] + 3\sigma^2_{\rms 0}}$ or $\chi_\rms(r) \approx 1$, reduces to the following form:
\begin{align}
T_\rms(r) &\approx \frac{\kappa_\rms}{\kappa_\rms - 3/2} \left[T_{\rms 0} \, + \, \frac{m_\rms}{\kappa_\rms k_\rmB}\left[\Phi_{\rm eff}^{(\rms)}(r) - \Phi_{\rm eff}^{(\rms)}(r_0)\right] \right].
\end{align}
Evidently, $T_\rms$ diverges at $\kappa_\rms \to 3/2$ ($\gamma_\rms \to 0$), since the pressure diverges, provided that $v_{\rm max} \gg \sqrt{2\left[\Phi_{\rm eff}^{(\rms)}(r) - \Phi_{\rm eff}^{(\rms)}(r_0)\right] + 3\sigma^2_{\rms 0}}$. This divergence is, however, logarithmic in $v_{\rm max}/\sqrt{2\left[\Phi_{\rm eff}^{(\rms)}(r) - \Phi_{\rm eff}^{(\rms)}(r_0)\right] + 3\sigma^2_{\rms 0}}$, as we shall now see. The density, pressure and effective temperature for $\kappa_\rms = 3/2$ become
\begin{align}
\rho_\rms(r) &= 4\pi\calN_\rms m_\rms n_\rms\,\frac{\chi^3_\rms(r)}{3}\, {\left[1 + \frac{2}{3}\frac{\Phi_{\rm eff}^{(\rms)}(r) - \Phi_{\rm eff}^{(\rms)}(r_0)}{\sigma^2_{\rms 0}}\right]}^{-1},\nonumber\\
p_\rms(r) &= 4\pi\calN_\rms m_\rms n_\rms \sigma^2_{\rms 0}\, \left[ \frac{1}{2}\ln{\left|\frac{1 + \chi_\rms(r)}{1 - \chi_\rms(r)}\right|} - \left(\,\chi_\rms(r) + \frac{\chi^3_\rms(r)}{3}\right)\right],\nonumber\\
T_\rms(r) &= \frac{3}{\chi^3_\rms(r)}\left[ \frac{1}{2}\ln{\left|\frac{1 + \chi_\rms(r)}{1 - \chi_\rms(r)}\right|} - \left(\,\chi_\rms(r) + \frac{\chi^3_\rms(r)}{3}\right)\right]\nonumber\\
&\times \left[T_{\rms 0} \, + \, \frac{2 m_\rms}{3 k_\rmB}\left[\Phi_{\rm eff}^{(\rms)}(r) - \Phi_{\rm eff}^{(\rms)}(r_0)\right] \right].
\label{kappa_1.5_limit}
\end{align}
In the limit of $v_{\rm max} \gg \sqrt{2\left[\Phi_{\rm eff}^{(\rms)}(r) - \Phi_{\rm eff}^{(\rms)}(r_0)\right] + 3\sigma^2_{\rms 0}}$, we have $\chi_\rms(r) \approx 1 - \left[\Phi_{\rm eff}^{(\rms)}(r) - \Phi_{\rm eff}^{(\rms)}(r_0) + 3\sigma^2_{\rms 0}/2\right]/v^2_{\rm max}$, and the corresponding density is $\rho_\rms(r) = (4\pi/3)\,{\left[1 + 2{\left(\Phi_{\rm eff}^{(\rms)}(r) - \Phi_{\rm eff}^{(\rms)}(r_0)\right)}/{3\sigma^2_{\rms 0}}\right]}^{-1}$, while the pressure is $p_\rms(r)=2\pi \calN_\rms m_\rms n_\rms \sigma^2_{\rms 0}\ln{\left(4v^2_{\rm max}\,\rho_\rms(r)/3\sigma^2_{\rms 0}\,\rho_\rms(r_0)\right)}$. This limit applies to $\rho_\rms(r) \gg (3\sigma^2_{\rms 0}/v^2_{\rm max})\,\rho_\rms(r_0)$, i.e., near the coronal base where the density is large. In the opposite limit of $v_{\rm max} \ll \sqrt{2\left[\Phi_{\rm eff}^{(\rms)}(r) - \Phi_{\rm eff}^{(\rms)}(r_0)\right] + 3\sigma^2_{\rms 0}}$, $\chi_\rms(r) \approx {v_{\rm max}}/{\sqrt{2\left[\Phi_{\rm eff}^{(\rms)}(r) - \Phi_{\rm eff}^{(\rms)}(r_0)\right] + 3\sigma^2_{\rms 0}}}$, which implies that $p_\rms(r)\propto \rho_\rms(r) \sim {\left[1 + {2\left(\Phi_{\rm eff}^{(\rms)}(r) - \Phi_{\rm eff}^{(\rms)}(r_0)\right)}/{3\sigma^2_{\rms 0}}\right]}^{-5/2}$. This holds for $\rho_\rms(r) \ll (3\sigma^2_{\rms 0}/v^2_{\rm max})\,\rho_\rms(r_0)$, i.e., a low density environment. Thus, for $\kappa_\rms = 3/2$, pressure increases very mildly (logarithmically) with density at high density, but linearly with density at low density.

Often, the DF, especially that of the ions, is a broken power-law, as shown in Fig.~\ref{fig:fi_vs_t}, in which case it can be expressed as a linear combination of different $\kappa_\rms$ distributions with different thermal speeds and cut-offs. If we assume
\begin{align}
f_{\rms 0}(v,r) &= (1 - x) f_{\rms 0}^{(1)}\left(v,r,\kappa_{\rms 1},\sigma_{\rms 0}^{(1)},v_{\rm max}^{(1)}\right) \nonumber\\
&+ x f_{\rms 0}^{(2)}\left(v,r,\kappa_{\rms 2},\sigma_{\rms 0}^{(2)},v_{\rm max}^{(2)}\right),
\label{double_kappa_DF}
\end{align}
with $0 \leq x \leq 1$, then the effective temperature can be expressed as
\begin{align}
&T_\rms\left(r,\kappa_{\rms 1},\kappa_{\rms 2},\sigma_{\rms 0}^{(1)},v_{\rm max}^{(1)},\sigma_{\rms 0}^{(2)},v_{\rm max}^{(2)}\right) \nonumber\\
&= \frac{m_\rms}{k_\rmB} \frac{(1 - x)\, p_{\rms 1}\left(r,\kappa_{\rms 1},\sigma_{\rms 0}^{(1)},v_{\rm max}^{(1)}\right) + x\,p_{\rms 2}\left(r,\kappa_{\rms 2},\sigma_{\rms 0}^{(2)},v_{\rm max}^{(2)}\right)}{(1 - x)\, \rho_{\rms 1}\left(r,\kappa_{\rms 1},\sigma_{\rms 0}^{(1)},v_{\rm max}^{(1)}\right) + x\,\rho_{\rms 2}\left(r,\kappa_{\rms 2},\sigma_{\rms 0}^{(2)},v_{\rm max}^{(2)}\right)},
\label{Ts_2_kappa}
\end{align}
where the density and pressure obtained from each individual $\kappa$ distribution are $\rho_{\rms i}(r,\kappa_{\rms i},\sigma_{\rms 0}^{(i)},v_{\rm max}^{(i)}) = 4\pi m_\rms \int \rmd v\, v^2 f_{\rms 0}^{(i)}(v,r,\kappa_{\rms i},\sigma_{\rms 0}^{(i)},v_{\rm max}^{(i)})$ and $p_{\rms i}(r,\kappa_{\rms i},\sigma_{\rms 0}^{(i)},v_{\rm max}^{(i)}) = (4\pi m_\rms/3) \int \rmd v\, v^4 f_{\rms 0}^{(i)}(v,r,\kappa_{\rms i},\sigma_{\rms 0}^{(i)},v_{\rm max}^{(i)})$, with $i = 1,2$. Since both pressure and density fall off (radially) faster for larger $\kappa_\rms$, the temperature is determined by the smaller value of $\kappa_\rms$ at large radii. This is the essence of velocity filtration \citep[][]{Scudder.92a,Scudder.92b,Meyer-Vernet.07}, the fact that the attractive potential only allows the harder/shallower power-law component to escape.

\subsection{Velocity filtration, steep transition, and temperature inversion}\label{sec:vel_filt}

Let us now discuss the behavior of the thermodynamic quantities in the solar atmosphere. As shown above, the profiles for the density and temperature for a single $\kappa_\rms$ distribution are given by equations~(\ref{p_rho_app}) and~(\ref{T_profile}) for $\kappa_\rms > 3/2$. The $\kappa_\rms = 3/2$ quantities are given by equations~(\ref{kappa_1.5_limit}). Substituting $\Phi_\rmG(r) = -G M_\odot/r$ and $\Phi_{\rm eff}^{(\rms)}(r) = \left(1-{Z_\rms}/{2A_\rms}\right)\Phi_\rmG(r)$ in the above yields the following radial dependence:
\begin{align}
\rho_\rms(r) &= 4\pi \calN_\rms m_\rms n_\rms \; B\left(\chi^2_\rms(r),\frac{3}{2},\kappa_\rms-\frac{1}{2}\right) \nonumber\\
&\times {\left[1 + \left(A_\rms-\frac{Z_\rms}{2}\right)\frac{T_{\rm vir}}{\kappa_\rms T_{\rms 0}}\left(\frac{R_\odot}{r_0} - \frac{R_\odot}{r}\right)\right]}^{1/2-\kappa_\rms},\nonumber\\
T_\rms(r) &= \frac{2\kappa_\rms}{3} \frac{B\left(\chi^2_\rms(r),\frac{5}{2},\kappa_\rms-\frac{3}{2}\right)}{B\left(\chi^2_\rms(r),\frac{3}{2},\kappa_\rms-\frac{1}{2}\right)} \nonumber\\
&\times \left[T_{\rms 0} \, + \, \left(A_\rms-\frac{Z_\rms}{2}\right)\frac{T_{\rm vir}}{\kappa_\rms} \left(\frac{R_\odot}{r_0} - \frac{R_\odot}{r}\right) \right],
\label{rho_T_profile_sun}
\end{align}
with $R_\odot$ the solar radius, the virial temperature $T_{\rm vir}$ given by
\begin{align}
T_{\rm vir} = \frac{G M_\odot m_p}{k_\rmB R_\odot},
\end{align}
and
\begin{align}
\chi_\rms(r) &= \sqrt{\dfrac{T_{{\rm max},\rms}}{T_{{\rm max},\rms} + 2\kappa_\rms T_{\rms 0} + 2\left(A_\rms-\dfrac{Z_\rms}{2}\right)T_{\rm vir}\left(\dfrac{R_\odot}{r_0} - \dfrac{R_\odot}{r}\right)}}\,,\nonumber\\
T_{{\rm max},\rms} &= \frac{m_\rms v^2_{\rm max}}{k_\rmB}.
\end{align}
We have assumed that the temperature at the reference radius $r_0$ of the upper chromosphere is $T_{\rms 0} = m_\rms \sigma^2_{\rms 0}/k_\rmB$. In other words, we have assumed that the plasma is nearly Maxwellian (large $\kappa$) in the collisional environment of the upper chromosphere. An important point to note is that $T_{\rms 0}$ is not equal to $T_\rms(r\to r_0^+)$, which is given by
\begin{align}
T_\rms(r\to r_0^+) &= \frac{2\kappa_\rms}{3} \frac{B\left(\chi^2_\rms(r\to r_0^+),\frac{5}{2},\kappa_\rms-\frac{3}{2}\right)}{B\left(\chi^2_\rms(r\to r_0^+),\frac{3}{2},\kappa_\rms-\frac{1}{2}\right)} T_{\rms 0},\nonumber\\
&\xrightarrow[]{T_{{\rm max},\rms}/2\kappa_\rms T_{\rms 0}\to \infty} \frac{\kappa_\rms}{\kappa_\rms - 3/2} T_{\rms 0},\nonumber\\
\chi_\rms(r\to r_0^+) &= \chi_{\rms 0} = \sqrt{\dfrac{T_{{\rm max},\rms}}{T_{{\rm max},\rms} + 2\kappa_\rms T_{\rms 0}}}.
\end{align}
Note that, for $\kappa_\rms \to 3/2$, $T_\rms(r\to r^+_0)$ heavily exceeds $T_{\rms 0}$, since the pressure diverges (but the density remains finite) in this limit. For large $\kappa_\rms$, $T_\rms \approx T_{\rms 0}$, i.e., the plasma is nearly isothermal, as expected. This apparent discontinuity is not a physical inconsistency, rather it indicates that there exists a thin layer where the non-thermal tail is generated from an initially Maxwellian DF. We estimate the thickness of this section based on our quasilinear analysis towards the end of this section.

The above scalings can also be obtained by solving the equation for hydrostatic equilibrium,
\begin{align}
\frac{\rmd p_\rms}{\rmd r} + \rho_\rms\frac{\rmd \Phi_{\rm eff}^{(\rms)}}{\rmd r} = 0,
\end{align}
with a macroscopic equation of state (EOS), $p_\rms = p_\rms\left(\rho_\rms\right)$, obtained from equations~(\ref{p_rho_app}) as shown above. For $\kappa_\rms > 1.5$, the EOS is approximately polytropic, i.e., $p_\rms \propto \rho_\rms^{\gamma_\rms}$ with $\gamma_\rms = \left(\kappa_\rms - 3/2\right)/\left(\kappa_\rms - 1/2\right)$ (as noted by \citet[][]{Scudder.92a}), while, for $\kappa_\rms = 1.5$, it is a much milder $p_\rms \propto \ln{\left(4v^2_{\rm max}\,\rho_\rms(r)/3\sigma^2_{\rms 0}\,\rho_\rms(r_0)\right)}$.

\begin{figure}
\centering
\includegraphics[width=0.48\textwidth]{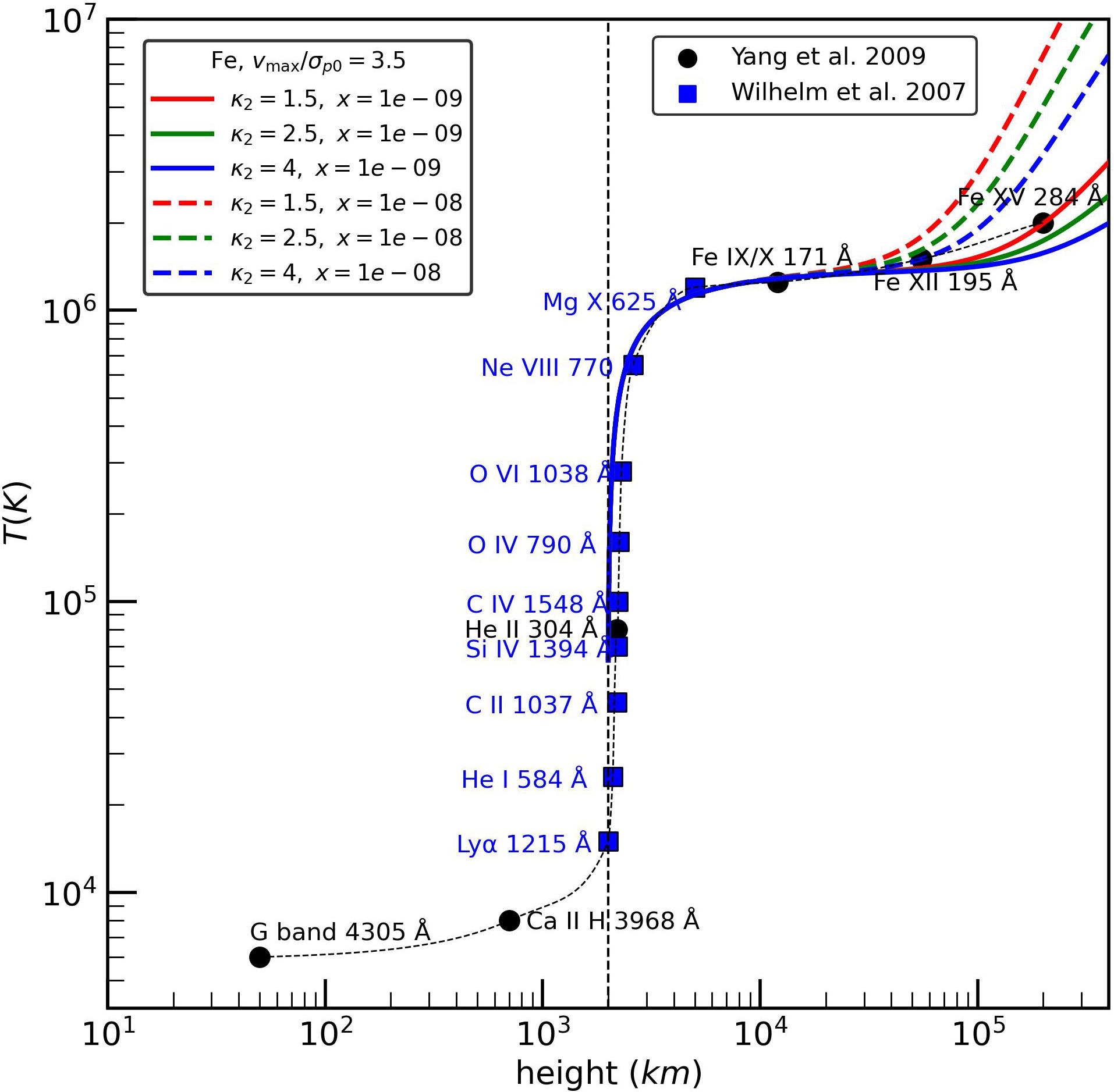}
\caption{Effective temperature profile in the corona as a function of height above the photosphere, obtained from equation~(\ref{Ts_2_kappa}) using the double-$\kappa$ distribution in equation~(\ref{double_kappa_DF}) for Fe, assuming $v_{\rm max}/\sigma_{p0}=3.5$. The solid and dashed curves correspond to $x=10^{-9}$ and $10^{-8}$, respectively, while the colors indicate $\kappa_2=1.5,2.5$ and $4$. The vertical dashed line marks the reference height $r_0-R_\odot=2\times10^3$ km, corresponding to the upper end of the chromosphere. The symbols denote characteristic effective temperatures inferred from UV line-width diagnostics of ionic species \citep[][]{Peter.01}, adapted from \citet[][]{Wilhelm.etal.07} and \citet[][]{Yang.etal.09}, and the black dashed curve shows a smooth interpolation through these points. As in Fig.~\ref{fig:T_corona}, the narrow transition layer connecting the upper-chromospheric temperature $T_{\rms 0}$ to the pre-filtration value $T_\rms(r\to r_0^+)$ is not resolved in the plotted profiles. Credit for data adapted from \citet[][]{Yang.etal.09}: Yang et al., ``Response of the solar atmosphere to magnetic field evolution in a coronal hole region'', A\&A, 501, 745--753, 2009, reproduced with permission \textcopyright~ESO. Credit for data adapted from \citet[][]{Wilhelm.etal.07}: Wilhelm et al., ``Observations of the Sun at Vacuum-Ultraviolet Wavelengths from Space. Part II: Results and Interpretations'', Space Science Reviews, 133, 103--179, 2007, Springer Nature.}
\label{fig:T_corona_Fe}
\end{figure}

It is evident from equations~(\ref{rho_T_profile_sun}) that the temperature (density) is an increasing (decreasing) function of $r$ in the corona. This is because the collisionless nature of the plasma yields a softer than isothermal EOS. As $\rho_\rms$ decreases outwards due to gravity, $p_\rms$ does so too but not quite enough, so that $T_\rms\sim p_\rms/\rho_\rms$ rises outwards. As $\kappa_\rms$ approaches $1.5$, the EOS gets softer and the temperature rises dramatically. This temperature inversion due to a non-thermal distribution function is known as velocity filtration \citep[][]{Scudder.92a,Scudder.92b,Meyer-Vernet.07}, which refers to the fact that the gravitational potential well of the sun filters out the fast suprathermal particles and lets them escape.

Before we describe the theoretical $T_\rms$ profiles predicted by filtration, let us provide a brief summary of the relevant observations. In the upper chromosphere and lower corona ($h\sim10^3$--$10^5$ km above the photosphere), electron temperatures inferred from line-ratio or ionization diagnostics are of order $T_e\sim 0.8$--$1.5\times10^6$ K \citep[][]{Fludra.etal.99}, whereas HI/Ly$\alpha$-based diagnostics near the coronal base indicate neutral-hydrogen (and proton-proxy) kinetic temperatures of order $10^5$--$2\times10^5$ K \citep[][]{Marsch.etal.99}. By contrast, UV/EUV line widths of heavier ions in this same height range imply effective or kinetic temperatures of order $10^5$--$10^6$ K \citep[][]{Peter.01,Wilhelm.etal.07}. These linewidth-based temperatures should be interpreted as second-moment measures of the emitting-ion velocity distribution and generally include unresolved non-thermal broadening unless modeled separately \citep[][]{Wilhelm.etal.07,Cranmer.09}. Farther out, in the extended corona ($r\sim1.5$--$3\,R_\odot$), UVCS/SOHO measurements show that H$^0$ Ly$\alpha$ widths (used as a proxy for proton properties below $\sim 3\,R_\odot$) correspond to kinetic temperatures of order a few MK, although part of this broadening may be due to unresolved waves \citep[][]{Cranmer.09}, while O$^{5+}$ reaches perpendicular kinetic temperatures $\gtrsim10^8$ K and exhibits strong anisotropy, $T_\perp\gg T_\parallel$ \citep[][]{Kohl.etal.98,Cranmer.etal.99,Cranmer.etal.08}. Thus the observations point to a strong separation between the comparatively modest light-species temperatures and the much larger effective temperatures of heavy ions, with the latter becoming strongly anisotropic in the extended corona \citep[][]{Cranmer.09}.

How does the velocity filtration model with a single $\kappa$ distribution compare to the observed temperature profile of the solar corona? Assuming the reference radius $r_0$ to be the upper end of the chromosphere, i.e., $r_0 = R_\odot + h$ with $R_\odot = 7\times 10^5$ km and $h = 2\times 10^3$ km, $T_{\rms 0} = 10^4$ K, and $M_\odot = 2\times 10^{30}$ kg, the corresponding temperature profile is computed as a function of the height above the photosphere, and plotted for the case of Fe $(m_\rms = 56)$ for different values of $\kappa$ in the left panel Fig.~\ref{fig:T_corona}, adopting $v_{\rm max} = 3.5\,\sigma_{p0} = 3.5\sqrt{m_\rms/m_p}\,\sigma_{\rms 0}$, and for different species in the right panel, assuming $\kappa_\rms = 1.5$. In the latter case, we show the $v_{\rm max}=3.5\,\sigma_{p0}$ case as solid lines, together with fainter comparison curves for $v_{\rm max}=3\,\sigma_{p0}$ and $6\,\sigma_{p0}$. The symbols denote the temperature inferred from the observed spectral (UV) linewidths of heavy ionic species in the upper chromosphere and lower corona \citep[][]{Peter.01,Wilhelm.etal.07,Yang.etal.09}. These should be interpreted as effective temperatures inferred from linewidths, which combine thermal broadening with non-thermal motions\citep[][]{Wilhelm.etal.07}. The choice $v_{\rm max}\sim {\rm few}\,\sigma_{p0}$ reproduces the coronal temperature of Fe reasonably well and is consistent with the fact that the first break in the ion DF is expected to lie between $\sigma_{p0}$ and $\sigma_{e0}$, if the drive is not too strong (see Fig.~\ref{fig:fi_vs_t}). While $T_\rms$ increases outwards in all cases, only for $\kappa_\rms \approx 1.5-3$ it rises as drastically as the observed profile, with a pressure scale height less than $10^3$ km. For a Maxwellian-like (large $\kappa_\rms$) DF, the EOS is nearly isothermal and $T_\rms$ gradually rises, with a large scale height $\gtrsim 10^4$ km. For $\kappa_\rms$ closer to $3/2$, the base temperature exceeds the chromospheric temperature $T_{\rms 0} = 10^4$ K by a greater degree. $T_{\rms}$ rises rapidly to a steady value that roughly scales linearly with the atomic mass through the ambipolar factor $A_\rms-Z_\rms/2$ ($\approx 3A_\rms/4$ for heavier elements, since $Z_\rms\approx A_\rms/2$ in an H-dominated plasma), increases with increasing $v_{\rm max}$, and is as large as $10^6$ K for heavier elements. As shown in the right panel of Fig.~\ref{fig:T_corona}, the proton temperature in the lower corona remains of order a few $\times 10^4$ to $10^5$ K (roughly consistent with the observations of \citet[][]{Marsch.etal.99}), whereas heavier ions can already reach temperatures of order $10^5$-$10^6$ K just above the transition region. This strong mass dependence is an important prediction of the velocity filtration model. Leaving aside the deviations due to the details of ionization physics not incorporated in the current model, the predictions of velocity filtration are in reasonable agreement with the data. We emphasize, however, that the comparison shown here is intended only as a proof of principle rather than as a precision fit to the observations. Although Section~\ref{sec:corona_timescales} shows that, for fiducial low-coronal parameters, the quasilinear tail-formation (heating) time is typically much shorter than the characteristic radiative cooling or transport times, a more quantitative confrontation with the data would still require incorporating sophisticated prescriptions for non-equilibrium ionization and radiative transfer/cooling, which are beyond the scope of the present paper.

Velocity filtration with a single $\kappa\approx 1.5-3$ distribution explains the observed drastic transition in the temperature fairly well, but the agreement with the Fe observations farther out in the corona is not as good. The left panel of Fig.~\ref{fig:T_corona} shows that increasing $\kappa$ improves the agreement at larger heights, but fails to reproduce the steep transition region. So far, we have assumed the ion DF to be a single $\kappa$ distribution that mimics the power-law fall-off beyond the ion thermal speed. As we showed in Fig.~\ref{fig:fi_vs_t}, though, the ion DF is typically a broken power-law, with one power-law between $\sigma_{i 0}$ and $\sigma_{e 0}$ and another beyond $\sigma_{e 0}$. Such a DF can be expressed as a linear combination of two $\kappa$ distributions as $f_{\rms 0}(v,r) = (1 - x) f_{\rms 0}^{(1)}\left(v,r,\kappa_{\rms 1},\sigma_{\rms 0}^{(1)},v_{\rm max}^{(1)}\right) + x f_{\rms 0}^{(2)}\left(v,r,\kappa_{\rms 2},\sigma_{\rms 0}^{(2)},v_{\rm max}^{(2)}\right)$, with $\sigma_{\rms 0}^{(1)}$ and $\sigma_{\rms 0}^{(2)}$ respectively equal to the ion thermal speed $\sigma_{i 0}$ and the electron thermal speed $\sigma_{e 0}$ at the upper end of the chromosphere. The corresponding temperature profile is given by equation~(\ref{Ts_2_kappa}). We compute this for Fe, adopting $\kappa_1 = 1.5$, $\kappa_2 = 1.5,\,2.5,$ and $4$, $x = 10^{-9}$ and $10^{-8}$ ($x$ is small since the second $\kappa$ distribution mimics the power-law fall-off beyond $\sigma_{e0}$ and is therefore heavily suppressed relative to the first), $v^{(1)}_{\rm max} = 3.5\sigma_{p0}$, and $v^{(2)}_{\rm max} = 1.5\sigma_{e0}$, and plot the resulting profiles in Fig.~\ref{fig:T_corona_Fe}. The choice $v_{\rm max}^{(1)} = 3.5\sigma_{p0}$ is motivated by the quasilinear ion DF itself: for realistic ion drive-to-BL diffusion ratios of order $10^{-3}$--$10^{-2}$, the first break typically appears at a few $\sigma_{p0}$ (see Fig.~\ref{fig:fi_vs_t}). The adopted value of $v^{(2)}_{\rm max} = 1.5\sigma_{e0}$ matches the escape velocity in the chromosphere. As shown in Figs.~\ref{fig:T_corona} and Fig.~\ref{fig:T_corona_Fe}, increasing $v_{\rm max}$ raises the filtered coronal temperature profile, as expected, without changing the basic conclusion that hard tails with $\kappa\approx 1.5$--$3$ are the most effective at producing a sharp transition. The symbols denote the UV spectroscopic observations \citep[][]{Peter.01,Wilhelm.etal.07,Yang.etal.09}. The double $\kappa$ velocity filtration model seems to agree better with the data than a single $\kappa$ one. The curves show that the first component with $\kappa_1=1.5$ controls the abrupt transition, while the second component determines how rapidly the temperature continues to rise in the outer corona. Smaller $x$ and larger $\kappa_2$ yield a flatter outer profile.

The basic physics behind coronal temperature inversion within the scope of the filtration model is as follows. Gravity acts as a filter and allows the high energy suprathermal particles to escape outwards. The virial temperature of the sun, $G M_\odot m_\rmP/k_\rmB R_\odot$, is $\sim 10^7$ K, which indicates that the filtering effect of gravity can invert the temperature profile and raise it to a million degrees K in the corona. However, this occurs only in a non-Maxwellian kinetic plasma with a shallow enough power-law DF, i.e., small enough $\kappa$. The limiting value of $\kappa$ at which the pressure and temperature diverge (for $v_{\rm max} \gg \sigma_{\rms 0}$) is $1.5$, which corresponds to a $v^{-5}$ DF and an $E^{-2}$ energy distribution. Such a shallow non-thermal tail works together with gravitational filtering to dramatically raise the temperature within a few $100$ km. Interestingly, this is exactly the power-law that is predicted by QLT for a kinetic plasma driven by super-Debye turbulence. Photospheric or chromospheric convection, which manifests as a large number of small-scale reconnection events or nano-flares (footprint motion of the field lines) \citep[][]{Parker.88}, can itself serve as a turbulent drive that heats the plasma to a non-thermal distribution, which can then enable temperature inversion via velocity filtration\citep[][]{Scudder.92a,Scudder.92b}. As we showed earlier, Maxwellianization through collisions cannot suppress the high $v$ suprathermal tail as long as super-Debye turbulence keeps stirring the plasma. This alleviates the concerns raised \citep[][]{Anderson.94,Landi.Pantellini.01} about the interference of collisional effects with velocity filtration. Even if collisions Maxwellianize the plasma, it happens over scales much larger than the width of the transition zone between the chromosphere and the corona.

The proportionality of $T_\rms$ to the mass of the species $m_\rms$ implies that velocity filtration predicts a much more gradual rise of the temperature of electrons than that of ions. The theory also predicts that the ion to electron temperature ratio $T_i/T_e$ in the corona is always larger than $1$ and can be as high as $10^2$. This is because the velocities of non-thermal ions and electrons in the weakly collisional corona are determined by the solar gravitational potential, which is the same for all masses, and are therefore more comparable in the corona than in the strongly collisional plasma of the solar interior. Higher $T_i/T_e$ might also be a consequence of the fact that radiative losses cool electrons more directly, whereas the ions cool mainly through slower Coulomb coupling to the electrons. However, a proper analysis of $T_i/T_e$ likely requires finite Larmor radius effects in a magnetized plasma along with radiative cooling, a topic that is beyond the scope of this paper and is left for future work.

The gradual rise of electron temperature predicted by filtration cannot account for its observed steep rise by a factor of $\sim 10^6 \rmK/10^4 \rmK = 100$ from the upper chromosphere to the coronal base. In the current quasilinear model, such a steep pre-filtration rise would require an electron DF with a shallower tail than $v^{-5}$, i.e. $\kappa_e<1.5$. Such tails are allowed in the quasilinear heating model, in the presence of (1) turbulent drives with shallow spatial spectra, $\alpha<5$, (2) turbulent drives with Gaussian temporal correlation, when $\omega_{\rmP e}t_{\rmc}$ is sufficiently large (see Fig.~\ref{fig:D_vs_v_alpha}), and (3) electron-cyclotron waves that resonantly interact with the electrons (see Appendix~\ref{App:DF_aniso_app}). If the chromospheric temperature $T_{e0}$ is defined as the temperature of a Maxwellian DF truncated at the (gravitational) escape speed, while the coronal temperature $T_e$ is computed from a truncated $\kappa$ distribution with a cutoff velocity $v_{\max,e}$, then
\begin{align}
&\frac{T_e}{T_{e0}}
=
\frac{\kappa_e\,
B\!\left(\chi^2_{e 0},\frac52,\kappa_e-\frac32\right)\Big/B\!\left(\chi^2_{e 0},\frac32,\kappa_e-\frac12\right)}
{
\gamma\!\left(\frac52,\frac{v^2_{\rm esc}}{2\sigma^2_{e0}}\right)\Big/\gamma\!\left(\frac32,\frac{v^2_{\rm esc}}{2\sigma^2_{e0}}\right)},\nonumber\\
&\chi_{e 0} = \sqrt{\frac{v^2_{\max,e}}{v^2_{\max,e}+2\kappa_e \sigma^2_{e0}}},
\end{align}
where $\gamma$ denotes the lower incomplete gamma function. This implies that a factor $\sim 100$ rise typically requires $v_{\max,e}/\sigma_{e0}\sim 20$--$30$ for $\kappa_e\lesssim0.5$, and $\sim 70$ for $\kappa_e=1$, whereas $\kappa_e\geq1.5$ cannot produce such a jump for any reasonable cutoff. In this context, we would like to note that, the whistler and electron-cyclotron waves that can efficiently heat electrons via wave-particle (Landau) resonances, result in an anisotropic steady-state DF that scales as $v_\parallel^{-n}$ with $n<2$ (see equation~[\ref{eq:fpar_main}] and Appendix~\ref{App:wave_drive}for a detailed derivation).

Even for a very shallow $\kappa_e\lesssim 0.5$ tail, one needs $v_{\max,e}$ well above the gravitational escape speed $v_{\rm esc}$, so an additional confinement mechanism is required to hold the electrons. The ambipolar potential between the electrons and ions would then be $e\Delta\Phi_{\rm amb}=\frac12m_e(v_{\max,e}^2-v_{\rm esc}^2)$, which is typically of order a few $10^2$--$10^3$ V for the required $v_{\max,e}/\sigma_{e0}$, comparable as an integrated potential to ambipolar potentials inferred from Parker Solar Probe electron distributions in the near-Sun solar wind \citep[][]{Bercic.etal.21,Horaites.Boldyrev.22}; in the present picture, the same potential would, however, have to be established much closer to the coronal base. Magnetic mirroring along a closed field line or coronal loop offers an alternative confinement mechanism: with $\tan\alpha_0=v_\perp/v_\parallel$ ($\alpha_0$ is the pitch angle), the mirror condition is $\sin^2\alpha_0>1/R_m$, where $R_m\equiv B_{\max}/B_{\min}$ is the footpoint-to-apex field ratio. For a characteristic anisotropic population with $v_\perp^2/v_\parallel^2\sim T_\perp/T_\parallel$, this gives the rule of thumb $R_m\gtrsim 1+T_\parallel/T_\perp$, so $R_m\gtrsim2$ suffices for isotropic electrons, while more field-aligned populations require larger $R_m$. Observationally inferred loop mirror ratios are of order a few to a few tens \citep[][]{Simoes.etal.13,Nakariakov.etal.21}, so magnetic confinement of such electrons in closed coronal structures is possible.

What gives rise to the remarkably narrow transition zone? The quasilinear picture developed here suggests the following sequence. As the plasma passes from the collisional chromosphere into the kinetic corona, direct wave or turbulent heating acts much more efficiently on the electrons than on the protons (see section~\ref{sec:criteria} and Appendix~\ref{App:Edrive_Eint}). Assuming that the drive diffusion coefficient is comparable to the (electron-electron) BL coefficient at $v\lesssim \sigma_e$, the electrons develop a non-thermal tail and diffuse away over a distance
\begin{align}
l_{{\rm diff}}^{(e)}\approx \frac{v_{\rm esc}^3}{D^{(e)}(\sigma_{e0})}\approx \frac{v_{\rm esc}^3}{\calD_2^{(e)}(\sigma_{e0})}
\approx \lambda_{\rm De}\frac{\Lambda}{\ln\Lambda}
\left(\frac{v_{\rm esc}}{\sigma_{e0}}\right)^3,
\end{align}
where $v_{\rm esc}=\sqrt{2GM_\odot/R_\odot}\approx 618\,{\rm km\,s^{-1}}$ is the solar escape speed. Since $v_{\rm esc}$ is only modestly larger than the upper-chromospheric electron thermal speed, this electron heating layer is extremely thin:
\begin{align}
l_{{\rm diff}}^{(e)}\approx 3.6\,
\frac{10^9\,{\rm cm}^{-3}}{n_e}
\sqrt{\frac{T_{e0}}{10^4\,{\rm K}}}\ {\rm m},
\end{align}
up to the weak logarithmic dependence through $\ln\Lambda$. As the electrons become displaced relative to the ions, a turbulent ambipolar electric field is generated. This field acts in addition to, and can be much stronger than, the small DC ambipolar field that maintains the quasi-neutral background plasma, and can act as a dominant accelerator of protons and heavier ions. If the drive is strong enough to beat the (proton-proton) BL coefficient, then the proton diffusion length turns out to be
\begin{align}
l_{{\rm diff}}^{(p)}= \frac{v_{\rm esc}^3}{D^{(p)}(\sigma_{p0})}&\approx \frac{v_{\rm esc}^3}{\calD_2^{(p)}(\sigma_{p0})}\nonumber\\
&\approx \lambda_{\rm Dp}\frac{\Lambda}{\ln\Lambda}
\left(\frac{v_{\rm esc}}{\sigma_{p0}}\right)^3 \nonumber\\
&= \lambda_{\rm De}\sqrt{\frac{T_{p0}}{T_{e0}}}\,
\frac{\Lambda}{\ln\Lambda}
\left(\frac{v_{\rm esc}}{\sigma_{p0}}\right)^3 \nonumber\\
&\approx 300\,
\left(\frac{10^{9}\,{\rm cm}^{-3}}{n_e}\right)
\left(\frac{T_{e0}}{10^4\,{\rm K}}\right)^{3/2}
\left(\frac{10^4\,{\rm K}}{T_{p0}}\right)\,
{\rm km},
\label{trans_zone_width}
\end{align}
again up to the weak logarithmic dependence through $\ln\Lambda$. Thus the electron heating layer is predicted to be much thinner than the proton heating layer. Interestingly, this characteristic proton diffusion scale for upper-chromospheric parameters is comparable to the observed $\sim 10^2$ km width of the transition zone. Equation~(\ref{trans_zone_width}) presents an approximate prediction of QLT: once the local density and temperature near the top of the chromosphere are specified, a narrow ion-heating layer of this order is predicted. Note that, by heating layer or transition zone we mean the narrow physical layer over which the plasma is heated (develops a non-thermal population of particles) and the temperature rises from the upper-chromospheric value of $T_{p0}\sim 10^4\,{\rm K}$ to the pre-filtration coronal-base value of $T_p(r\to r_0^+)\sim 10^6\,{\rmK}$. In the filtration model, $T_{p0}\neq T_p(r\to r_0^+)$, so the temperature profiles shown in Figs.~\ref{fig:T_corona} and \ref{fig:T_corona_Fe} do not resolve this narrow layer and instead begin immediately above it. The broader rise of the proton curve at larger heights in Fig.~\ref{fig:T_corona} should not be identified with the transition-zone width estimated by equation~(\ref{trans_zone_width}); it reflects the subsequent gradual variation of the filtered coronal profile. Note that the width predicted above is smaller than that found in the idealized 1D simulations of \citet[][]{Barbieri.etal.24a,Barbieri.etal.24b}, which hints at the need for calculations with realistic plasma turbulence, preferably in 3D, to resolve the issue of abrupt transition.

\subsection{Quasilinear heating versus transport and radiative cooling: a timescale comparison}\label{sec:corona_timescales}

Because the coronal plasma is both radiative and magnetically structured, any viable heating mechanism must operate faster than both radiative cooling and escape from the local heating region. We therefore compare the quasilinear tail-formation time associated with the relevant drive diffusion coefficient to characteristic transport and radiative cooling times in the low corona.

\subsubsection{Quasilinear diffusion}

In general, the quasilinear diffusion or tail-formation time is $t_{\rm diff}^{(\rms)}(v)\equiv v^2/D_{\rm drive}^{(\rms)}(v)$. For a broad-band turbulent drive normalized at the outer scale $k_{\min}^{-1}$, with $\langle E^2\rangle$ the rms electric-field variance and $E_{\rm int}\equiv m_e\omega_{\rmP e}\sigma_e/e$ the internal electric field associated with Coulomb collisions, the diffusion coefficient is given by equation~(\ref{eq:Dp_subturnover_final}).

A detailed calculation of the diffusion timescale for solar parameters is provided in Appendix~\ref{App:QL_diff_timescale}. Here we highlight the main estimates. Let us first consider the case of a turbulent drive. For upper-chromospheric parameters, the electron diffusion time at the thermal speed is
\begin{align}
t_{\rm diff}^{(e)}(\sigma_e)
\approx 1.27\times10^{-9}\,{\rm s}\,
\left(\frac{E_{\rm int}}{E_{\rm drive}}\right)^2
\left(\frac{B_0}{100\,{\rm G}}\right)^{-1},
\quad \alpha=5.
\end{align}
At the ion thermal speed, using $D_{\rm drive}^{(i)}(\sigma_i)\approx D_{\rm drive}^{(i)}(\sigma_e)$, one obtains
\begin{align}
t_{\rm diff}^{(i)}(\sigma_i)
\approx 1.27\times10^{-9}\,{\rm s}\,
Z_i^{-2}\left(\frac{T_i}{T_e}\right)\left(\frac{m_i}{m_e}\right)
\left(\frac{E_{\rm int}}{E_{\rm drive}}\right)^2
&\left(\frac{B_0}{100\,{\rm G}}\right)^{-1},\nonumber\\
&\alpha=5.
\end{align}
For protons with $T_p\simeq T_e$, this becomes
$t_{\rm diff}^{(p)}(\sigma_p)\approx 2.3\times10^{-6}\,{\rm s}\,
(E_{\rm int}/E_{\rm drive})^2(B_0/100\,{\rm G})^{-1}$. While protons are harder to excite than electrons due to higher mass, the above timescale estimates show that both can be energized by a turbulent drive quite rapidly. If $E_{\rm drive}/E_{\rm int}$ exceeds the threshold given in Appendix~\ref{App:threshold_turb}, which is of order $1$ $(10)$ for electrons (protons) then the diffusion time is shorter than the BL (collisional) relaxation time.

Let us now explore the case of a coherent wave drive. For the $R$-branch (electron-cyclotron and whistler) waves, which are the only ones with enough power to heat electrons under chromospheric conditions (see Appendix~\ref{App:wave_drive}), the tail-formation time is
$t_{{\rm diff},R}^{(e)}(v_\parallel)\equiv v_\parallel^2/
D_{\parallel\parallel,R}^{(e)}(v_\parallel)$, where
$D_{\parallel\parallel,R}^{(e)}$ includes the sum of the whistler and
electron-cyclotron resonances below the cut-off velocity
$v_\parallel=\omega_{\rmc e}d_e/2=\sigma_e/\sqrt{2\beta_e}$, at which diffusion spikes and beyond which it falls off precipitously. Using $(E_{\rm drive}/E_{\rm int})_{R,e}\approx
(\omega_{\rmc e}/\omega_{\rmP e})\,2(\beta_e+2)^{-1}\,\delta B/B_0$, which is characteristic of these waves, gives, for a representative spectral index $p_\parallel=\alpha_\parallel + \delta_\parallel=5$,
\begin{align}
t_{{\rm diff},R}^{(e)}(\sigma_e)
&\approx
3.1\times10^{-4}\,{\rm s}\,
\left(\frac{n_e}{10^{10}\,{\rm cm^{-3}}}\right)^2
\left(\frac{B_0}{10\,{\rm G}}\right)^{-5}
\left(\frac{\delta B}{B_0}\right)^{-2}.
\end{align}
For steeper spectra, with fiducial parameters, the same estimate becomes
$t_{{\rm diff},R}^{(e)}(\sigma_e)\approx
1.0\times10^{-4}\,{\rm s}$ for $p_\parallel=6$, and
$2.7\times10^{-7}\,{\rm s}$ for $p_\parallel=7$. These depend sensitively on $B_0$. Evidently, this timescale of wave-driven diffusion is longer than the turbulent diffusion time. However, the $R$ branch waves can drive very rapid electron heating through Landau resonance, with the timescale decreasing further as the resonant spike near $v_\parallel=\sigma_e/\sqrt{2\beta_e}$ is approached from below.

\subsubsection{Transport}

The spatial transport time, i.e., the time taken by the particles to stream along a pressure scale height or the length of a magnetic loop, $L_\parallel$, is given by
\begin{align}
t_{\rm tr}^{(\rms)} \sim \frac{L_\parallel}{\sigma_\rms},
\quad
\sigma_\rms=\sqrt{\frac{k_{\rm B}T_\rms}{m_\rms}}
\approx 90.9\,\mu_\rms^{-1/2}\left(\frac{T_\rms}{10^6\,{\rm K}}\right)^{1/2}\,{\rm km\,s^{-1}},
\label{eq:tau_tr}
\end{align}
\begin{align}
t_{\rm tr}^{(\rms)} \approx 1.1\times10^3\,{\rm s}\,
\left(\frac{L_\parallel}{10^5\,{\rm km}}\right)
\mu_\rms^{1/2}
\left(\frac{10^6\,{\rm K}}{T_\rms}\right)^{1/2}.
\end{align}
For a representative low-coronal pressure scale height, $H\sim 5\times10^4\,{\rm km}$, this gives $t_{\rm tr}^{(p)}\sim 5.5\times10^2\,{\rm s}$ and $t_{\rm tr}^{(e)}\sim 13\,{\rm s}$; for a loop of length $L_{\rm loop}\sim 10^5\,{\rm km}$, one finds $t_{\rm tr}^{(p)}\sim 1.1\times10^3\,{\rm s}$ and $t_{\rm tr}^{(e)}\sim 26\,{\rm s}$. Since this free-streaming estimate is conservative, trapping, mirroring, or scattering would only increase the residence time. Thus, for $E_{\rm drive}\sim E_{\rm int}$ or even weaker, both electron and proton tails form essentially instantaneously relative to transport timescales.

\subsubsection{Radiative cooling}

Let us now compare $t_{\rm diff}^{(\rms)}$ to radiative cooling, which proceeds through thermal bremsstrahlung (free-free) and line (bound-bound and free-bound) losses. The optically thin free-free cooling time of electrons is $t_{\rm ff}^{(e)} = 3 n_e k_\rmB T_e/2\epsilon_{\rm ff}$ \citep[][]{Karzas.Latter.61}, with
\begin{align}
\epsilon_{\rm ff}\approx 1.4\times10^{-27} g_{\rm B}\,n_e\left(n_{\rm H}+4n_{\rm He}\right)T_e^{1/2}\,{\rm erg\,cm^{-3}\,s^{-1}}.
\end{align}
For a fully ionized H/He plasma with $n_{\rm H}\approx 3n_{\rm He}$, this becomes $\epsilon_{\rm ff}\approx 2.0\times10^{-27} g_{\rmB} n_e^2 T_e^{1/2}\,{\rm erg\,cm^{-3}\,s^{-1}}$, so
\begin{align}
t_{\rm ff}^{(e)}&\approx 1.1\times10^{11}\,
\frac{T_e^{1/2}}{g_{\rm B}\,n_e}\ {\rm s}\nonumber\\
&= 8.8\times10^5\,{\rm s}\,
\left(\frac{g_{\rm B}}{1.2}\right)^{-1}
\left(\frac{T_e}{10^6\,{\rm K}}\right)^{1/2}
\left(\frac{10^8\,{\rm cm}^{-3}}{n_e}\right),
\label{eq:tff}
\end{align}
where $g_{\rm B}$ is the Gaunt factor. Since free-free losses are controlled mainly by the electrons, the formal cooling time associated with species $\rms$ scales as
\begin{align}
t_{\rm ff}^{(\rms)} \approx \frac{n_\rms T_\rms}{n_e T_e}\, t_{\rm ff}^{(e)}.
\end{align}

Line cooling is faster in the low corona and transition region. Although the line photons are emitted by ions, the relevant excited and recombined ionic states are populated mainly by electrons: bound-bound excitation and free-bound recombination are driven predominantly by electron impacts, whereas direct ion-impact processes are much weaker because ions are much heavier and hence slower at comparable temperature. Thus line emission also drains energy primarily from the electrons. Ions cool by Coulomb coupling with electrons. Writing the optically thin line-loss rate as $n_e^2\Lambda_{\rm line}(T_e)$ therefore yields the electron cooling time,
\begin{align}
t_{\rm line}^{(e)}&= \frac{3k_\rmB T_e}{2 n_e\Lambda_{\rm line}(T_e)}\nonumber\\
&\approx 2\times10^4\,{\rm s}\,
\left(\frac{T_e}{10^6\,{\rm K}}\right)
\left(\frac{10^8\,{\rm cm}^{-3}}{n_e}\right)
\left(\frac{10^{-22}\,{\rm erg\,cm^3\,s^{-1}}}{\Lambda_{\rm line}}\right),
\label{eq:tline}
\end{align}
where $\Lambda_{\rm line}\sim 10^{-22}\,{\rm erg\,cm^3\,s^{-1}}$ is a representative low-coronal value dominated by metal lines, especially Fe \citep[][]{Raymond.etal.76,Rosner.etal.78,Dere.etal.97}. The corresponding formal cooling time of species $\rms$ scales as $t_{\rm line}^{(\rms)}\sim (n_\rms T_\rms/n_eT_e)\,t_{\rm line}^{(e)}$. However, CHIANTI-based loss curves show that the loss function is a strong function of temperature and abundance set: near the upper transition region it can rise to a few $\times 10^{-22}$ or even $\sim 10^{-21}\,{\rm erg\,cm^3\,s^{-1}}$, whereas at several MK it drops to a few $\times 10^{-23}\,{\rm erg\,cm^3\,s^{-1}}$, so equation~(\ref{eq:tline}) should be regarded as an order-of-magnitude estimate. Using the velocity-filtration estimate $T_\rms/T_p \sim A_\rms-Z_\rms/2 \approx 3A_\rms/4$, one finds
\begin{align}
\frac{n_\rms T_\rms}{n_pT_p}\sim \frac{X_\rms}{X_p}\left(1-\frac{Z_\rms}{2A_\rms}\right)\approx \frac{3X_\rms}{4X_p},
\end{align}
where $X_\rms$ is the mass fraction. Helium therefore has a cooling time comparable to that of protons and electrons, whereas for heavier minor ions the formal radiative times are shorter because their mass fractions are smaller. Their effective cooling time, however, cannot fall below the electron--ion energy-exchange time, $t_{e\rms}^{(\rm E)}\approx 3\times10^2\,{\rm s}\,(A_\rms/Z_\rms^2)(T_e/10^6\,{\rm K})^{3/2}(10^8\,{\rm cm}^{-3}/n_e)(20/\ln\Lambda)\times(1+m_e T_\rms/m_\rms T_e)^{3/2}$, since otherwise electrons and ions would thermally decouple and the ions would no longer cool radiatively through the electrons. Under fiducial low-coronal conditions this equilibration time remains shorter than the radiative cooling times, so electrons and ions stay thermally coupled (on cooling timescales) to leading order even though the coupling weakens in more tenuous environments. 

In the upper chromosphere, the dominant hydrogen coolant is optically thick Ly$\alpha$. A more physical estimate than the optically thin form $n_e^2\Lambda(T)$ is obtained from the net Ly$\alpha$ cooling rate in semi-empirical radiative-transfer models. In general this rate depends on the local density, temperature, and Ly$\alpha$ escape probability $P_{\rm esc}$, schematically as $Q_{\rm Ly\alpha}\sim n_e n_1 q_{12}(T)\,h\nu_\alpha\,P_{\rm esc}$ ($q_{12}$ is the collisional reaction rate $\left<\sigma_{12} v\right>$ with $\sigma_{12}$ the collision cross section between species 1 and 2 moving with relative speed $v$), so it increases with density but more weakly than the optically thin $n_e^2$ scaling because photon trapping strengthens as the optical depth grows. Quiet-Sun models give a Ly$\alpha$ radiative flux of order $2\times10^5\,{\rm erg\,cm^{-2}\,s^{-1}}$, corresponding to a local net cooling rate $Q_{\rm Ly\alpha}\sim 0.1\,{\rm erg\,cm^{-3}\,s^{-1}}$ in the Ly$\alpha$-forming layer of thickness $\sim 20$ km. For $T\sim 10^4\,{\rm K}$ and $n_e\sim 10^{10}\,{\rm cm^{-3}}$ in the upper chromosphere, with total particle density $n_{\rm tot}\sim 3\times10^{10}\,{\rm cm^{-3}}$, the thermal energy density is $u_{\rm th}\sim 6\times10^{-2}\,{\rm erg\,cm^{-3}}$. The corresponding local Ly$\alpha$ cooling time is therefore
\[
t_{\rm cool,Ly\alpha}\sim \frac{u_{\rm th}}{Q_{\rm Ly\alpha}}\sim 0.6\,{\rm s},
\]
i.e. of order one second. This should be understood as a local upper-chromospheric estimate based on a semi-empirical radiative-transfer model.

Overall, for low-coronal and upper chromospheric conditions the drive (turbulence or waves) can exceed BL diffusion if $E_{\rm drive}\gtrsim E_{\rm int}$, in which case one typically finds $t_{\rm diff}^{(\rms)}\ll t_{\rm tr}^{(\rms)}<t_{\rm line}^{(\rms)}<t_{\rm ff}^{(\rms)}$ or $t_{\rm diff}^{(\rms)}\ll t_{\rm line}^{(\rms)}<t_{\rm tr}^{(\rms)}<t_{\rm ff}^{(\rms)}$ for both electrons and ions. Therefore, particle heating is very efficient in the upper chromosphere for reasonable values of $\delta B/B_0$. As soon as the drive exceeds the BL threshold, it is able to heat the plasma and generate a non-thermal tail; radiative cooling (even in the dense chromosphere) and field-aligned free-streaming occur on much longer timescales and are not inhibiting factors. What heats the particles? While generic electrostatic (e.g., Langmuir) turbulence can heat both electrons and ions, resonant interactions with waves can only heat electrons (in the current unmagnetized treatment of the plasma response). A self-consistent magnetized treatment would also include heating due to cyclotron resonances, which could be an efficient source of ion-heating; we leave this for future investigation. Under the quasilinear paradigm of this paper, with unmagnetized plasma response, we find that right circularly polarized whistler and electron cyclotron waves are the only ones powerful enough (for $\delta B/B_0\sim 1$) to overcome Coulomb collisions and heat the electrons (see Appendix~\ref{App:threshold_wave} for a detailed estimate of the threshold for various EM waves). This heating occurs via Landau resonance or wave-particle interactions; in more familiar terms, this is electron Landau damping in action.

\section{\label{sec:conclusion}Conclusion}

We have formulated a novel theory for NTPA in electromagnetically driven, multi-species, kinetic (weakly collisional) plasmas. The EM drive can either be of external origin, e.g., large-scale EM turbulence or waves, or describe the forces exerted by non-linear coherent structures (e.g., BGK holes and plasmoids) on the bulk plasma. A crucial ingredient of the theory is self-consistency through the Maxwell equations, manifested through the Debye screening of large-scale EM fluctuations, something that is ignored in earlier test particle treatments \citep[][]{Parker.65,Jokipii.Lee.10}. Using the quasilinear framework for the Vlasov-Maxwell equations, we have derived a general quasilinear transport equation for the relaxation of the mean coarse-grained DF of each charged species under the action of the drive, modulated by self-consistent fields, as well as self-generated small-scale fluctuations and Coulomb collisions. The quasilinear transport equation is a Fokker-Planck equation with diffusion and friction/drag coefficients that describe the energy and momentum exchange between the particles and fields. While the Balescu-Lenard (BL) coefficients describe the relaxation due to self-generated small-scale fluctuations and Coulomb collisions, the drive diffusion coefficient represents the heating of particles either directly by the drive (turbulent or wave-like in nature) or by the waves excited by it or both. We derive the expressions for the transport coefficients in stable, unstable and marginally stable plasmas, assuming that the instability is not violent enough to negate the quasilinear approximation (e.g., a classic bump-on-tail instability). We find that the direct heating of particles by a turbulent drive generally dominates over that mediated by the waves it excites, as long as the plasma is stable or marginally stable (once the instability has saturated). 

The drive diffusion coefficient typically scales as a power-law in the particle velocity $v$ over an extended range of $v$. Direct dressed particle diffusion typically dominates over indirect wave-mediated diffusion for moderately steep drive power-spectra. If the plasma is forced by an isotropic, large-scale (super-Debye), turbulent drive with a correlation time $t_\rmc>1/\omega_{\rmP e}$ and a spatial power-spectrum of the driving electric field that scales as $k^{-\alpha}$ with $\alpha \geq 5$ and $k_{\rm min}\lambda_{\rmD e} < 1$, then $D^{(\rms)}(v)$ scales as $\sim v^4$ between the electron thermal speed $\sigma_e$ and $1/k_{\rm min} t_\rmc$, and as $v^{\alpha-1}$ beyond $1/k_{\rm min} t_\rmc$ and up to the phase-velocity of the large-scale Langmuir waves, $\omega_{\rmP e}/k_{\rm min}$. The $v^4$ scaling arises from the universal $1 - \omega^2_{\rmP e}/{(\bk\cdot\bv)}^2$ scaling of the longitudinal component of the dielectric tensor in the range $\sigma_e < v < \omega_{\rmP e}/k$, and yields a universal $v^{-5}$ scaling (in 3D) for the steady-state DF $f_{\rms 0}$, with a corresponding $E^{-2}$ energy distribution, in the range $\sigma_e < v < 1/k_{\rm min} t_\rmc$. At $1/k_{\rm min} t_\rmc < v < \omega_{\rmP e}/k_{\rm min}$, $D^{(\rms)}(v)\sim v^{\alpha-1}$ yields $f_{\rms 0}(v) \sim v^{-\alpha}$. The steep spectra required to generate these tails are encountered in several settings, e.g., in strong Langmuir turbulence\citep[][]{Sun.etal.22}, and in near-Debye-scale plasma turbulence \citep[][]{Ewart.etal.25,Nastac.etal.23,Nastac.etal.25,Ginat.etal.25}. The power-law tail develops in both electron and ion DFs even in the presence of weak collisions, since the drive diffusion coefficient always exceeds the BL coefficients at large enough $v$, as long as the plasma is subject to large-scale (super-Debye) EM turbulence. Such large scale fields are Debye shielded, even more so for particles at low $v$, which implies that the slower particles are less heated while the faster particles are unscreened and readily accelerated, unaffected by collisions. Ultimately it is Debye screening and the consequent suppression of particle heating at low $v$ that generates the universal $v^{-5}$ tail in (non-relativistic, unmagnetized) kinetic plasmas. 

For shallower power-spectra with $\alpha < 5$, the universality is broken and $D^{(\rms)}(v)$ scales as $v^{\alpha - 1}$, which yields a $v^{-\alpha}$ scaling for $f_{\rms 0}$ over the entire range, $\sigma_e<v<\omega_{\rmP e}/k_{\rm min}$. At $v>\omega_{\rmP e}/k_{\rm min}$, $D^{(\rms)}(v)$ scales as $v^{-3}$, similar to the BL diffusion coefficient, which, together with the $v^{-2}$ scaling of the BL drag coefficient, yields a Maxwellian high $v$ cut-off for the DF. This yields all moments finite. We have also extended our analysis to a drive consisting of coherent anisotropic EM waves, in which case the power-law exponent of the tail depends on the wave branch and the anisotropic spectral indices. Therefore, a generic prediction of our theory is that kinetic plasmas driven by large-scale EM fluctuations, be it broad-band/turbulent or narrow-band/wave-like, harbor power-law tails in their DFs, with a particular preference for the $v^{-5}$ tail, even in the presence of collisions and instabilities. Our transport equation is more general than the conventional Parker transport equation \citep[][]{Parker.65}, since, unlike the latter, we incorporate the effects of self-consistent EM fields as well as collisions on particle acceleration.

Our theory provides a general description of NTPA in hot and tenuous kinetic plasmas, be that in fusion plasmas, e.g., inertial confinement fusion devices, or in astrophysical and space plasmas, e.g., the solar corona and solar wind, circumstellar coronae, collsionless shocks from supernova explosions, the hot ionized interstellar medium, etc. As an astrophysical application, we discuss the implications of the theory for the solar transition region and coronal temperature inversion. The velocity filtration model \citep[][]{Scudder.92a,Scudder.92b,Scudder.94} predicts that the presence of a non-thermal power-law tail in the DF of a plasma confined in an attractive potential raises its effective temperature (but decreases its density) towards the region of shallower potential. In the presence of such a non-thermal DF, an attractive gravitation potential filters out the high energy suprathermal particles \citep[][]{Scudder.92a,Scudder.92b}. This offers a plausible mechanism for coronal temperature inversion through the filtering action of the solar gravitational potential: hotter particles can go further up. However, reproducing the sharp temperature rise, especially for heavy ionic species, suggested by spectroscopic line-width diagnostics \citep[][]{Peter.01} in the transition region between the upper chromosphere and the lower corona appears to require very shallow power-law tails with $\kappa \approx 1.5-3$ if velocity filtration is to play an important role; a similar range of $\kappa$ required for filtration was also inferred by \citet[][]{Scudder.94}, although they did not offer a robust mechanism for forming such tails, something we do in this paper. The presence of such tails in the corona was, however, deemed implausible by \citet[][]{Anderson.94,Landi.Pantellini.01}, since, due to the smallness of the collisional mean free path relative to the pressure scale height, the plasma was believed to Maxwellianize on large scales even if kinetic power-law tails exist over small scales at the coronal base. 

Our quasilinear analysis shows that this assumption of Maxwellianization is not quite correct when it comes to the suprathermal tail. Even in the presence of collisions, a plasma driven by large-scale (super-Debye) EM fluctuations generically develops a power-law tail, with a particular preference for $\kappa = 1.5$ for sufficiently steep spectra, $\alpha\geq 5$. Although the mean free path is short, the free-streaming scale of the fast non-thermal population significantly exceeds the scale height, indicating that these particles accelerate and diffuse away without collisional losses. Moreover, the timescale comparison in Section~\ref{sec:corona_timescales} shows that, for fiducial chromospheric or low-coronal parameters, the quasilinear diffusion/tail-formation time is much shorter than the characteristic transport time across one pressure scale height or a coronal loop as well as the radiative cooling time, so a $\kappa\sim 1.5-3$ tail can develop locally, way before the particles can escape or cool. We find that resonant wave heating (via whistler and electron-cyclotron waves) is much more efficient in heating electrons than protons and heavier ions (under an unmagnetized treatment of plasma response that ignores cyclotron resonances), so ion energization proceeds predominantly through a turbulent drive, which could be a turbulent ambipolar electric field generated by electron--ion separation.

Hard power-law tails with $\kappa \approx 1.5-3$ can produce a steep transition of the ion temperature and density from the chromosphere to the corona over a scale of order $100$ km. The electron temperature undergoes a steeper transition and requires a $\kappa<1.5$. This can arise for a shallower turbulent spectrum, $\alpha<5$, or via resonant-wave particle interactions with whistler and electron-cyclotron waves (electron Landau damping). The abrupt rise in the effective temperature is an outcome of two effects: (i) pressure greatly increases for $\kappa \leq 1.5$ and is sensitive to the tail end of the distribution, and (ii) the width of the transition zone, within the scope of QLT, is set by the diffusion length of the non-thermal particles (see equation~[\ref{trans_zone_width}]), which is quite small. The simple requirements for generating a non-thermal tail, together with its resilience against collisions, suggest that it may be a widespread feature of weakly collisional plasmas in general. More broadly, if hard suprathermal tails are present, abrupt transition and temperature inversion through velocity filtration may be relevant in other weakly collisional, gravitationally stratified media, e.g., circumstellar and circumplanetary coronae \citep[][]{Scudder.92b,Meyer-Vernet.07} and those around black holes and active galactic nuclei.

It is useful to contrast this kinetic picture with the standard conduction--radiation interpretation of the solar transition region, in which a downward field-aligned heat flux from the hot corona is radiated away as the plasma enters the denser transition region and upper chromosphere, typically using the Spitzer--H\"arm conductivity or a flux-limited extension of it \citep[][]{Spitzer.Harm.53,Klimchuk.06}. For $\kappa$ above the critical value at which the electron heat flux changes sign, kinetic calculations indicate that the magnitude of the conductive heat flux generally increases as $\kappa$ decreases; for a given downward heat flux, this implies a smaller $|dT/dr|$ and hence a broader transition layer \citep[][]{Dorelli.Scudder.99,Landi.Pantellini.01}. In this local sense, conduction--radiation balance and velocity filtration pull in opposite directions: the former favors larger $\kappa$ for a sharper transition, whereas the latter favors smaller $\kappa$ because harder suprathermal tails are filtered more efficiently by gravity. For sufficiently hard tails, however, with $\kappa \lesssim 5$--$6$, the electron heat flux can reverse and flow outward along the outwardly increasing temperature profile \citep[][]{Dorelli.Scudder.99,Landi.Pantellini.01}. In that regime, the usual Klimchuk-style interpretation of the transition region as a layer that radiates heat conducted down from the hot corona is no longer self-consistent: radiation can still drain the conductive flux but the flux is now outward rather than inward. The sign reversal signals a breakdown of the local conductivity closure itself, especially in the transition region where the temperature gradient is strongest and nonlocal (beyond moment-closure) or flux-limited transport is expected to matter \citep[][]{Luciani.etal.83,Schurtz.etal.00,Arber.etal.23}. Our kinetic calculation is aimed precisely at this regime, where the classical picture becomes questionable. However, the present treatment still omits non-equilibrium ionization and a self-consistent coupling to radiative cooling, which would be important for a more direct confrontation with transition-region observations. In addition, UVCS/SOHO observations of heavy ions in the extended corona indicate strong anisotropy, $T_\perp \gg T_\parallel$ \citep[][]{Kohl.etal.98,Cranmer.etal.08}, which likely points to significant ion-cyclotron heating that we do not investigate here but intend to do in a follow-up paper. Despite these caveats, our theory makes two broad predictions: first, hard suprathermal tails arise naturally in weakly collisional plasmas driven by super-Debye EM turbulence or waves; second, such tails lead to (1) an abrupt rise in the effective temperature, and (2) a further rise through velocity filtration in the solar gravitational potential, even more so for heavier ions. One question remains though: why does the heating happen where it happens? In other words, what is so special about the transition region?

It is important to note that our theory for NTPA only applies to non-relativistic unmagnetized plasmas, although the drive can be electromagnetic in nature. In a follow up paper, we will present a self-consistent theory for the relaxation of magnetized plasmas with anisotropic turbulence, incorporate relativistic effects, and derive a generalized version of the Parker transport equation for NTPA in the context of the solar wind as well as cosmic ray transport. To this end we will also explore the energy partition between electrons and ions, especially the role of helicity barrier in imbalanced turbulence\citep[][]{Squire.etal.22,Zhang.etal.25,Adkins.etal.25}, and cyclotron heating. While our current framework for unmagnetized plasmas predicts similar non-thermal power-laws in the electron and ion DFs, the scenario might change in magnetized plasmas due to their significantly different Larmor radii. The appearance of similar power-law tails $(\sim E^{-2})$ in relativistic or strongly magnetized PIC simulations to what has been predicted by the quasilinear treatment in this paper does not necessarily imply the same underlying acceleration mechanism. In the present theory, the non-thermal tail is generated by the (Debye shielded) drive diffusion coefficient and corresponds to a constant-flux steady state in velocity space. By contrast, \citet[][]{Wong.etal.25} infer from an effective Fokker--Planck description of relativistic PIC turbulence (calculating the transport coefficients from simulation-based particle trajectories) that the power-law tail may arise from a balance between advection and diffusion, i.e., a zero-flux steady state. Preliminary calculations indicate that, in magnetized plasmas where cyclotron heating dominates, an advection coefficient may indeed appear alongside diffusion, although the physical reason for this is not yet clear. What power-laws arise in magnetized plasma DFs in general remains to be investigated.

\begin{acknowledgments}
The authors are thankful to Steven Cranmer, Srijan Das, Mihir Desai, Robert Ewart, Barry Ginat, Matthew Kunz, Nuno Loureiro, Michael Nastac, Alex Schekochihin, Anatoly Spitkovsky, Dmitri Uzdensky and Vladimir Zhdankin for stimulating discussions and valuable suggestions. The first author (UB) is especially thankful to Nuno Loureiro for illuminating exchanges and the Plasma Science Fusion Center at MIT for providing an intellectually stimulating environment. This research is supported by the National Science Foundation Award 2206607 at the Multi-Messenger Plasma Physics Center (MPPC), Princeton University, and the Bezos Membership grant at the IAS.
\end{acknowledgments}

\section*{Data Availability Statement}

\begin{center}
\renewcommand{\arraystretch}{1.2}
\begin{tabular}{|p{0.3\linewidth}|p{0.65\linewidth}|}
\hline
\textbf{AVAILABILITY OF DATA} & \textbf{STATEMENT OF DATA AVAILABILITY}\\
\hline
{\raggedright Data available on request from the authors}
&
{\raggedright The data that support the findings of this study are available from the corresponding author upon reasonable request.}
\\ \hline
\end{tabular}
\end{center}

\appendix

\section{Computation of $D_p^{(\rms)}(v)$}\label{App:Dp_calc}

The dressed particle diffusion coefficient $D_p^{(\rms)}(v)$ is given by

\begin{align}
D^{(\rms)}_p(v) &= \frac{32\pi^5 q^2_\rms}{m^2_\rms V} \nonumber\\
&\times \int_0^{\infty} \rmd k\, k^2 \calE(k) \int_0^1 \rmd\cos{\theta}\,\cos^2\theta\, \frac{ \calC_{\omega}\left(k v \cos{\theta}\right)}{{\left|\varepsilon_{k\parallel}\left(kv\cos\theta\right)\right|}^2}.
\label{Dp_iso_app}
\end{align}
In the range $\sigma_e < v < \omega_{\rmP e}/k$, the dielectric constant $\varepsilon_{k\parallel}(k v \cos\theta)$ is approximately equal to $1 - \omega^2_{\rmP e}/k^2v^2\cos^2\theta$, when $k\lambda_{\rmD e}<1$. The real part dominates over the imaginary part in this range; as we shall see shortly, the imaginary part is going to play a crucial role in another interval. Substituting $\varepsilon_{k\parallel}(k v \cos\theta)$ in the above, and adopting $\calC_{\omega}\left(k v \cos{\theta}\right) = 1/\left(1 + {\left(kv t_\rmc \cos\theta\right)}^2\right)$, $D_p^{(\rms)}(v)$ is given by the following in the range $\sigma_e < v < \omega_{\rmP e}/k$:

\begin{align}
&D^{(\rms)}_p(v) = \frac{32\pi^5 q^2_\rms v^4}{m^2_\rms V} \nonumber\\
&\times \int_0^{\infty} \rmd k\, k^6 \calE(k) \int_0^1 \rmd\cos{\theta} \frac{1}{1 + {\left(kv t_\rmc \cos\theta\right)}^2} \frac{\cos^6\theta}{{\left(k^2v^2\cos^2\theta - \omega^2_{\rmP e}\right)}^2}
\end{align}
The $\cos\theta$ integral can be evaluated as follows:

\begin{align}
&I = \int_0^1 \rmd\cos{\theta} \frac{1}{1 + {\left(kv t_\rmc \cos\theta\right)}^2} \frac{\cos^6\theta}{{\left(k^2v^2\cos^2\theta - \omega^2_{\rmP e}\right)}^2} \nonumber\\
&= \frac{I_1 - I_2}{{\left(k v t_\rmc\right)}^2},\nonumber\\
&I_1 = \int_0^1\rmd\cos\theta \frac{\cos^4\theta}{{\left(k^2v^2\cos^2\theta - \omega^2_{\rmP e}\right)}^2},\nonumber\\
&I_2 = \int_0^1 \rmd\cos{\theta} \frac{1}{1 + {\left(kv t_\rmc \cos\theta\right)}^2} \frac{\cos^4\theta}{{\left(k^2v^2\cos^2\theta - \omega^2_{\rmP e}\right)}^2}.
\end{align}
The $I_1$ and $I_2$ integrals can be evaluated to yield:

\begin{align}
&I_1 = \frac{1}{{\left(k v\right)}^4}\left[1 - \frac{3}{4}\frac{\omega_{\rmP e}}{k v}\ln{\left(\left|\frac{\omega_{\rmP e} + kv}{\omega_{\rmP e} - kv}\right|\right)} + \frac{1}{2} \frac{\omega^2_{\rmP e}}{\omega^2_{\rmP e}-k^2 v^2} \right],\nonumber\\
&I_2 = \frac{1}{{\left(k v\right)}^4 k v t_\rmc} \nonumber\\
&\times \left[\frac{\tan^{-1}\left(k v t_\rmc\right)}{{\left(1+{\omega_{\rmP e}^2 t_\rmc^2}\right)}^2} - \frac{\omega_{\rmP e} t_\rmc}{4}\frac{3+{\omega_{\rmP e}^2t_\rmc^2}}{{\left(1+{\omega_{\rmP e}^2 t_\rmc^2}\right)}^2}\ln{\left(\left|\frac{\omega_{\rmP e} + k v}{\omega_{\rmP e} - k v}\right|\right)} \right.\nonumber\\
&\left.+ \frac{1}{2}\frac{k v t_\rmc}{1+{\omega_{\rmP e}^2 t_\rmc^2}}\frac{\omega^2_{\rmP e}}{\omega^2_{\rmP e}-{k^2 v^2}} \right].
\end{align}
Therefore, $I$ can be expressed as

\begin{align}
&I = \frac{1}{{\left(k v\right)}^4{\left(k v t_\rmc\right)}^2} \left[1 - \frac{1}{{\left(1+{\omega_{\rmP e}^2 t_\rmc^2}\right)}^2} \frac{\tan^{-1}\left(k v t_\rmc\right)}{k v t_\rmc} \right.\nonumber\\
&\left.+ \frac{\omega^2_{\rmP e}t^2_\rmc}{1 + \omega^2_{\rmP e}t^2_\rmc} \left(\frac{1}{2} \frac{\omega^2_{\rmP e}}{\omega^2_{\rmP e}-k^2 v^2} - \frac{5 + 3\omega^2_{\rmP e}t^2_\rmc}{4\left(1 + \omega^2_{\rmP e}t^2_\rmc\right)}\frac{\omega_{\rmP e}}{k v}\ln{\left(\left|\frac{\omega_{\rmP e} + kv}{\omega_{\rmP e} - kv}\right|\right)} \right)\, \right].
\end{align}
Asymptotic analysis shows that, for $v \ll \omega_{\rmP e}/k \ll 1/k t_\rmc$ and $v \ll 1/k t_\rmc \ll \omega_{\rmP e}/k$, $I \approx 1/7$, and for $1/k t_\rmc \ll v \ll \omega_{\rmP e}/k$ (when $\omega_{\rmP e}t_\rmc > 1$), $I \approx 1/5 {\left(k v t_\rmc\right)}^2$. Therefore, the asymptotic scalings of $D^{(\rms)}_p(v)$ (for $\omega_{\rmP e}t_\rmc > 1$) are:

\begin{align}
&D^{(\rms)}_p(v) \approx \dfrac{32\pi^5 q^2_\rms}{m^2_\rms V} \nonumber\\
&\times
\begin{cases}
\dfrac{v^4}{7 \omega^4_{\rmP e}} \bigintsss_{\;0}^{\infty}\rmd k\, k^6\, \calE(k), & \sigma \ll v \ll \dfrac{1}{k\tau_\rmc},\\ \\
\dfrac{v^2}{5 \omega^4_{\rmP e} t^2_\rmc} \bigintsss_{\;0}^{\infty}\rmd k\, k^4\, \calE(k), & \dfrac{1}{k t_\rmc} \ll v \ll \dfrac{\omega_{\rmP e}}{k},\\ \\
\dfrac{1}{1 + \omega^2_{\rmP e}t^2_\rmc}\dfrac{\omega_{\rmP e}}{2 v^3}\bigintsss_{\;0}^{\infty} \rmd k\, \dfrac{k^5 {\calE(k)}}{\left|\left.{\partial F_0}/{\partial v}\right|_{\omega_{\rmP e}/k}\right|}, & v \gg \dfrac{\omega_{\rmP e}}{k}.
\label{Dp_asymptotic}
\end{cases}
\end{align}

The derivation of the last scaling is subtle. To see this, we need to take into account the imaginary part of $\varepsilon_{k\parallel}(kv\cos\theta)$. For $k\lambda_{\rmD e}<1$, it can be written as $\varepsilon_{k\parallel}(kv\cos\theta) = 1 - \omega^2_{\rmP e}/k^2v^2\cos^2\theta + i\varepsilon_\rmI$, where $\varepsilon_\rmI = -\pi(\omega^2_{\rmP e}/k^2)\left.\partial F_0/\partial v\right|_{v\cos\theta}$, where $F_0$ is the 1D DF.  The $\cos\theta$ integral can then be evaluated as follows:

\begin{align}
&I = \int_0^1 \rmd\cos{\theta} \frac{1}{1 + {\left(kv t_\rmc \cos\theta\right)}^2} \frac{\cos^6\theta}{{\left(\cos^2\theta - \omega^2_{\rmP e}/k^2 v^2\right)}^2 + \varepsilon^2_\rmI\cos^4\theta} \nonumber\\
&= \int_0^1 \rmd z\, \frac{1}{1 + y^2 z^2} \frac{z^6}{{\left(z^2 - x^2\right)}^2 + \varepsilon^2_\rmI z^4},\quad x = \omega_{\rmP e}/kv,\; y = kv t_\rmc\nonumber\\
&\approx\int_0^1 \rmd z\, \frac{1}{1 + y^2 z^2} \frac{z^4}{4{\left(z - x\right)}^2 + \varepsilon^2_\rmI z^2},
\end{align}
where we have substituted $z = \cos\theta$. In the last line, we have used the fact that the dominant contribution to the integral comes from $z\approx x$, which implies that $z^2-x^2 = (z+x)(z-x)\approx 2z(z-x)$, and canceled $z^2$ from the numerator and denominator of the second factor. This is possible because $x = \omega_{\rmP e}/kv \ll 1$, since we are computing in the regime, $v \gg \omega_{\rmP e}/k$. And, since the dominant contribution comes from $z\approx x$, i.e., $\cos\theta \approx \omega_{\rmP e}/kv$, we can safely take $\varepsilon_\rmI = -\pi(\omega^2_{\rmP e}/k^2)\left.\partial F_0/\partial v\right|_{v\cos\theta}\approx -\pi(\omega^2_{\rmP e}/k^2)\left.\partial F_0/\partial v\right|_{\omega_{\rmP e}/k}$, which is of course a small number if $F_0(v)$ decays fast enough with $v$. We can now rearrange $I$ to write it as

\begin{align}
I \approx\int_0^1 \rmd z\, \frac{z^2}{1 + y^2 z^2} \frac{1}{4{\left(1 - x/z\right)}^2 + \varepsilon^2_\rmI}.
\end{align}
Now we can divide and multiply by $\varepsilon_\rmI$ to rewrite it as

\begin{align}
I\approx \frac{1}{\varepsilon_\rmI}\int_0^1 \rmd z\, \frac{z^2}{1 + y^2 z^2} \frac{\varepsilon_\rmI}{4{\left(1 - x/z\right)}^2 + \varepsilon^2_\rmI}.
\end{align}
Since, in the limit of $v\gg \omega_{\rmP e}/k$, i.e., $x\to 0$, $\varepsilon_\rmI\propto F'_0(1/x)\to 0$, we can write the asymptotic limit of $I$ as

\begin{align}
\lim_{x\to 0} I &\approx \lim_{\varepsilon_\rmI \to 0}\frac{1}{\varepsilon_\rmI} \int_0^1 \rmd z\, \frac{z^2}{1 + y^2 z^2} \frac{\varepsilon_\rmI}{4{\left(1 - x/z\right)}^2 + \varepsilon^2_\rmI}\nonumber\\
&\approx \frac{1}{\varepsilon_\rmI} \int_0^1 \rmd z\, \frac{z^2}{1 + y^2 z^2} \lim_{\varepsilon_\rmI \to 0}\frac{\varepsilon_\rmI}{4{\left(1 - x/z\right)}^2 + \varepsilon^2_\rmI},
\end{align}
which, upon using the identity, $\lim_{\epsilon\to 0}\epsilon/(w^2 + \epsilon^2) = \pi\delta(w)$, reduces to

\begin{align}
I &\approx \frac{\pi}{2\left|\varepsilon_\rmI\right|} \int_0^1 \rmd z\, \frac{z^2}{1 + y^2 z^2}\,\delta\left(1-\frac{x}{z}\right)\nonumber\\
&\approx \frac{\pi}{2\left|\varepsilon_\rmI\right| x} \int_0^1 \rmd z\, \frac{z^4}{1 + y^2 z^2}\,\delta\left(z-x\right)\nonumber\\
&\approx \frac{\pi}{2\left|\varepsilon_\rmI\right|}\frac{x^3}{1 + y^2 x^2}.
\end{align}
Plugging $x$ and $y$ back, we get

\begin{align}
I &\approx \frac{\pi}{2\left|\varepsilon_\rmI\right|} {\left(\frac{\omega_{\rmP e}}{kv}\right)}^3 \frac{1}{1 + \omega^2_{\rmP e}t^2_\rmc}.
\end{align}
Finally, substituting the form for $\varepsilon_\rmI$ yields

\begin{align}
I = \frac{1}{1 + \omega^2_{\rmP e}t^2_\rmc}\frac{1}{\left|\left.{\partial F_0}/{\partial v}\right|_{\omega_{\rmP e}/k}\right|}\frac{\omega_{\rmP e}}{2 k v^3}.
\end{align}
Since $\left.{\partial F_0}/{\partial v}\right|_{\omega_{\rmP e}/k}$ is a small number, the value of $I$ is large at $v = \omega_{\rmP e}/k$. This is nothing but a manifestation of the wave-particle resonance $\omega_{\rmP e} = k v$, i.e., the particle velocity is equal to the phase-space velocity of Langmuir waves. These resonant particles respond the most to any electric field perturbations. However, the response is not infinite. These resonant particles are the ones that gain energy from the Langmuir waves, causing Landau damping of these waves. The net result of this is a loss of phase-coherence between the particles and the wave, which ultimately {\it {broadens}} the resonance. A finite gradient of the DF ensures that there {\it {is}} a net energy exchange, which implies a net loss of coherence and therefore a finite broadening. On the other hand, a zero gradient would mean no damping and therefore an infinitely thin Dirac-delta resonance.

\section{Landau damping vs quasilinear relaxation}\label{App:Landau_vs_QL}
The wave-mediated diffusion can be neglected relative to direct diffusion by the drive in the stable regime if the Landau damping timescale is shorter than that of quasilinear relaxation. In the large mean free path case, $\nu_{\rmc \rms} \ll k\sigma_\rms$, and on super-Debye scales ($k\lambda_{\rmD e} \ll 1$), the electron and ion Langmuir waves damp away at a rate faster than the quasilinear relaxation rate $\sim {\left(k\lambda_{\rmD e}\right)}^4\sigma^2_{\rms}/2 D^{(\rms)}_0$ ($D^{(\rms)}_0$ is the diffusion coefficient at $v \gtrsim \omega_{\rmP e}/k$), at which the power-law tail develops at intermediate $v$, by a factor of $\sim {\left(k\lambda_{\rmD e}\right)}^{-2}$. The ion-acoustic waves damp away even faster, at a rate $\sim {\left(k\lambda_{\rmD e}\right)}^{-3}$ higher than the quasilinear relaxation rate. On sub-Debye ($k\lambda_{\rmD e} \gtrsim 1$) scales, Landau damping is efficient and rapidly damps away the waves on a timescale $\sim {\left(k\sigma_\rms\right)}^{-1} \sim {\left(k\lambda_{\rmD e}\right)}^{-1}{\omega}^{-1}_{\rmP\rms}$. Therefore, the longitudinal modes do not contribute to quasilinear diffusion at long time. The transverse light waves do not Landau damp away, though, and can potentially contribute. However, for an isotropic plasma subject to isotropic perturbations, the transverse components do not show up in the quasilinear transport equation. Under anisotropic conditions, they do, and the undamped light waves contribute, albeit with a strength suppressed by a factor $\sim {\left(\sum_\rms \omega^2_{\rmP\rms}/c^2k^2\right)}^2$ relative to direct diffusion for sub-skin depth ($k\gtrsim \sqrt{\sum_\rms \omega^2_{\rmP\rms}}/c$) perturbations.

In the small mean free path limit $(\nu_{\rmc \rms} \gg k \sigma_\rms)$, the least damped Landau mode (the Lenard-Bernstein mode \citep[][]{Lenard.Bernstein.58,Banik.Bhattacharjee.24b}), damps at the rate, ${\left(k\sigma_\rms\right)}^2/\nu_{\rmc \rms} \sim {\left(k\lambda_{\rmD e}\right)}^2 \omega_{\rmP\rms}\,\Lambda/\ln\Lambda$. The ratio of this damping rate to the quasilinear diffusion rate is equal to $\sim {\left(k\lambda_{\rmD e}\right)}^{-2} \Lambda/\ln\Lambda$. Hence, the Landau modes damp faster than the power-law tail develops for super-Debye $(k\lambda_{\rmD e} \ll 1)$ perturbations, even more so in a more collisionless environment (larger $\Lambda$). Based on the above considerations, the waves/Landau term can be neglected in the diffusion coefficient for stable plasmas (see also equation~[A19] of \citet[][]{Banik.Bhattacharjee.24a}).

\section{Entropy considerations}\label{App:entropy}
Can we construct an entropy functional for the effective collision operator in equation~(\ref{QL_eq}), that satisfies the H-theorem (never decreases with time) and yields the above DF as an extremal solution? It can be shown that $S = -\int \rmd^3 x\,\rmd^3v\,G(f)$ with any convex function $G(f)$ follows the H-theorem. Since the DF approaches a $\kappa$ distribution at long times in the presence of a large-scale EM drive, $G(f)\sim \left(f^{q}-1\right)/\left(q-1\right)$ with $q = \kappa/\left(1+\kappa\right)$ is the function for which the entropy $S$, when maximized with the constraints of total energy, momentum and particle number conservation, yields this DF as the extremal solution. The corresponding entropy $S$ is nothing but the Tsallis entropy \citep{Tsallis.88,Livadiotis.McComas.13,Zhdankin.22a,Zhdankin.22b}. And, since our DF typically scales as $v^{-5}$ at large $v$, $\kappa = 1.5$ and $q = 3/5$ are the preferred values.

Interestingly, \citet[][]{Ewart.etal.22} and \citet[][]{Ewart.etal.23} recast the entropy into a generalized Boltzmann-Shannon form, $S = \int \rmd \bv \int \rmd \eta\, P(\bv,\eta)\ln{P(\bv,\eta)}$, with $P(\eta)$ the probability that $f$ (treated as a random variable) takes the value $\eta$. Maximizing this with the constraints of the conservation of probability, total energy and phase-volume (waterbag content), $\rho(\eta) = \int \rmd^3v\,P(\bv,\eta) = \frac{1}{V}\int\rmd\bx\int\rmd\bv\,\delta\left(f(\bx,\bv)-\eta\right)$, they obtain a Fermi-Dirac distribution for $P(\bv,\eta)$, along the lines of \citet[][]{LyndenBell.67}. Noting that $\rho(\eta) = \int \rmd \bv\,\delta\left(f_\rmG-\eta\right)$ ($f_\rmG$ is the Gardener distribution that is a function of $E$ and not $\bx$ and has the same $\rho(\eta)$ as $f$) scales as $\eta^{-1}$ for exponentially truncated (or steeper) $f_\rmG$, they find that the Fermi-Dirac $P(\bv,\eta)$ yields $N(E)\sim E^{-2}$ in the non-degenerate limit. It is, however, a priori, not clear why the Gardener distribution would be of such a form. It is easy to see that, for $f_\rmG(v)\sim v^{-n}$, $\rho(\eta)\sim \eta^{-\left(1+1/n\right)}$ (as also pointed out by \citet[][]{Ewart.etal.22}), which would yield a shallower tail than $E^{-2}$. The derivation of the $E^{-2}$ tail from this generalized Lynden-Bell approach hinges on (i) a sufficient deviation of $f_{\rms 0}$ from $f_\rmG$ and (ii) the $\eta^{-1}$ scaling of $\rho(\eta)$. As shown by \citet[][]{Ewart.etal.23} using $\left(1x,1v\right)$ PIC simulations of two-stream instability in electrostatic plasmas, the former depends on the precise nature of Vlasov turbulence (e.g., presence or absence of BGK holes). Interestingly, the $E^{-2}$ tail appears in the spatially averaged energy distribution after the saturation of the two-stream instability, only when BGK holes appear in the phase-space, such as in an electron-ion plasma. The electron-positron case neither features large-scale holes nor the $E^{-2}$ tail. Therefore, it appears that the appearance of the tail is intimately related to the occurrence of a large-scale drive, which, in this case, consists of the electric fields exerted by the BGK holes themselves on the bulk plasma. And the BGK holes are an outcome of the self-consistent plasma response. In other words, self-consistency through the Poisson equation, something that the entropy approach does not explicitly take into account but our kinetic formalism does, is crucial for the emergence of the power-law tail. As we show, it is the Debye-shielding of large-scale EM fields (e.g., those exerted by coherent structures as in \citet[][]{Ewart.etal.23}'s simulation) and the consequent suppression of particle heating at low $v$, that fundamentally gives rise to the tail.

It is possible to incorporate self-consistency in the entropy-based approach. When the entropy is maximized using the conservation of energy and particle number as constraints, one would need to take into account the dependence of the energy distribution and particle number not only on the DF but also on the density of states. In an inhomogeneous plasma, the latter depends on the electrostatic potential, which itself depends on the DF through the Poisson equation. A self-consistent entropy maximization thus becomes significantly more complicated and would involve constrained optimization using functional derivatives, something we leave for future investigation.

\section{Drive as waves}
\label{App:wave_drive}

Here we estimate the direct drive diffusion coefficient for a narrow set of coherent waves in the long-time limit. We retain a guide field $\bB_0=B_0\hat{\bz}$ only to define parallel and perpendicular directions in velocity space, but we keep the \emph{plasma response} unmagnetized and isotropic, as in the main text. In this non-relativistic limit, the transverse light-wave response does not contribute to the secular resonant diffusion, since the resonance condition $\omega_\bk=\bk\cdot\bv$ cannot be satisfied on the light branch. Thus only the longitudinal projection of the drive electric field contributes. We therefore work directly with
\[
\calE_L^{(\rmP)}(\bk)=\hat{k}_i \hat{k}_j \,\calE^{(\rmP)}_{ij}(\bk),
\]
and assume that this projected spectrum follows the anisotropic power-law forms introduced below. For definiteness we restrict to $k_\parallel>0$, corresponding to one propagation direction along the guide field; the opposite direction contributes analogously.

Writing $\bv=(v_\perp,0,v_\parallel)$ and $\bk=(k_\perp\cos\phi,k_\perp\sin\phi,k_\parallel)$, so that $\bk\cdot\bv=k_\parallel v_\parallel+k_\perp v_\perp\cos\phi$, the long-time direct diffusion tensor is
\begin{align}
D^{(\rms)}_{ij}(\bv)
\approx
\frac{8\pi^5 q_\rms^2}{m_\rms^2 V}
&\int_0^\infty \rmd k_\parallel
\int_0^\infty \rmd k_\perp\,k_\perp
\int_0^{2\pi}\rmd\phi\,
\hat{k}_i\hat{k}_j\nonumber\\
&\times\frac{\calE^{(\rmP)}_L(\bk)}
{\left|\varepsilon_{\bk\parallel}(\omega_\bk)\right|^2}\,
\delta\!\left(\omega_\bk-\bk\cdot\bv\right).
\label{eq:Dij_wave_drive_app_mod2}
\end{align}
Performing the $\phi$ integral gives
\[
\bar{\delta}
\equiv
\int_0^{2\pi}\rmd\phi\,
\delta\!\left(
\omega_\bk-k_\parallel v_\parallel-k_\perp v_\perp\cos\phi
\right)
\]
\[
=
\frac{
2\,\Theta\!\left[
(k_\perp v_\perp)^2-(\omega_\bk-k_\parallel v_\parallel)^2
\right]
}{
\sqrt{
(k_\perp v_\perp)^2-(\omega_\bk-k_\parallel v_\parallel)^2
}
}\],
so that, with $C_\rms=8\pi^5 q_\rms^2/m_\rms^2V$,
\begin{align}
D^{(\rms)}_{\parallel\parallel}
&\approx
C_\rms
\int_0^\infty \rmd k_\perp\,k_\perp
\int_0^\infty \rmd k_\parallel\,
\frac{k_\parallel^2}{k^2}\,
\frac{\calE^{(\rmP)}_L(\bk)}
{\left|\varepsilon_{\bk\parallel}(\omega_\bk)\right|^2}\,
\bar{\delta},
\nonumber\\
D^{(\rms)}_{\perp\perp}
&\approx
\frac{C_\rms}{2}
\int_0^\infty \rmd k_\perp\,k_\perp
\int_0^\infty \rmd k_\parallel\,
\frac{k_\perp^2}{k^2}\,
\frac{\calE^{(\rmP)}_L(\bk)}
{\left|\varepsilon_{\bk\parallel}(\omega_\bk)\right|^2}\,
\bar{\delta},
\nonumber\\
D^{(\rms)}_{\parallel\perp}
&=
D^{(\rms)}_{\perp\parallel}\nonumber\\
&\approx
C_\rms
\int_0^\infty \rmd k_\perp\,k_\perp
\int_0^\infty \rmd k_\parallel\,
\frac{k_\parallel k_\perp}{k^2}\,
\frac{\omega_\bk-k_\parallel v_\parallel}{k_\perp v_\perp}\,
\frac{\calE^{(\rmP)}_L(\bk)}
{\left|\varepsilon_{\bk\parallel}(\omega_\bk)\right|^2}\,
\bar{\delta}.
\label{eq:Dcoeffs_wave_drive_app_mod2}
\end{align}

The corresponding axisymmetric diffusion equation is
\begin{align}
\frac{\partial f_{\rms0}}{\partial t}
&=
\frac{\partial}{\partial v_\parallel}
\left(
D^{(\rms)}_{\parallel\parallel}\frac{\partial f_{\rms0}}{\partial v_\parallel}
+
D^{(\rms)}_{\parallel\perp}\frac{\partial f_{\rms0}}{\partial v_\perp}
\right)
\nonumber\\
&\quad+
\frac{1}{v_\perp}\frac{\partial}{\partial v_\perp}
\left[
v_\perp
\left(
D^{(\rms)}_{\perp\parallel}\frac{\partial f_{\rms0}}{\partial v_\parallel}
+
D^{(\rms)}_{\perp\perp}\frac{\partial f_{\rms0}}{\partial v_\perp}
\right)
\right].
\label{eq:FP2D_wave_drive_app_mod2}
\end{align}

We consider two idealized families of spectra. For nearly parallel waves,
\begin{align}
\calE^{(\rmP)}_{L,\parallel}(\bk)
&=
A_\parallel\,
k_\parallel^{-\alpha_\parallel}
k_\perp^{-\delta_\parallel}
\Theta\!\left(\xi_\parallel k_\parallel-k_\perp\right)
\Theta\!\left(k_\parallel-k_{\min,\parallel}\right)
\Theta\!\left(k_{\rmc,\parallel}-k_\parallel\right),
\end{align}
with $\xi_\parallel\ll 1$, while for nearly perpendicular waves,
\begin{align}
\calE^{(\rmP)}_{L,\perp}(\bk)
&=
A_\perp\,
k_\perp^{-\alpha_\perp}
k_\parallel^{-\delta_\perp}
\Theta\!\left(\xi_\perp k_\perp-k_\parallel\right)
\Theta\!\left(k_\perp-k_{\min,\perp}\right)
\Theta\!\left(k_{\rmc,\perp}-k_\perp\right),
\end{align}
with $\xi_\perp\ll 1$. Define
\[
p_\parallel\equiv \alpha_\parallel+\delta_\parallel,
\quad
p_\perp\equiv \alpha_\perp+\delta_\perp.
\]

Using Parseval's theorem,
\[
\frac{V\langle E_\parallel^2\rangle}{(2\pi)^3}
=
\int \rmd^3k\,\calE^{(\rmP)}_{L,\parallel}(\bk),
\quad
\frac{V\langle E_\perp^2\rangle}{(2\pi)^3}
=
\int \rmd^3k\,\calE^{(\rmP)}_{L,\perp}(\bk),
\]
one finds, provided $\delta_\parallel<2$, $\delta_\perp<1$, and
$p_\parallel\neq 3$, $p_\perp\neq 3$,
\begin{align}
A_\parallel
&=
\frac{V\langle E_\parallel^2\rangle}{(2\pi)^3}\,
\frac{(2-\delta_\parallel)(3-p_\parallel)}
{2\pi\,\xi_\parallel^{\,2-\delta_\parallel}
\left(k_{\rmc,\parallel}^{\,3-p_\parallel}-k_{\min,\parallel}^{\,3-p_\parallel}\right)},
\\
A_\perp
&=
\frac{V\langle E_\perp^2\rangle}{(2\pi)^3}\,
\frac{(1-\delta_\perp)(3-p_\perp)}
{2\pi\,\xi_\perp^{\,1-\delta_\perp}
\left(k_{\rmc,\perp}^{\,3-p_\perp}-k_{\min,\perp}^{\,3-p_\perp}\right)}.
\end{align}
The marginal cases $p_\parallel=3$, $p_\perp=3$, $\delta_\parallel=2$, and
$\delta_\perp=1$ are logarithmic, while $\delta_\parallel>2$ and
$\delta_\perp>1$ require additional lower cutoffs inside the cone.

Integrating over the narrow cone gives the effective one-dimensional spectra
\begin{align}
\calE^{\rm eff}_{\parallel}(k_\parallel)
&\equiv
2\pi\int_0^{\xi_\parallel k_\parallel}\rmd k_\perp\,k_\perp\,
\calE^{(\rmP)}_{L,\parallel}(\bk)
\nonumber\\
&=
\frac{V\langle E_\parallel^2\rangle}{(2\pi)^3}\,
\frac{3-p_\parallel}
{k_{\rmc,\parallel}^{\,3-p_\parallel}-k_{\min,\parallel}^{\,3-p_\parallel}}
\,k_\parallel^{\,2-p_\parallel},
\label{eq:Eeff_par_norm}
\\
\calE^{\rm eff}_{\perp}(k_\perp)
&\equiv
2\pi\int_0^{\xi_\perp k_\perp}\rmd k_\parallel\,
\calE^{(\rmP)}_{L,\perp}(\bk)
\nonumber\\
&=
\frac{V\langle E_\perp^2\rangle}{(2\pi)^3}\,
\frac{3-p_\perp}
{k_{\rmc,\perp}^{\,3-p_\perp}-k_{\min,\perp}^{\,3-p_\perp}}
\,k_\perp^{\,2-p_\perp},
\label{eq:Eeff_perp_norm}
\end{align}
with the usual logarithmic replacement when $p_\parallel=3$ or $p_\perp=3$.

The following electromagnetic wave families are considered. First, the parallel
$R$ (right-circularly polarized) branch,
\[
\omega_{\rm R} \approx \omega_{\rm c e}\frac{k_\parallel^2 d_e^2}{1+k_\parallel^2 d_e^2},
\]
which reduces to electron whistler (wh) waves,
$\omega_{\rm R} \approx \omega_{\rm c e}k_\parallel^2 d_e^2$, for
$k_\parallel d_e \ll 1$, and to electron-cyclotron (EC) waves,
$\omega_{\rm R} \approx \omega_{\rm c e}$, for $k_\parallel d_e \gg 1$, where
$d_e=c/\omega_{\rm P e}$ is the electron inertial scale. For low $\beta_e$ typical of upper chromosphere or lower corona, waves near the electron gyroscale $\lambda_{\rmc e}$ are EC-like since $k_\parallel d_e \sim k_\parallel\lambda_{\rmc e}\sqrt{2/\beta_e}\gg 1$, whereas those near the ion gyroscale $\lambda_{\rmc i}$ are a combination of both since $k_\parallel d_e\sim 1$. Second, the dispersive
shear-Alfv\'en branch,
\[
\omega_{\rm A} \approx k_\parallel v_{\rm A}
\left[
\frac{1+k_\perp^2\lambda_{\rm c i}^2/\left(\beta_i+2/(1+T_e/T_i)\right)}
{1+k_\perp^2 d_e^2}
\right]^{1/2}.
\]
The inertial Alfv\'en (InA) and kinetic-Alfv\'en waves (KAW) are the
corresponding limits of $\omega_{\rm A}$: for
$k_\perp d_e\gg 1$ and
$k_\perp^2\lambda_{\rm c i}^2/Q_i\ll1$,
\[
\omega_{\rm InA}\approx \frac{k_\parallel v_{\rm A}}{\sqrt{1+k_\perp^2 d_e^2}},
\]
while for
$k_\perp\lambda_{\rm c i}\gg1$ and
$k_\perp d_e\ll1$,
\[
\omega_{\rm KAW}\approx
k_\parallel v_{\rm A}\left(1+\frac{k_\perp^2\lambda_{\rm c i}^2}{Q_i}\right)^{1/2},
\]
where
\[
Q_i\equiv \beta_i+\frac{2}{1+T_e/T_i}.
\]
Finally, we consider the ion-cyclotron (IC) branch,
\[
\omega_{\rm IC} \approx
k_\parallel v_{\rm A}
\left[
\sqrt{1+\frac{k_\parallel^2 d_i^2}{4}}
-
\frac{k_\parallel d_i}{2}
\right],
\]
where $d_i=c/\omega_{\rm P i}$ is the ion inertial scale. For the resonant
particles of interest, $v\gg \sigma_e$, so that
$\omega_\bk=k_r v\gg k_r\sigma_e$ on the resonant manifold. In the super-Debye
regime,
\[
\left|\varepsilon_{\bk\parallel}(\omega_\bk)\right|^{-2}
\approx
\left(\frac{\omega_\bk}{\omega_{\rm P e}}\right)^4.
\]

\subsection{Nearly parallel and nearly perpendicular waves}

For $k_\perp\ll k_\parallel$, the dominant coefficient is
$D^{(\rms)}_{\parallel\parallel}$, with
$D^{(\rms)}_{\parallel\perp}\sim \xi_\parallel D^{(\rms)}_{\parallel\parallel}$
and
$D^{(\rms)}_{\perp\perp}\sim \xi_\parallel^2 D^{(\rms)}_{\parallel\parallel}$.
Using equation~(\ref{eq:Eeff_par_norm}), the leading coefficient can be written
as a sum over all resonant roots $k_{r,j}$ satisfying $\omega(k_{r,j})=k_{r,j}v_\parallel$:
\begin{align}
D^{(\rms)}_{\parallel\parallel}(v_\parallel)
&\approx
\pi^2 Z_\rms^2\left(\frac{m_e}{m_\rms}\right)^2
\sigma_e^2\omega_{\rm P e}
\left(\frac{\langle E_\parallel^2\rangle}{E_{\rm int}^2}\right)
\frac{3-p_\parallel}
{k_{\rmc,\parallel}^{\,3-p_\parallel}-k_{\min,\parallel}^{\,3-p_\parallel}}
\nonumber\\
&\quad\times
\sum_j
\frac{k_{r,j}^{\,2-p_\parallel}}{|v_\parallel-v_g(k_{r,j})|}
\left(\frac{\omega(k_{r,j})}{\omega_{\rm P e}}\right)^4,
\label{eq:Dpar_master_wave_drive_sum}
\end{align}
where $E_{\rm int}\equiv m_e\omega_{\rm P e}\sigma_e/e$.

For the parallel $R$ branch, take $k_{\rmc,\parallel}=\lambda_{\rmD e}^{-1}$ and
define $x\equiv k_\parallel d_e$. The resonance condition is
\[
v_\parallel=\omega_{\rm c e}d_e\,\frac{x}{1+x^2},
\quad
v_g(k_r)=\omega_{\rm c e}d_e\,\frac{2x}{(1+x^2)^2}.
\]
Because $v_\parallel(x)$ has a maximum at $x=1$, the root structure is:
\[
\begin{cases}
0<v_\parallel<\omega_{\rm c e}d_e/2: & \text{two roots } x_-<1,\ x_+>1,\ x_-x_+=1,\\
v_\parallel=\omega_{\rm c e}d_e/2: & \text{one degenerate root } x=1,\\
v_\parallel>\omega_{\rm c e}d_e/2: & \text{no real root}.
\end{cases}
\]
Thus $v_g=v_\parallel$ at $x=1$, i.e. at the transition between the whistler
and electron-cyclotron parts of the branch, and the diffusion coefficient
develops a narrow resonant spike there. For $0<v_\parallel<\omega_{\rm c e}d_e/2$,
the full coefficient is the sum of the two resonant contributions,
\begin{align}
D^{(\rms)}_{\parallel\parallel,{\rm R}}(v_\parallel)
&\approx
\pi^2 Z_\rms^2\left(\frac{m_e}{m_\rms}\right)^2
\sigma_e^2\omega_{\rm P e}
\left(\frac{\langle E_\parallel^2\rangle}{E_{\rm int}^2}\right)
\frac{(p_\parallel-3)(k_{\min,\parallel}d_e)^{p_\parallel-3}}
{1-(k_{\min,\parallel}\lambda_{\rmD e})^{p_\parallel-3}}
\nonumber\\
&\quad\times
\left(\frac{\omega_{\rm c e}}{\omega_{\rm P e}}\right)^3 \left[
\frac{x_-^{\,9-p_\parallel}}
{(1+x_-^2)^2(1-x_-^2)}
+
\frac{x_+^{\,9-p_\parallel}}
{(1+x_+^2)^2(x_+^2-1)}
\right].
\label{eq:DR_full_norm}
\end{align}
Using $x_+=x_-^{-1}$, this may be written as
\begin{align}
D^{(\rms)}_{\parallel\parallel,{\rm R}}(v_\parallel)
&\approx
\pi^2 Z_\rms^2\left(\frac{m_e}{m_\rms}\right)^2
\sigma_e^2\omega_{\rm P e}
\left(\frac{\langle E_\parallel^2\rangle}{E_{\rm int}^2}\right)
\frac{(p_\parallel-3)(k_{\min,\parallel}d_e)^{p_\parallel-3}}
{1-(k_{\min,\parallel}\lambda_{\rmD e})^{p_\parallel-3}}
\nonumber\\
&\quad\times\left(\frac{\omega_{\rm c e}}{\omega_{\rm P e}}\right)^3
\frac{x_-^{\,9-p_\parallel}+x_-^{\,p_\parallel-3}}
{(1+x_-^2)^2(1-x_-^2)}.
\label{eq:DR_full_norm_reduced}
\end{align}
For $x_-\ll1$,
\[
x_-\approx u_\parallel,\quad
x_+\approx u_\parallel^{-1},\quad
u_\parallel\equiv \sqrt{\frac{\beta_e}{2}}\frac{v_\parallel}{\sigma_e},
\]
so away from the narrow spike the total $R$-branch coefficient scales as
\begin{align}
D^{(\rms)}_{\parallel\parallel,{\rm R}}(v_\parallel)
&\propto
u_\parallel^{\,9-p_\parallel}
+
u_\parallel^{\,p_\parallel-3}.
\label{eq:DR_asym_norm}
\end{align}
Thus, below the tangency, the whistler-like contribution scales as
$v_\parallel^{\,9-\alpha_\parallel-\delta_\parallel}$, the
electron-cyclotron-like contribution scales as
$v_\parallel^{\,\alpha_\parallel+\delta_\parallel-3}$, and the dominant one at $v_\parallel<\omega_{\rm c e}d_e/2$ is
whistler-like for $p_\parallel>6$ and electron-cyclotron-like for
$4<p_\parallel<6$ (the term with smaller exponent wins since $x_{-}<1$). For $v_\parallel>\omega_{\rm c e}d_e/2$, there is no real
root and hence no resonant contribution from this one-sided ($k_\parallel>0$)
$R$ branch.

For the ion-cyclotron branch, take $k_{\rmc,\parallel}=\lambda_{\rmD e}^{-1}$ and
define $y_r\equiv k_{r\parallel}d_i$. The resonance condition is
\[
v_\parallel=v_{\rm A}\left[\sqrt{1+\frac{y_r^2}{4}}-\frac{y_r}{2}\right].
\]
This function is monotonic decreasing for $y_r>0$, so there is exactly one root
for $0<v_\parallel<v_{\rm A}$ and no root for $v_\parallel>v_{\rm A}$. Also,
\[
\frac{v_g(k_r)}{v_\parallel}=1-\frac{y_r}{\sqrt{y_r^2+4}}<1
\quad (y_r>0),
\]
so there is no resonant spike on the cyclotron branch itself. In the cyclotron
limit $y_r\gg1$, one obtains
\begin{align}
D^{(\rms)}_{\parallel\parallel,{\rm IC}}(v_\parallel)
&\approx
\pi^2 Z_\rms^2\left(\frac{m_e}{m_\rms}\right)^2
\sigma_e^2\omega_{\rm P e}
\left(\frac{\langle E_\parallel^2\rangle}{E_{\rm int}^2}\right)
\frac{(p_\parallel-3)(k_{\min,\parallel}d_i)^{p_\parallel-3}}
{1-(k_{\min,\parallel}\lambda_{\rmD e})^{p_\parallel-3}}\nonumber\\
&\times\left(\frac{\omega_{\rm c i}}{\omega_{\rm P e}}\right)^3
y_r^{\,3-p_\parallel},
\label{eq:DIC_asym_norm}
\end{align}
so that
$D^{(\rms)}_{\parallel\parallel,{\rm IC}}\propto
v_\parallel^{\,\alpha_\parallel+\delta_\parallel-3}$.

For $k_\parallel\ll k_\perp$, the dominant coefficient is
$D^{(\rms)}_{\perp\perp}$, with
$D^{(\rms)}_{\parallel\perp}\sim \xi_\perp D^{(\rms)}_{\perp\perp}$ and
$D^{(\rms)}_{\parallel\parallel}\sim \xi_\perp^2 D^{(\rms)}_{\perp\perp}$.
Using equation~(\ref{eq:Eeff_perp_norm}), the leading coefficient can be written
as a sum over all resonant roots $k_{r,j}$ satisfying $\omega(k_{r,j})=k_{r,j}v_\perp$:
\begin{align}
D^{(\rms)}_{\perp\perp}(v_\perp)
&\approx
\frac{\pi^2}{2}\,
Z_\rms^2\left(\frac{m_e}{m_\rms}\right)^2
\sigma_e^2\omega_{\rm P e}
\left(\frac{\langle E_\perp^2\rangle}{E_{\rm int}^2}\right)
\frac{3-p_\perp}
{k_{\rmc,\perp}^{\,3-p_\perp}-k_{\min,\perp}^{\,3-p_\perp}}
\nonumber\\
&\quad\times
\sum_j
\frac{k_{r,j}^{\,2-p_\perp}}{|v_\perp-v_{g,\perp}(k_{r,j})|}
\left(\frac{\omega(k_{r,j})}{\omega_{\rm P e}}\right)^4.
\label{eq:Dperp_master_wave_drive_sum}
\end{align}

For KAWs, use a cutoff
$k_{\rmc,\perp}^{\rm KAW}$ satisfying
\[
\lambda_{\rm c i}^{-1}\ll k_{\rmc,\perp}^{\rm KAW}\ll d_e^{-1},
\]
so that the spectrum lies inside the genuine KAW window
$k_\perp\lambda_{\rm c i}\gg1$ and $k_\perp d_e\ll1$. Writing
$x_r\equiv k_{r\perp}\lambda_{\rm c i}$, the resonance condition becomes
\[
v_\perp=\xi_\perp v_{\rm A}\left(1+\frac{x_r^2}{Q_i}\right)^{1/2}.
\]
This is monotonic increasing for $x_r>0$, so there is exactly one root for
$v_\perp>\xi_\perp v_{\rm A}$ and no root for $v_\perp<\xi_\perp v_{\rm A}$.
Moreover,
\[
\frac{v_{g,\perp}(k_r)}{v_\perp}
=
\frac{1+2x_r^2/Q_i}{1+x_r^2/Q_i}>1
\quad (x_r>0),
\]
so there is no resonant spike in the KAW regime. In the sub-Larmor limit
$x_r\gg1$, one obtains
\begin{align}
&D^{(\rms)}_{\perp\perp,{\rm KAW}}(v_\perp)
\approx
\frac{\pi^2}{2}\,
Z_\rms^2\left(\frac{m_e}{m_\rms}\right)^2
\sigma_e^2\omega_{\rm P e}
\left(\frac{\langle E_\perp^2\rangle}{E_{\rm int}^2}\right)\nonumber\\
&\times\frac{3-p_\perp}
{(k_{\rmc,\perp}^{\rm KAW}\lambda_{\rm c i})^{3-p_\perp}
-(k_{\min,\perp}\lambda_{\rm c i})^{3-p_\perp}}
\left(\frac{\xi_\perp v_{\rm A}}{\sqrt{Q_i}\,\lambda_{\rm c i}\omega_{\rm P e}}\right)^3
x_r^{\,9-p_\perp},
\label{eq:DKAW_asym_norm}
\end{align}
so that
$D^{(\rms)}_{\perp\perp,{\rm KAW}}\propto
v_\perp^{\,9-\alpha_\perp-\delta_\perp}$.

For inertial-Alfv\'en waves, use a cutoff
$k_{\rmc,\perp}^{\rm InA}$ satisfying
\[
d_e^{-1}\ll k_{\rmc,\perp}^{\rm InA}\ll \sqrt{Q_i}\,\lambda_{\rm c i}^{-1},
\]
so that the spectrum lies inside the genuine inertial-Alfv\'en window
$k_\perp d_e\gg1$ and $k_\perp^2\lambda_{\rm c i}^2/Q_i\ll1$.
Writing $x_r\equiv k_{r\perp}d_e$, one has
\[
v_\perp=\frac{\xi_\perp v_{\rm A}}{\sqrt{1+x_r^2}}.
\]
This is monotonic decreasing for $x_r>0$, so there is exactly one root for
$0<v_\perp<\xi_\perp v_{\rm A}$ and no root for $v_\perp>\xi_\perp v_{\rm A}$.
Also,
\[
\frac{v_{g,\perp}(k_r)}{v_\perp}=\frac{1}{1+x_r^2}<1
\quad (x_r>0),
\]
so there is no resonant spike in the inertial-Alfv\'en regime. In the
sub-Larmor limit $x_r\gg1$, one obtains
\begin{align}
D^{(\rms)}_{\perp\perp,{\rm InA}}(v_\perp)
&\approx
\frac{\pi^2}{2}\,
Z_\rms^2\left(\frac{m_e}{m_\rms}\right)^2
\sigma_e^2\omega_{\rm P e}
\left(\frac{\langle E_\perp^2\rangle}{E_{\rm int}^2}\right)\nonumber\\
&\times\frac{3-p_\perp}
{(k_{\rmc,\perp}^{\rm InA}d_e)^{3-p_\perp}
-(k_{\min,\perp}d_e)^{3-p_\perp}}
\left(\frac{\xi_\perp v_{\rm A}}{d_e\omega_{\rm P e}}\right)^3
x_r^{\,3-p_\perp},
\label{eq:DInA_asym_norm}
\end{align}
so that
$D^{(\rms)}_{\perp\perp,{\rm InA}}\propto
v_\perp^{\,\alpha_\perp+\delta_\perp-3}$.

\subsection{Constant-flux steady states and resonant plateaus}\label{App:DF_aniso_app}

When nearly parallel waves dominate, the steady state is controlled by
$J^{(\rms)}_\parallel\approx
-D^{(\rms)}_{\parallel\parallel}\,\partial f_{\rms0}/\partial v_\parallel$.
Thus $J^{(\rms)}_\parallel={\rm const}$ implies
$f_{\rms0}(v_\parallel)\propto v_\parallel^{\,1-p_\parallel}$ if
$D^{(\rms)}_{\parallel\parallel}\propto v_\parallel^{p_\parallel}$.

For the $R$ branch below the tangency, the exact coefficient is the sum of a
whistler-like term and an electron-cyclotron-like term,
\[
D^{(\rms)}_{\parallel\parallel,{\rm R}}(v_\parallel)
\propto
v_\parallel^{\,9-p_\parallel}
+
v_\parallel^{\,p_\parallel-3},
\quad
0<v_\parallel<\frac{\omega_{\rm c e}d_e}{2},
\]
so the steady state is not a single pure power law across the whole sub-resonant
range. At $v_\parallel<\omega_{\rmc e}d_e/2$, one recovers the dominant scaling,
\begin{align}
f_{\rms0,{\rm wh}}(v_\parallel)
&\propto
v_\parallel^{-(8-\alpha_\parallel-\delta_\parallel)},
\quad \alpha_\parallel+\delta_\parallel>6,
\nonumber\\
f_{\rms0,{\rm EC}}(v_\parallel)
&\propto
v_\parallel^{\,4-\alpha_\parallel-\delta_\parallel},
\qquad 4<\alpha_\parallel+\delta_\parallel<6.
\label{eq:fpar_wave_drive_app_mod3}
\end{align}
Either way, the power-law index is more than $-2$, which indicates a very hard tail. At tangency,
\[
v_\parallel=\frac{\omega_{\rm c e}d_e}{2}
=
\frac{\sigma_e}{\sqrt{2\beta_e}},
\]
the $R$-branch diffusion coefficient develops its resonant spike, so the steady
state develops a narrow plateau there. In practice, the singularity in the
diffusion coefficient is regularized, and the plateau in $f_{\rms0}$ is
shortened by the finite resonance width. For
$v_\parallel>\omega_{\rm c e}d_e/2$, there is no resonant contribution from the one-sided ($k_\parallel>0$) $R$ branch, so $f_{\rms 0}$ drops off as a Maxwellian beyond $\omega_{\rm c e}d_e/2=\sigma_e/\sqrt{2\beta_e}$ (due to the interplay of BL diffusion and drag).

For the ion-cyclotron branch, which has only one root for $0<v_\parallel<v_A$,
one obtains
\[
f_{\rms0,{\rm IC}}(v_\parallel)
\propto
v_\parallel^{\,4-\alpha_\parallel-\delta_\parallel}.
\]

When nearly perpendicular waves dominate,
\[
\frac{\partial}{\partial v_\perp}
\left(
v_\perp D^{(\rms)}_{\perp\perp}
\frac{\partial f_{\rms0}}{\partial v_\perp}
\right)=0,
\]
so that
$v_\perp D^{(\rms)}_{\perp\perp}\,\partial f_{\rms0}/\partial v_\perp={\rm const}$.
Hence, if $D^{(\rms)}_{\perp\perp}\propto v_\perp^{p_\perp}$, then
$f_{\rms0}(v_\perp)\propto v_\perp^{-p_\perp}$. Away from any resonant spike,
\begin{align}
f_{\rms0,{\rm KAW}}(v_\perp)
&\propto
v_\perp^{-(9-\alpha_\perp-\delta_\perp)},
\nonumber\\
f_{\rms0,{\rm InA}}(v_\perp)
&\propto
v_\perp^{\,3-\alpha_\perp-\delta_\perp}.
\label{eq:fperp_wave_drive_app_mod3}
\end{align}
Since neither the KAW nor the inertial-Alfv\'en branch admits
$v_{g,\perp}(k_r)=v_\perp$ in its respective asymptotic regime, no resonant plateau is expected there.

Finally, because these wave families are tied to a guide field, an isotropic spectrum is generally not appropriate for them. The isotropic case is more naturally associated with unmagnetized electrostatic turbulence or disordered wave fields without a preferred direction, and is therefore not pursued further here.

\section{Estimation of the drive strength for solar parameters}\label{App:corona}

Here we estimate the amplitude of electric field fluctuations exerted by a coherent set of (narrow-band) waves or broad-band EM turbulence, as well as evaluate the expansion parameter $\epsilon$ used in our perturbative treatment. We check if the perturbations are weak enough so that QLT is valid, and also strong enough to overcome collisional relaxation.

\subsection{Estimate of the quasilinear expansion parameter}
\label{App:epsilon_corona}

Here we estimate the small expansion parameter controlling the validity of QLT in the coronal application. For a species $\rms$, an appropriate order-of-magnitude measure of the perturbation strength is
\begin{align}
\epsilon \approx g_\epsilon \frac{e E_{\rm drive} L}{m_e \sigma_e^2},
\qquad
g_\epsilon = \max_\rms\left\{\frac{Z T_e}{T_\rms}\right\},
\label{eq:eps_def_app_wave}
\end{align}
where $L$ is the coarse-graining scale, $E_{\rm drive}$ is the characteristic drive electric field, and $\sigma_\rms=\sqrt{k_\rmB T_\rms/m_\rms}$ is the thermal speed. For electron heating we take $L\approx \lambda_{\rmc e}=\sigma_e/\omega_{\rmc e}$, while for ion heating we take $L\approx \lambda_{\rmc i}=\sigma_i/\omega_{\rmc i}$, with $\omega_{\rmc i}=ZeB_0/m_i c$.

We estimate the wave electric field from Faraday's law, $E_{\rm drive}\approx (\omega/k_{\rm eff}c)\,\delta B$, with $k_{\rm eff}$ the component of the wavevector associated with the dominant spatial gradient of the electric field. Using the branch dispersions summarized in Appendix~\ref{App:wave_drive} and evaluating them at the appropriate electron or ion gyroscale gives the following order-of-magnitude estimates.

For the electron-scale $R$ branch,
\begin{align}
\epsilon_{{R},e}
\approx
g_\epsilon\,\frac{2}{\beta_e+2}\,\frac{\delta B}{B_0},
\label{eq:eps_R_e_app_wave_short}
\end{align}
where $\beta_e=8\pi n_e k_\rmB T_e/B_0^2$. In the low corona, $\beta_e\ll1$, so $\epsilon_{R,e}\approx g_\epsilon\,\delta B/B_0$.

For electron-scale inertial-Alfv\'en waves,
\begin{align}
\epsilon_{{\rm InA},e}
&\approx
g_\epsilon
\frac{v_\rmA}{\sigma_e}
\frac{k_\parallel}{k_\perp}
\frac{1}{\sqrt{1+2/\beta_e}}
\frac{\delta B_\perp}{B_0}
\nonumber\\
&\approx
g_\epsilon
\sqrt{\frac{m_e}{m_i}}
\frac{k_\parallel}{k_\perp}
\frac{\delta B_\perp}{B_0},
\qquad
(\beta_e\ll 1),
\label{eq:eps_IA_e_app_wave_short}
\end{align}
so inertial-Alfv\'en heating is parametrically weaker than $R$-branch heating because of the anisotropy factor $k_\parallel/k_\perp$ and the factor $\sqrt{m_e/m_i}$.

For ion-scale KAWs,
\begin{align}
\epsilon_{{\rm KAW},i}
&\approx
g_\epsilon
\frac{k_\parallel}{k_\perp}
\frac{1}{Z}
\frac{\sqrt{2T_i/(\beta_e T_e)}}
{\sqrt{\beta_i+2/(1+T_e/T_i)}}
\frac{\delta B_\perp}{B_0},
\label{eq:eps_KAW_i_app_wave_short}
\end{align}
so ion heating by KAWs is intrinsically anisotropy dependent. This is the sub-Larmor KAW estimate evaluated at the ion gyroscale, and therefore should be regarded as an order-of-magnitude approximation.

For ion-cyclotron waves,
\begin{align}
\epsilon_{{\rm IC},i}
\approx
g_\epsilon
\frac{T_i}{Z T_e}
\frac{\delta B}{B_0},
\label{eq:eps_IC_i_app_wave_short}
\end{align}
and, more generally, for a minor ion species $\rms$,
\begin{align}
\epsilon_{{\rm IC},\rms}
\approx
g_\epsilon
\frac{T_\rms}{Z T_e}
\frac{\delta B}{B_0}.
\label{eq:eps_IC_s_app_wave_short}
\end{align}
Unlike the KAW estimate, these are not suppressed by $k_\parallel/k_\perp$.

For typical low-coronal parameters, $n_e=10^8\,{\rm cm^{-3}}$, $T_e=10^6\,{\rm K}$, and $B_0=10\,{\rm G}$, one has $\beta_e\approx 3.5\times10^{-3}$ and $\omega_{\rmc e}/\omega_{\rmP e}\approx 0.31$. Hence
\begin{align}
\epsilon_{{R},e}\approx g_\epsilon\,\frac{\delta B}{B_0},
\qquad
\epsilon_{{\rm InA},e}\approx
2.3\times10^{-2}\,
g_\epsilon
\frac{k_\parallel}{k_\perp}
\frac{\delta B_\perp}{B_0},
\end{align}
while for protons with $T_i\approx T_e$ and $\beta_i\approx \beta_e$,
\begin{align}
\epsilon_{{\rm KAW},p}
\approx
24\,g_\epsilon
\frac{k_\parallel}{k_\perp}
\frac{\delta B_\perp}{B_0},
\qquad
\epsilon_{{\rm IC},p}
\approx
g_\epsilon\,\frac{\delta B}{B_0}.
\end{align}
Thus, at the level of the wave electric field, direct electron heating is most naturally provided by the electron-scale $R$ branch, while ion heating is driven by KAW or ion-cyclotron fluctuations at ion scales.

\subsection{Estimate of $E_{\rm drive}/E_{\rm int}$ and comparison with the threshold}
\label{App:Edrive_Eint}

Here we estimate the $E_{\rm drive}/E_{\rm int}$ for the different EM waves discussed above, under typical chromospheric conditions. We also compute the threshold level of $E_{\rm drive}/E_{\rm int}$ required for the wave or the turbulent drive to heat the plasma and generate a non-thermal power-law tail before it Maxwellianizes via Coulomb collisions. For this to occur, the drive diffusion coefficient has to become comparable to or exceed the BL diffusion coefficient.

\subsubsection{Estimating $E_{\rm drive}/E_{\rm int}$ for wave drive}

The internal Debye-scale electric field associated with Coulomb collisions is given by
\begin{align}
E_{\rm int}=\frac{m_e\omega_{\rmP e}\sigma_e}{e}.
\end{align}
Estimating the wave electric field from Faraday's law, $E_{\rm drive}\approx (\omega/k_{\rm eff}c)\,\delta B$, one finds for the branches of interest
\begin{align}
\left(\frac{E_{\rm drive}}{E_{\rm int}}\right)_{{R},e}
&\approx
\frac{\omega_{\rmc e}}{\omega_{\rmP e}}\,
\frac{2}{\beta_e+2}\,
\frac{\delta B}{B_0},
\nonumber\\
\left(\frac{E_{\rm drive}}{E_{\rm int}}\right)_{{\rm InA},e}
&\approx
\frac{\omega_{\rmc e}}{\omega_{\rmP e}}\,
\sqrt{\frac{m_e}{m_i}}\,
\frac{k_\parallel}{k_\perp}\,
\frac{\delta B_\perp}{B_0},
\nonumber\\
\left(\frac{E_{\rm drive}}{E_{\rm int}}\right)_{{\rm KAW},i}
&\approx
\frac{\omega_{\rmc e}}{\omega_{\rmP e}}\,
\frac{k_\parallel}{k_\perp}\,
\frac{\sqrt{2T_i/(\beta_e T_e)}}
{\sqrt{\beta_i+2/(1+T_e/T_i)}}\,
\frac{\delta B_\perp}{B_0},
\nonumber\\
\left(\frac{E_{\rm drive}}{E_{\rm int}}\right)_{{\rm IC},\rms}
&\approx
\frac{\omega_{\rmc e}}{\omega_{\rmP e}}\,
\sqrt{\frac{m_e T_\rms}{m_\rms T_e}}\,
\frac{\delta B}{B_0}.
\label{eq:Edrive_Eint_branches_short}
\end{align}
For fiducial upper-chromospheric parameters, $n_e=10^{10}\,{\rm cm^{-3}}$, $T_e\sim T_i=10^4\,{\rm K}$, and $B_0=10\,{\rm G}$, one has $\beta_e\approx \beta_i\approx 3.5\times10^{-3}$ and $\omega_{\rmc e}/\omega_{\rmP e}\approx 3.1\times10^{-2}$. Thus $(E_{\rm drive}/E_{\rm int})_{{R},e}\approx 3.1\times10^{-2}\,\delta B/B_0$, while $(E_{\rm drive}/E_{\rm int})_{{\rm InA},e}\approx 7.2\times10^{-4}(k_\parallel/k_\perp)\,\delta B_\perp/B_0$, $(E_{\rm drive}/E_{\rm int})_{{\rm KAW},i}\approx 1.2\times10^{-3}(k_\parallel/k_\perp)\,\delta B_\perp/B_0$, and, for protons, $(E_{\rm drive}/E_{\rm int})_{{\rm IC},p}\approx 7.2\times10^{-4}\,\delta B/B_0$. For fiducial low-coronal parameters, $n_e=10^8\,{\rm cm^{-3}}$, $T_e\sim T_i=10^6\,{\rm K}$, and $B_0=10\,{\rm G}$, one again has $\beta_e\approx \beta_i\approx 3.5\times10^{-3}$, but now $\omega_{\rmc e}/\omega_{\rmP e}\approx 0.31$. Hence $(E_{\rm drive}/E_{\rm int})_{{R},e}\approx 0.31\,\delta B/B_0$, while $(E_{\rm drive}/E_{\rm int})_{{\rm InA},e}\approx 7.2\times10^{-3}(k_\parallel/k_\perp)\,\delta B_\perp/B_0$, $(E_{\rm drive}/E_{\rm int})_{{\rm KAW},i}\approx 1.2\times10^{-2}(k_\parallel/k_\perp)\,\delta B_\perp/B_0$, and $(E_{\rm drive}/E_{\rm int})_{{\rm IC},p}\approx 7.2\times10^{-3}\,\delta B/B_0$.

\subsubsection{Threshold estimation for wave drive}\label{App:threshold_wave}

For coherent waves, the threshold electric field required to overcome BL diffusion is obtained by comparing the
$v$-dependent wave-drive coefficient from Appendix~\ref{App:wave_drive} with the
$v$-dependent BL coefficient,
\[
D_{\rm BL}^{(\rms)}(v)\approx \omega_{{\rm P}\rms}\sigma_\rms^2
\frac{\ln\Lambda}{\Lambda}\left(\frac{\sigma_\rms}{v}\right)^3,\quad v>\sigma_\rms
\]
which yields
\begin{align}
\left(\frac{E_{\rm drive}}{E_{\rm int}}\right)_{{\rm thr},\rms}^{\rm wave}(v)
\approx
\left[
\frac{D_{\rm BL}^{(\rms)}(v)}
{D_{\rm wave}^{(\rms)}(v;E_{\rm drive}=E_{\rm int})}
\right]^{1/2}.
\label{eq:Edrive_Eint_thr_wave_short3}
\end{align}
Here $E_{\rm drive}\equiv \sqrt{\langle E_\parallel^2\rangle}$ for nearly
parallel families and
$E_{\rm drive}\equiv \sqrt{\langle E_\perp^2\rangle}$ for nearly perpendicular
ones.

In the upper chromosphere, taking
$n_e=10^{10}\,{\rm cm^{-3}}$, $T_e=10^4\,{\rm K}$, and $B_0=10\,{\rm G}$, one has
\[
\Lambda\approx 3.3\times10^3,\quad
\ln\Lambda\approx 8.1,\quad
\sqrt{\frac{\ln\Lambda}{\Lambda}}\approx 4.96\times10^{-2},
\]
\[
\frac{\omega_{\rm c e}}{\omega_{\rm P e}}\approx 3.1\times10^{-2},\quad
\beta_e\approx 3.5\times10^{-3},\quad
\sigma_e\approx 390\,{\rm km\,s^{-1}}.
\]
For ions, we write
\[
\sigma_i=\sigma_e\left(\frac{T_i}{T_e}\right)^{1/2}\left(\frac{m_e}{m_i}\right)^{1/2},
\quad
Q_i\equiv \beta_i+\frac{2}{1+T_e/T_i}.
\]

Writing
$p_\parallel\equiv \alpha_\parallel+\delta_\parallel$ and
$p_\perp\equiv \alpha_\perp+\delta_\perp$, the corresponding thresholds become
\begin{widetext}
\begin{align}
\left(\frac{E_{\rm drive}}{E_{\rm int}}\right)_{{\rm thr},e}^{R}(v_\parallel)
&\approx
2.9\,
\left[
\frac{(d_e/\lambda_{\rmD e})^{3-p_\parallel}
-(k_{\min,\parallel}d_e)^{3-p_\parallel}}
{3-p_\parallel}
\right]^{1/2}
\left(\frac{\sigma_e}{v_\parallel}\right)^{3/2}
\left[
\frac{(1+x_-^2)^2(1-x_-^2)}
{x_-^{\,9-p_\parallel}+x_-^{\,p_\parallel-3}}
\right]^{1/2},
\label{eq:Edrive_Eint_thr_R_exact_num}
\\
\left(\frac{E_{\rm drive}}{E_{\rm int}}\right)_{{\rm thr},e}^{\rm InA}(v_\perp)
&\approx
4.8\times10^2\,
\left[
\frac{(k_{\rmc,\perp}^{\rm InA}d_e)^{3-p_\perp}
-(k_{\min,\perp}d_e)^{3-p_\perp}}
{3-p_\perp}
\right]^{1/2}
\left(\frac{\xi_\perp v_{\rm A}}{\sigma_e}\right)^{p_\perp/2-3}
\left(\frac{\sigma_e}{v_\perp}\right)^{p_\perp/2},
\quad 0<v_\perp<\xi_\perp v_{\rm A},
\label{eq:Edrive_Eint_thr_InA_num}
\\
\left(\frac{E_{\rm drive}}{E_{\rm int}}\right)_{{\rm thr},i}^{\rm IC}(v_\parallel)
&\approx
2.9\,
Z_i^{-9/4}
\left(\frac{T_i}{T_e}\right)^{1/2}
\left(\frac{m_i}{m_e}\right)^{7/4}
\left[
\frac{(d_i/\lambda_{\rmD e})^{3-p_\parallel}
-(k_{\min,\parallel}d_i)^{3-p_\parallel}}
{3-p_\parallel}
\right]^{1/2}\nonumber\\
&\quad\times
\left(\frac{v_{\rm A}}{\sigma_i}\right)^{(p_\parallel-3)/2}
\left(\frac{\sigma_i}{v_\parallel}\right)^{p_\parallel/2},
\quad 0<v_\parallel<v_{\rm A},
\label{eq:Edrive_Eint_thr_IC_num}
\\
\left(\frac{E_{\rm drive}}{E_{\rm int}}\right)_{{\rm thr},i}^{\rm KAW}(v_\perp)
&\approx
4.1\,
Z_i^{-9/4}
\left(\frac{T_i}{T_e}\right)^{1/2}
\left(\frac{m_i}{m_e}\right)^{7/4}
\left[
\frac{(k_{\rmc,\perp}^{\rm KAW}\lambda_{\rm c i})^{3-p_\perp}
-(k_{\min,\perp}\lambda_{\rm c i})^{3-p_\perp}}
{3-p_\perp}
\right]^{1/2}\nonumber\\
&\quad\times
Q_i^{(p_\perp-6)/4}
\left(\frac{\xi_\perp v_{\rm A}}{\sigma_i}\right)^{(6-p_\perp)/2}
\left(\frac{\sigma_i}{v_\perp}\right)^{6-p_\perp/2},
\quad v_\perp>\xi_\perp v_{\rm A}.
\label{eq:Edrive_Eint_thr_KAW_num}
\end{align}
\end{widetext}
Here $x_-$ is the sub-unity root of
\[
v_\parallel=\omega_{\rm c e}d_e\,\frac{x}{1+x^2},
\quad 0<x_-<1,
\]
and the second root is $x_+=x_-^{-1}>1$. The exact $R$-branch threshold above
already includes the sum of both resonant contributions. It is therefore valid
only for
\[
0<v_\parallel<\frac{\omega_{\rm c e}d_e}{2},
\]
while for $v_\parallel>\omega_{\rm c e}d_e/2$ there is no real root and hence
no resonant contribution from this one-sided ($k_\parallel>0$) $R$ branch.

Away from the narrow resonant spike at $x_\pm=1$, the $R$-branch result reduces
to a piecewise asymptotic form:
\begin{align}
\left(\frac{E_{\rm drive}}{E_{\rm int}}\right)_{{\rm thr},e}^{R}
&\approx
2.9\,
\left[
\frac{1-\left[(k_{\min,\parallel}\lambda_{\rmc e})
(\omega_{\rmc e}/\omega_{\rmP e})\right]^{p_\parallel-3}}
{(p_\parallel-3)(k_{\min,\parallel}\lambda_{\rmc e})^{p_\parallel-3}}
\right]^{1/2}\nonumber\\
&\times
\begin{cases}
\left(\dfrac{\sigma_e}{v_\parallel}\right)^{p_\parallel/2},
& 3<p_\parallel<6,\\[10pt]
\dfrac{1}{\sqrt{2}}
\left(\dfrac{\sigma_e}{v_\parallel}\right)^3,
& p_\parallel=6,\\[10pt]
\left(\dfrac{\beta_e}{2}\right)^{(p_\parallel-6)/2}
\left(\dfrac{\sigma_e}{v_\parallel}\right)^{6-p_\parallel/2},
& p_\parallel>6.
\end{cases}
\label{eq:Edrive_Eint_thr_wave_R_asym_num_lce}
\end{align}
where the branch is electron-cyclotron-dominated for $3<p_\parallel<6$,
whistler-dominated for $p_\parallel>6$, and receives comparable contributions
from both roots at $p_\parallel=6$.

The exact $R$-branch threshold is sharply reduced near the tangency
$x_-=x_+=1$, i.e. at
\[
v_\parallel=\frac{\omega_{\rm c e}d_e}{2}
=
\frac{\sigma_e}{\sqrt{2\beta_e}},
\]
where the diffusion coefficient develops its resonant spike. By contrast, the
IC, KAW, and InA branches are monotonic in their respective asymptotic regimes,
so each admits at most one resonant root, their group-velocity factors remain
order unity, and no analogous resonant reduction occurs.

For the fiducial upper-chromospheric parameters, the thresholds for KAWs,
inertial Alfv\'en waves, and ion-cyclotron waves remain quite large. Thus,
among the wave families considered here, only the $R$ branch---through
whistlers and electron-cyclotron waves, especially the narrow resonant region near $x_\pm=1$---appears capable of crossing the threshold required to overcome BL diffusion.

\subsubsection{Threshold estimation for turbulent drive}\label{App:threshold_turb}

Just as in the wave drive, the minimal turbulent electric field required to overcome BL diffusion is obtained by comparing the BL coefficient with the drive diffusion coefficient. For broad-band turbulent forcing, the direct dressed-particle diffusion coefficient typically dominates over the indirect wave-mediated one and is given by
\[
D_p^{(\rms)}(v)=\left(\frac{E_{\rm drive}}{E_{\rm int}}\right)^2
\widehat D_p^{(\rms)}(v),
\]
with 
\begin{widetext}
\begin{align}
\widehat D_p^{(\rms)}(v)
&\equiv D_p^{(\rms)}(v)\big|_{E_{\rm drive}=E_{\rm int}} =
\frac{\pi}{5}\,
Z_\rms^2\left(\frac{m_e}{m_\rms}\right)^2
\sigma_e^2\omega_{\rmP e}\times
\begin{cases}
\dfrac{1}{1-\left(k_{\min}\lambda_{\rmD e}\right)^{3-\alpha}}
\left(\dfrac{v}{\sigma_e}\right)^{\alpha-1},
& 0<\alpha<3,\\[8pt]
\dfrac{\ln(\omega_{\rmP e}t_\rmc)}
{\ln\!\big[(k_{\min}\lambda_{\rmD e})^{-1}\big]}
\left(\dfrac{v}{\sigma_e}\right)^2,
& \alpha=3,\\[10pt]
\dfrac{\left(k_{\min}\lambda_{\rmD e}\right)^{\alpha-3}}
{1-\left(k_{\min}\lambda_{\rmD e}\right)^{\alpha-3}}
\left(\omega_{\rmP e}t_\rmc\right)^{\alpha-4}
\left(\dfrac{v}{\sigma_e}\right)^{\alpha-1},
& 3<\alpha<5,\\[10pt]
\dfrac{\left(k_{\min}\lambda_{\rmD e}\right)^{\alpha-3}}
{1-\left(k_{\min}\lambda_{\rmD e}\right)^{\alpha-3}}
\left(\omega_{\rmP e}t_\rmc\right)^{\alpha-4}
\left(\dfrac{v_{\rm tr}}{\sigma_e}\right)^{\alpha-1}
\left(\dfrac{v}{v_{\rm tr}}\right)^4,
& \alpha>5,\ \sigma_e<v<v_{\rm tr},\\[10pt]
\dfrac{\left(k_{\min}\lambda_{\rmD e}\right)^{\alpha-3}}
{1-\left(k_{\min}\lambda_{\rmD e}\right)^{\alpha-3}}
\left(\omega_{\rmP e}t_\rmc\right)^{\alpha-4}
\left(\dfrac{v}{\sigma_e}\right)^{\alpha-1},
& \alpha>5,\ v_{\rm tr}<v<v_*,
\end{cases}
\label{eq:Dhat_turb}
\end{align}
\end{widetext}
the threshold is
\[
\left(\frac{E_{\rm drive}}{E_{\rm int}}\right)_{{\rm thr},\rms}^{\rm turb}(v)
=
\left[
\frac{D_{\rm BL}^{(\rms)}(v)}
{\widehat D_p^{(\rms)}(v)}
\right]^{1/2}.
\]

For electrons, using equation~(\ref{eq:Dhat_turb}) gives
\begin{widetext}
\begin{align}
\left(\frac{E_{\rm drive}}{E_{\rm int}}\right)_{{\rm thr},e}^{\rm turb}(v)
&\approx
\begin{cases}
\left[
\dfrac{5}{\pi}\dfrac{\ln\Lambda}{\Lambda}
\Bigl(1-\left(k_{\min}\lambda_{\rmD e}\right)^{3-\alpha}\Bigr)
\right]^{1/2}
\left(\dfrac{\sigma_e}{v}\right)^{(\alpha+2)/2},
& 0<\alpha<3,\\[10pt]
\left[
\dfrac{5}{\pi}\dfrac{\ln\Lambda}{\Lambda}
\dfrac{\ln[(k_{\min}\lambda_{\rmD e})^{-1}]}
{\ln(\omega_{\rmP e}t_\rmc)}
\right]^{1/2}
\left(\dfrac{\sigma_e}{v}\right)^{5/2},
& \alpha=3,\\[12pt]
\left[
\dfrac{5}{\pi}\dfrac{\ln\Lambda}{\Lambda}
\dfrac{1-\left(k_{\min}\lambda_{\rmD e}\right)^{\alpha-3}}
{\left(k_{\min}\lambda_{\rmD e}\right)^{\alpha-3}
(\omega_{\rmP e}t_\rmc)^{\alpha-4}}
\right]^{1/2}
\left(\dfrac{\sigma_e}{v}\right)^{(\alpha+2)/2},
& 3<\alpha<5,\\[12pt]
\left[
\dfrac{5}{\pi}\dfrac{\ln\Lambda}{\Lambda}
\dfrac{1-\left(k_{\min}\lambda_{\rmD e}\right)^{\alpha-3}}
{\left(k_{\min}\lambda_{\rmD e}\right)^{\alpha-3}
(\omega_{\rmP e}t_\rmc)^{\alpha-4}}
\right]^{1/2}
\left(\dfrac{\sigma_e}{v_{\rm tr}}\right)^{(\alpha-1)/2}
\left(\dfrac{v_{\rm tr}}{v}\right)^{7/2},
& \alpha>5,\ \sigma_e<v<v_{\rm tr},\\[12pt]
\left[
\dfrac{5}{\pi}\dfrac{\ln\Lambda}{\Lambda}
\dfrac{1-\left(k_{\min}\lambda_{\rmD e}\right)^{\alpha-3}}
{\left(k_{\min}\lambda_{\rmD e}\right)^{\alpha-3}
(\omega_{\rmP e}t_\rmc)^{\alpha-4}}
\right]^{1/2}
\left(\dfrac{\sigma_e}{v}\right)^{(\alpha+2)/2},
& \alpha>5,\ v_{\rm tr}<v<v_* .
\end{cases}
\label{eq:Ethr_turb_e_exact_final}
\end{align}
\end{widetext}
For $\alpha>5$, the transition speed is
$v_{\rm tr}\approx (k_{\min}t_\rmc)^{-1}[5(\alpha-3)/7(\alpha-5)]^{1/(\alpha-5)}$.

For an ion species $i$, at a fixed velocity $v\gtrsim \sigma_e$ one finds
\[
\left(\frac{E_{\rm drive}}{E_{\rm int}}\right)_{{\rm thr},i}^{\rm turb}(v)
\approx
Z_i^{-3/4}
\left(\frac{T_i}{T_e}\right)^{5/4}
\left(\frac{m_e}{m_i}\right)^{1/2}
\left(\frac{E_{\rm drive}}{E_{\rm int}}\right)_{{\rm thr},e}^{\rm turb}(v),
\]
where we used $\sigma_i/\sigma_e=(T_i/T_e)^{1/2}(m_e/m_i)^{1/2}$ and
$\omega_{{\rm P}i}/\omega_{\rmP e}=\sqrt{Z_i m_e/m_i}$ for a quasi-neutral single-ion plasma. At the ion thermal speed, however, the turbulent coefficient should be evaluated at $v\sim \sigma_e$, i.e. $D_{\rm turb}^{(i)}(\sigma_i)\approx D_{\rm turb}^{(i)}(\sigma_e)$, so
\[
\left(\frac{E_{\rm drive}}{E_{\rm int}}\right)_{{\rm thr},i}^{\rm turb}(\sigma_i)
\approx
Z_i^{-3/4}
\left(\frac{T_i}{T_e}\right)^{1/2}
\left(\frac{m_i}{m_e}\right)^{1/4}
\left(\frac{E_{\rm drive}}{E_{\rm int}}\right)_{{\rm thr},e}^{\rm turb}(\sigma_e).
\]

For upper-chromospheric estimates, take $k_{\min}=\lambda_{\rmc e}^{-1}$ and
$t_\rmc=\omega_{\rmc e}^{-1}$ so that $k_{\min}\lambda_{\rmD e}=\omega_{\rmc e}/\omega_{\rmP e}$ and $\omega_{\rmP e}t_\rmc=\omega_{\rmP e}/\omega_{\rmc e}$. This is because a turbulent power-spectrum is known to significantly steepen around $\lambda_{\rmc e}$. Then the $\alpha=3$ logarithmic factor is exactly unity, while for $\alpha>3$,
\[
\frac{\left(k_{\min}\lambda_{\rmD e}\right)^{\alpha-3}}
{1-\left(k_{\min}\lambda_{\rmD e}\right)^{\alpha-3}}
(\omega_{\rmP e}t_\rmc)^{\alpha-4}
=
\frac{\omega_{\rmc e}/\omega_{\rmP e}}
{1-\left(\omega_{\rmc e}/\omega_{\rmP e}\right)^{\alpha-3}}.
\]
With $\Lambda\simeq n_e\lambda_{\rmD e}^3$ and
$\lambda_{\rmD e}=(k_B T_e/4\pi n_e e^2)^{1/2}$, one has
\[
\Lambda \approx 1.04\times10^3
\left(\frac{T_e}{10^4\,{\rm K}}\right)^{3/2}
\left(\frac{n_e}{10^{11}\,{\rm cm}^{-3}}\right)^{-1/2},
\]
so that
\[
\left(\frac{\ln\Lambda}{\Lambda}\right)^{1/2}
\approx 8.18\times10^{-2}
\left(\frac{\ln\Lambda}{6.95}\right)^{1/2}
\left(\frac{T_e}{10^4\,{\rm K}}\right)^{-3/4}
\left(\frac{n_e}{10^{11}\,{\rm cm}^{-3}}\right)^{1/4}.
\]
Also,
\[
\frac{\omega_{\rmc e}}{\omega_{\rmP e}}
\approx 9.85\times10^{-2}
\left(\frac{B_0}{100\,{\rm G}}\right)
\left(\frac{n_e}{10^{11}\,{\rm cm}^{-3}}\right)^{-1/2}.
\]

Evaluated at the electron thermal speed $v=\sigma_e$, this gives
\begin{widetext}
\begin{align}
\left(\frac{E_{\rm drive}}{E_{\rm int}}\right)_{{\rm thr},e}^{\rm turb}(\sigma_e)
&\approx
0.06\,
\Bigl[1-\left(\omega_{\rmc e}/\omega_{\rmP e}\right)^{3-\alpha}\Bigr]^{1/2}
\left(\frac{\ln\Lambda}{6.95}\right)^{1/2}
\left(\frac{T_e}{10^4\,{\rm K}}\right)^{-3/4}
\left(\frac{n_e}{10^{10}\,{\rm cm}^{-3}}\right)^{1/4},
&& 0<\alpha<3,\\[6pt]
\left(\frac{E_{\rm drive}}{E_{\rm int}}\right)_{{\rm thr},e}^{\rm turb}(\sigma_e)
&\approx
0.06\,
\left(\frac{\ln\Lambda}{6.95}\right)^{1/2}
\left(\frac{T_e}{10^4\,{\rm K}}\right)^{-3/4}
\left(\frac{n_e}{10^{10}\,{\rm cm}^{-3}}\right)^{1/4},
&& \alpha=3,\\[6pt]
\left(\frac{E_{\rm drive}}{E_{\rm int}}\right)_{{\rm thr},e}^{\rm turb}(\sigma_e)
&\approx
0.18\,
\Bigl[1-\left(\omega_{\rmc e}/\omega_{\rmP e}\right)^{\alpha-3}\Bigr]^{1/2}
\left(\frac{\ln\Lambda}{6.95}\right)^{1/2}
\left(\frac{B_0}{100\,{\rm G}}\right)^{-1/2}
\left(\frac{T_e}{10^4\,{\rm K}}\right)^{-3/4}
\left(\frac{n_e}{10^{10}\,{\rm cm}^{-3}}\right)^{1/2},
&& 3<\alpha<5,\\[6pt]
\left(\frac{E_{\rm drive}}{E_{\rm int}}\right)_{{\rm thr},e}^{\rm turb}(\sigma_e)
&\approx
0.22\,
\left(\frac{\alpha-5}{\alpha-3}\right)^{1/2}
\Bigl[1-\left(\omega_{\rmc e}/\omega_{\rmP e}\right)^{\alpha-3}\Bigr]^{1/2}
\left(\frac{\ln\Lambda}{6.95}\right)^{1/2}
\left(\frac{B_0}{100\,{\rm G}}\right)^{-1/2}
\left(\frac{T_e}{10^4\,{\rm K}}\right)^{-3/4}
\left(\frac{n_e}{10^{10}\,{\rm cm}^{-3}}\right)^{1/2},
&& \alpha>5.
\end{align}
\end{widetext}
Since $\omega_{\rmc e}/\omega_{\rmP e}\sim 0.1$ in the upper chromosphere, the
factor $1-(\omega_{\rmc e}/\omega_{\rmP e})^{\alpha-3}$ is nearly unity for
$\alpha>3$ unless $\alpha$ is extremely close to $3$.

At the ion thermal speed, one finds
\begin{widetext}
\begin{align}
\left(\frac{E_{\rm drive}}{E_{\rm int}}\right)_{{\rm thr},i}^{\rm turb}(\sigma_i)
&\approx
0.06\,
Z_i^{-3/4}
\left(\frac{T_i}{T_e}\right)^{1/2}
\left(\frac{m_i}{m_e}\right)^{1/4}
\left(\frac{\ln\Lambda}{6.95}\right)^{1/2}
\left(\frac{T_e}{10^4\,{\rm K}}\right)^{-3/4}
\left(\frac{n_e}{10^{10}\,{\rm cm}^{-3}}\right)^{1/4},
&& 0<\alpha<3,\\[6pt]
\left(\frac{E_{\rm drive}}{E_{\rm int}}\right)_{{\rm thr},i}^{\rm turb}(\sigma_i)
&\approx
0.06\,
Z_i^{-3/4}
\left(\frac{T_i}{T_e}\right)^{1/2}
\left(\frac{m_i}{m_e}\right)^{1/4}
\left(\frac{\ln\Lambda}{6.95}\right)^{1/2}
\left(\frac{T_e}{10^4\,{\rm K}}\right)^{-3/4}
\left(\frac{n_e}{10^{10}\,{\rm cm}^{-3}}\right)^{1/4},
&& \alpha=3,\\[6pt]
\left(\frac{E_{\rm drive}}{E_{\rm int}}\right)_{{\rm thr},i}^{\rm turb}(\sigma_i)
&\approx
0.18\,
Z_i^{-3/4}
\left(\frac{T_i}{T_e}\right)^{1/2}
\left(\frac{m_i}{m_e}\right)^{1/4}
\left(\frac{\ln\Lambda}{6.95}\right)^{1/2}
\left(\frac{B_0}{100\,{\rm G}}\right)^{-1/2}
\left(\frac{T_e}{10^4\,{\rm K}}\right)^{-3/4}
\left(\frac{n_e}{10^{10}\,{\rm cm}^{-3}}\right)^{1/2},
&& 3<\alpha<5,\\[6pt]
\left(\frac{E_{\rm drive}}{E_{\rm int}}\right)_{{\rm thr},i}^{\rm turb}(\sigma_i)
&\approx
0.22\,
Z_i^{-3/4}
\left(\frac{T_i}{T_e}\right)^{1/2}
\left(\frac{m_i}{m_e}\right)^{1/4}
\left(\frac{\alpha-5}{\alpha-3}\right)^{1/2}
\left(\frac{\ln\Lambda}{6.95}\right)^{1/2}
\left(\frac{B_0}{100\,{\rm G}}\right)^{-1/2}
\left(\frac{T_e}{10^4\,{\rm K}}\right)^{-3/4}
\left(\frac{n_e}{10^{10}\,{\rm cm}^{-3}}\right)^{1/2},
&& \alpha>5.
\end{align}
\end{widetext}
For protons, $Z_i=1$ and $(m_i/m_e)^{1/4}\simeq 6.54$, so for $T_i\simeq T_e$ the ion threshold is about $6.5$ times the electron threshold at the respective thermal speeds.

\section{Quasilinear diffusion timescale}\label{App:QL_diff_timescale}

Here we compute the quasilinear diffusion time,
\begin{align}
t_{\rm diff}^{(\rms)}(v)\equiv \frac{v^2}{D_{\rm drive}^{(\rms)}(v)},
\end{align}
for wave as well as turbulent drives.

\subsection{Wave drive}

We showed earlier, that, of the various EM waves, the only ones powerful enough to beat BL relaxation and heat electrons under typical chromospheric conditions, are the R branch ones. Thus we will only compute the diffusion timescale for these waves. These waves heat electrons through resonant wave-particle interactions. The resonance condition, $\omega_\bk = k_\parallel v_\parallel$, i.e.,
\begin{align}
v_\parallel=\omega_{\rmc e}d_e\,\frac{x}{1+x^2}
\label{eq:R_res_condition}
\end{align}
where $x=k_\parallel d_e$, has two real roots for
$0<v_\parallel<\omega_{\rmc e}d_e/2$, which we denote by
$x_-<1$ and $x_+>1$, with $x_-x_+=1$.
The full diffusion coefficient is therefore the sum of the two resonant
contributions,
\begin{align}
D_{\parallel\parallel,R}^{(e)}(v_\parallel)
&\approx
\pi^2\sigma_e^2\omega_{\rmP e}
\left(\frac{E_{\rm drive}}{E_{\rm int}}\right)^2_{R,e}
\frac{(p_\parallel-3)(k_{\min,\parallel}d_e)^{p_\parallel-3}}
{1-(k_{\min,\parallel}\lambda_{\rmD e})^{p_\parallel-3}}
\left(\frac{\omega_{\rmc e}}{\omega_{\rmP e}}\right)^3 \nonumber\\
&\quad\times
\left[
\frac{x_-^{\,9-p_\parallel}}
{(1+x_-^2)^2(1-x_-^2)}
+
\frac{x_+^{\,9-p_\parallel}}
{(1+x_+^2)^2(x_+^2-1)}
\right].
\label{eq:DR_two_roots}
\end{align}
Using $x_+=x_-^{-1}$, this reduces to
\begin{align}
D_{\parallel\parallel,R}^{(e)}(v_\parallel)
&\approx
\pi^2\sigma_e^2\omega_{\rmP e}
\left(\frac{E_{\rm drive}}{E_{\rm int}}\right)^2_{R,e}
\frac{(p_\parallel-3)(k_{\min,\parallel}d_e)^{p_\parallel-3}}
{1-(k_{\min,\parallel}\lambda_{\rmD e})^{p_\parallel-3}}
\left(\frac{\omega_{\rmc e}}{\omega_{\rmP e}}\right)^3 \nonumber\\
&\quad\times
\frac{x_-^{\,9-p_\parallel}+x_-^{\,p_\parallel-3}}
{(1+x_-^2)^2(1-x_-^2)}.
\label{eq:DR_two_roots_reduced}
\end{align}
Hence the diffusion time is
\begin{align}
t_{{\rm diff},R}^{(e)}(v_\parallel)
&\equiv \frac{v_\parallel^2}{D_{\parallel\parallel,R}^{(e)}(v_\parallel)}
\nonumber\\
&\approx
\frac{2}{\pi^2\beta_e\omega_{\rmP e}}\,
\frac{1-(k_{\min,\parallel}\lambda_{\rmD e})^{p_\parallel-3}}
{(p_\parallel-3)(k_{\min,\parallel}d_e)^{p_\parallel-3}}
\left(\frac{\omega_{\rmP e}}{\omega_{\rmc e}}\right)^3
\left(\frac{E_{\rm int}}{E_{\rm drive}}\right)^2_{R,e}
\nonumber\\
&\quad\times
\frac{x_-^2(1-x_-^2)}
{x_-^{\,9-p_\parallel}+x_-^{\,p_\parallel-3}},
\qquad
0<v_\parallel<\frac{\omega_{\rmc e}d_e}{2}.
\label{eq:tR_general}
\end{align}
At the tangency $v_\parallel=\omega_{\rmc e}d_e/2=\sigma_e/\sqrt{2\beta_e}$,
the two roots merge at $x_-=x_+=1$ and the diffusion coefficient develops its
resonant spike. For $v_\parallel>\omega_{\rmc e}d_e/2$, there is no real root
for $k_\parallel>0$, so the one-sided resonant $R$-branch contribution vanishes.

For upper-chromospheric parameters, thermal electrons lie in the two-root regime
and one may evaluate the coefficient at $v_\parallel=\sigma_e$. It is convenient
to write the result in terms of $k_{\min,\parallel}\lambda_{\rmc e}$, noting that
\[
k_{\min,\parallel}d_e
=
(k_{\min,\parallel}\lambda_{\rmc e})\sqrt{\frac{2}{\beta_e}},
\qquad
k_{\min,\parallel}\lambda_{\rmD e}
=
(k_{\min,\parallel}\lambda_{\rmc e})
\left(\frac{\omega_{\rmc e}}{\omega_{\rmP e}}\right),
\]
and that the lower root at the thermal speed satisfies
\[
\frac{x_-}{1+x_-^2}=\sqrt{\frac{\beta_e}{2}},
\qquad
x_-(\sigma_e)\simeq \sqrt{\frac{\beta_e}{2}}
\quad (\beta_e\ll1).
\]
Substituting
\[
\left(\frac{E_{\rm drive}}{E_{\rm int}}\right)_{R,e}
\approx
\frac{\omega_{\rmc e}}{\omega_{\rmP e}}\,
\frac{2}{\beta_e+2}\,
\frac{\delta B}{B_0},
\]
one obtains
\begin{align}
t_{{\rm diff},R}^{(e)}(\sigma_e)
&\approx
\frac{1}{\pi^2\omega_{\rmP e}}\,
\frac{1-\left[(k_{\min,\parallel}\lambda_{\rmc e})
(\omega_{\rmc e}/\omega_{\rmP e})\right]^{p_\parallel-3}}
{(p_\parallel-3)(k_{\min,\parallel}\lambda_{\rmc e})^{p_\parallel-3}}
\left(\frac{\omega_{\rmP e}}{\omega_{\rmc e}}\right)^5\nonumber\\
&\times\left(\frac{\beta_e}{2}\right)^{(p_\parallel-3)/2}
\frac{(1+x_-^2)^2(1-x_-^2)}
{x_-^{\,9-p_\parallel}+x_-^{\,p_\parallel-3}}
\left(\frac{\beta_e+2}{2}\right)^2
\left(\frac{\delta B}{B_0}\right)^{-2}.
\label{eq:tR_sigma_general}
\end{align}

For $\beta_e\ll1$, this simplifies. At the electron thermal speed, the
electron-cyclotron contribution dominates for $3<p_\parallel<6$, the two roots
contribute comparably at $p_\parallel=6$, and the whistler contribution dominates
for $p_\parallel>6$. For the fiducial upper-chromospheric values
$n_e=10^{10}\,{\rm cm^{-3}}$, $T_e=10^4\,{\rm K}$, and $B_0=10\,{\rm G}$, one finds
\begin{align}
t_{{\rm diff},R}^{(e)}(\sigma_e)
&\approx
3.1\times10^{-4}\,{\rm s}\,
\frac{1-\left[(k_{\min,\parallel}\lambda_{\rmc e})
(\omega_{\rmc e}/\omega_{\rmP e})\right]^2}
{(k_{\min,\parallel}\lambda_{\rmc e})^2}
\left(\frac{\delta B}{B_0}\right)^{-2}\nonumber\\
&\times \left(\frac{n_e}{10^{10}\,{\rm cm^{-3}}}\right)^2
\left(\frac{B_0}{10\,{\rm G}}\right)^{-5}, \qquad \qquad p_\parallel=5,
\\[4pt]
t_{{\rm diff},R}^{(e)}(\sigma_e)
&\approx
1.0\times10^{-4}\,{\rm s}\,
\frac{1-\left[(k_{\min,\parallel}\lambda_{\rmc e})
(\omega_{\rmc e}/\omega_{\rmP e})\right]^3}
{(k_{\min,\parallel}\lambda_{\rmc e})^3}
\left(\frac{\delta B}{B_0}\right)^{-2}\nonumber\\
&\times\left(\frac{n_e}{10^{10}\,{\rm cm^{-3}}}\right)^2
\left(\frac{B_0}{10\,{\rm G}}\right)^{-5}, \qquad \qquad p_\parallel=6,
\\[4pt]
t_{{\rm diff},R}^{(e)}(\sigma_e)
&\approx
2.7\times10^{-7}\,{\rm s}\,
\frac{1-\left[(k_{\min,\parallel}\lambda_{\rmc e})
(\omega_{\rmc e}/\omega_{\rmP e})\right]^4}
{(k_{\min,\parallel}\lambda_{\rmc e})^4}
\left(\frac{\delta B}{B_0}\right)^{-2}\nonumber\\
&\times\left(\frac{n_e}{10^{10}\,{\rm cm^{-3}}}\right)^3
\left(\frac{T_e}{10^4\,{\rm K}}\right)
\left(\frac{B_0}{10\,{\rm G}}\right)^{-7}, \qquad p_\parallel=7.
\end{align}
These estimates assume a one-sided ($k_\parallel>0$) resonance and
therefore apply only below the tangency,
$v_\parallel<\omega_{\rmc e}d_e/2$.

\subsection{Turbulent drive}

Now we calculate the quasilinear diffusion time for the isotropic turbulent drive, only considering the direct dressed particle diffusion coefficient $D_p^{(\rms)}(v)$, since we showed that it typically dominates over the indirect wave-mediated one in stable and marginally stable plasmas. Using equation~(\ref{eq:Dp_subturnover_final}), for $\sigma_e<v<v_*=\omega_{\rmP e}/k_{\min}$ one finds
\begin{widetext}
\begin{align}
t_{\rm diff}^{(\rms)}(v)
&\approx
\frac{5}{\pi}\,
Z_\rms^{-2}\left(\frac{m_\rms}{m_e}\right)^2
\omega_{\rmP e}^{-1}
\left(\frac{E_{\rm int}}{E_{\rm drive}}\right)^2
\times
\begin{cases}
\Bigl[1-\left(k_{\min}\lambda_{\rmD e}\right)^{3-\alpha}\Bigr]
\left(\dfrac{v}{\sigma_e}\right)^{3-\alpha},
& 0<\alpha<3,\\[8pt]
\dfrac{\ln[(k_{\min}\lambda_{\rmD e})^{-1}]}
{\ln(\omega_{\rmP e}t_\rmc)},
& \alpha=3,\\[10pt]
\dfrac{1-\left(k_{\min}\lambda_{\rmD e}\right)^{\alpha-3}}
{\left(k_{\min}\lambda_{\rmD e}\right)^{\alpha-3}
(\omega_{\rmP e}t_\rmc)^{\alpha-4}}
\left(\dfrac{v}{\sigma_e}\right)^{3-\alpha},
& 3<\alpha<5,\\[10pt]
\dfrac{1-\left(k_{\min}\lambda_{\rmD e}\right)^{\alpha-3}}
{\left(k_{\min}\lambda_{\rmD e}\right)^{\alpha-3}
(\omega_{\rmP e}t_\rmc)^{\alpha-4}}
\left(\dfrac{\sigma_e}{v_{\rm tr}}\right)^{\alpha-5}
\left(\dfrac{\sigma_e}{v}\right)^2,
& \alpha>5,\ \sigma_e<v<v_{\rm tr},\\[10pt]
\dfrac{1-\left(k_{\min}\lambda_{\rmD e}\right)^{\alpha-3}}
{\left(k_{\min}\lambda_{\rmD e}\right)^{\alpha-3}
(\omega_{\rmP e}t_\rmc)^{\alpha-4}}
\left(\dfrac{v}{\sigma_e}\right)^{3-\alpha},
& \alpha>5,\ v_{\rm tr}<v<v_* .
\end{cases}
\label{eq:tdiff_turb_general}
\end{align}
\end{widetext}
Here the $\alpha=3$ line is the exact $\alpha\to3$ limit, and for $\alpha>5$,
$v_{\rm tr}\approx (k_{\min}t_\rmc)^{-1}[5(\alpha-3)/7(\alpha-5)]^{1/(\alpha-5)}$.

For upper-chromospheric estimates, take $k_{\min}=\lambda_{\rmc e}^{-1}$ and
$t_\rmc=\omega_{\rmc e}^{-1}$, so that
$k_{\min}\lambda_{\rmD e}=\omega_{\rmc e}/\omega_{\rmP e}$,
$\omega_{\rmP e}t_\rmc=\omega_{\rmP e}/\omega_{\rmc e}$, and
$v_{\rm tr}\approx C(\alpha)\sigma_e$ with
$C(\alpha)=[5(\alpha-3)/7(\alpha-5)]^{1/(\alpha-5)}$. The electron thermal speed therefore lies in the $v^4$ branch when $\alpha>5$.
Using
\[
\omega_{\rmP e}^{-1}\approx 5.61\times10^{-11}
\left(\frac{n_e}{10^{11}\,{\rm cm}^{-3}}\right)^{-1/2}\,{\rm s},
\]
\[
\omega_{\rmc e}^{-1}\approx 5.69\times10^{-10}
\left(\frac{B_0}{100\,{\rm G}}\right)^{-1}\,{\rm s},
\]
the diffusion time at the electron thermal speed becomes
\begin{widetext}
\begin{align}
t_{\rm diff}^{(e)}(\sigma_e)
&\approx
8.9\times10^{-11}
\Bigl[1-(\omega_{\rmc e}/\omega_{\rmP e})^{3-\alpha}\Bigr]
\left(\frac{E_{\rm int}}{E_{\rm drive}}\right)^2
\left(\frac{n_e}{10^{11}\,{\rm cm}^{-3}}\right)^{-1/2}\ {\rm s},
&& 0<\alpha<3,\\[4pt]
t_{\rm diff}^{(e)}(\sigma_e)
&\approx
8.9\times10^{-11}
\left(\frac{E_{\rm int}}{E_{\rm drive}}\right)^2
\left(\frac{n_e}{10^{11}\,{\rm cm}^{-3}}\right)^{-1/2}\ {\rm s},
&& \alpha=3,\\[4pt]
t_{\rm diff}^{(e)}(\sigma_e)
&\approx
9.0\times10^{-10}
\Bigl[1-(\omega_{\rmc e}/\omega_{\rmP e})^{\alpha-3}\Bigr]
\left(\frac{E_{\rm int}}{E_{\rm drive}}\right)^2
\left(\frac{B_0}{100\,{\rm G}}\right)^{-1}\ {\rm s},
&& 3<\alpha<5,\\[4pt]
t_{\rm diff}^{(e)}(\sigma_e)
&\approx
1.27\times10^{-9}\,
\frac{\alpha-5}{\alpha-3}
\Bigl[1-(\omega_{\rmc e}/\omega_{\rmP e})^{\alpha-3}\Bigr]
\left(\frac{E_{\rm int}}{E_{\rm drive}}\right)^2
\left(\frac{B_0}{100\,{\rm G}}\right)^{-1}\ {\rm s},
&& \alpha>5.
\end{align}
\end{widetext}
For $\alpha>3$, the factor $1-(\omega_{\rmc e}/\omega_{\rmP e})^{\alpha-3}$ is close to unity in the upper chromosphere unless $\alpha$ is extremely close to $3$.

For ions, the expression above applies at $v\gtrsim \sigma_e$. At the ion thermal speed, however, one should use
$D_{\rm drive}^{(i)}(\sigma_i)\approx D_{\rm drive}^{(i)}(\sigma_e)$, so
\[
t_{\rm diff}^{(i)}(\sigma_i)\approx
\frac{\sigma_i^2}{D_{\rm drive}^{(i)}(\sigma_e)}
=
Z_i^{-2}\left(\frac{T_i}{T_e}\right)\left(\frac{m_i}{m_e}\right)
t_{\rm diff}^{(e)}(\sigma_e).
\]
Hence
\begin{widetext}
\begin{align}
t_{\rm diff}^{(i)}(\sigma_i)
&\approx
8.9\times10^{-11}\,
Z_i^{-2}\left(\frac{T_i}{T_e}\right)\left(\frac{m_i}{m_e}\right)
\Bigl[1-(\omega_{\rmc e}/\omega_{\rmP e})^{3-\alpha}\Bigr]
\left(\frac{E_{\rm int}}{E_{\rm drive}}\right)^2
\left(\frac{n_e}{10^{11}\,{\rm cm}^{-3}}\right)^{-1/2}\ {\rm s},
&& 0<\alpha<3,\\[4pt]
t_{\rm diff}^{(i)}(\sigma_i)
&\approx
8.9\times10^{-11}\,
Z_i^{-2}\left(\frac{T_i}{T_e}\right)\left(\frac{m_i}{m_e}\right)
\left(\frac{E_{\rm int}}{E_{\rm drive}}\right)^2
\left(\frac{n_e}{10^{11}\,{\rm cm}^{-3}}\right)^{-1/2}\ {\rm s},
&& \alpha=3,\\[4pt]
t_{\rm diff}^{(i)}(\sigma_i)
&\approx
9.0\times10^{-10}\,
Z_i^{-2}\left(\frac{T_i}{T_e}\right)\left(\frac{m_i}{m_e}\right)
\Bigl[1-(\omega_{\rmc e}/\omega_{\rmP e})^{\alpha-3}\Bigr]
\left(\frac{E_{\rm int}}{E_{\rm drive}}\right)^2
\left(\frac{B_0}{100\,{\rm G}}\right)^{-1}\ {\rm s},
&& 3<\alpha<5,\\[4pt]
t_{\rm diff}^{(i)}(\sigma_i)
&\approx
1.27\times10^{-9}\,
Z_i^{-2}\left(\frac{T_i}{T_e}\right)\left(\frac{m_i}{m_e}\right)
\frac{\alpha-5}{\alpha-3}
\Bigl[1-(\omega_{\rmc e}/\omega_{\rmP e})^{\alpha-3}\Bigr]
\left(\frac{E_{\rm int}}{E_{\rm drive}}\right)^2
\left(\frac{B_0}{100\,{\rm G}}\right)^{-1}\ {\rm s},
&& \alpha>5.
\end{align}
\end{widetext}
For protons with $T_i\simeq T_e$, this is simply larger than the electron value by a factor $m_p/m_e\simeq 1836$, i.e.
$t_{\rm diff}^{(p)}(\sigma_p)\approx 1.64\times10^{-7}\,{\rm s}$ for $\alpha\le3$,
$\approx 1.66\times10^{-6}\,{\rm s}$ for $3<\alpha<5$, and
$\approx 2.33\times10^{-6}[(\alpha-5)/(\alpha-3)]\,{\rm s}$ for $\alpha>5$, all multiplied by $(E_{\rm int}/E_{\rm drive})^2$ and the same $n_e$- or $B_0$-dependent factors as above.

At threshold, by construction $D_{\rm drive}^{(\rms)}=D_{\rm BL}^{(\rms)}$, so
$t_{\rm diff}^{(\rms)}=t_{\rm BL}^{(\rms)}$. The BL time is
\[
t_{\rm BL}^{(\rms)}(v)\equiv \frac{v^2}{D_{\rm BL}^{(\rms)}(v)}
\approx \frac{\Lambda}{\omega_{{\rm P}s}\ln\Lambda}\left(\frac{v}{\sigma_\rms}\right)^5,
\]
and at the respective thermal speeds this reduces to
\[
t_{\rm BL}^{(e)}(\sigma_e)\approx
8.4\times10^{-9}
\left(\frac{6.95}{\ln\Lambda}\right)
\left(\frac{T_e}{10^4\,{\rm K}}\right)^{3/2}
\left(\frac{n_e}{10^{11}\,{\rm cm}^{-3}}\right)^{-1}\ {\rm s},
\]
and
\begin{align}
t_{\rm BL}^{(i)}(\sigma_i)&\approx
\sqrt{\frac{m_i}{Z_i m_e}}\,
t_{\rm BL}^{(e)}(\sigma_e)
\approx
3.6\times10^{-7}\,Z_i^{-1/2}
\left(\frac{m_i}{m_p}\right)^{1/2}\nonumber\\
&\times\left(\frac{6.95}{\ln\Lambda}\right)
\left(\frac{T_e}{10^4\,{\rm K}}\right)^{3/2}
\left(\frac{n_e}{10^{11}\,{\rm cm}^{-3}}\right)^{-1}\ {\rm s}.
\end{align}
Substituting the turbulent threshold field derived above into the quasilinear diffusion time reproduces these same BL times, as it should.

%\nocite{*}
\bibliographystyle{aipnum4-1}
\bibliography{references_ub}% Produces the bibliography via BibTeX.

\end{document}